\documentclass[%
 aip,
 amsmath,amssymb,
reprint, onecolumn
]{revtex4-2}

\usepackage{graphicx}
\usepackage{subcaption}
\usepackage{dcolumn}
\usepackage{bm}

\usepackage[utf8]{inputenc}
\usepackage[T1]{fontenc}
\usepackage{mathptmx}
\usepackage{etoolbox}
\usepackage{url}
\usepackage{hyperref}
\usepackage{float}
\usepackage{xcolor}
\usepackage{lmodern}
\usepackage[utf8]{inputenc}
\usepackage[T1]{fontenc}
\usepackage{tikz}
\usepackage{pgfplots}
\pgfplotsset{compat=1.18}
\usetikzlibrary{calc}
\usetikzlibrary{external}
\DeclareMathOperator{\STr}{STr}
\DeclareMathOperator{\Tr}{Tr}

\DeclareMathOperator{\Det}{Det}

\newcommand{\rowfour}[4]{\left( \begin{array}{c }
		#1 \\
		#2 \\
		#3 \\
		#4
	\end{array}\right)}

\newcommand{\columnfour}[4]{\left( \begin{array}{c c c c}
		#1 & #2 & #3 & #4
	\end{array}\right)}

\newcommand{\ii}{\mathrm{i}}

\newcommand{\e}{\mathrm{e}}

\newcommand{\bbR}{\mathbb{R}}

\makeatletter
\def\@email#1#2{%
 \endgroup
 \patchcmd{\titleblock@produce}
  {\frontmatter@RRAPformat}
  {\frontmatter@RRAPformat{\produce@RRAP{*#1\href{mailto:#2}{#2}}}\frontmatter@RRAPformat}
  {}{}
}%
\makeatother
\begin{document}

\preprint{AIP/123-QED}

\title{Wegner model in high dimension: singular continuous spectrum from\\ $\mathrm{U}(1)$ symmetry breaking and AAT self-consistency \\}

\author{Julian Arenz}
\author{Martin R. Zirnbauer}
\affiliation{Institut f\"ur Theoretische Physik, Universit\"at zu K\"oln, Z\"ulpicher Str.\ 77a, 50937 K\"oln, Germany.}

\date{\today}


\begin{abstract}
We analyze the energy-spectral decomposition of the Wegner one-orbital model in high lattice dimension using ideas from field theory and statistical mechanics. Most notably, we identify a local spectral measure of singular continuous type for the disordered model at the Anderson transition. The key mechanism that leads to a singular continuous spectrum is a novel scenario of spontaneous symmetry breaking, intimately connected to the non-compactness of the global symmetry group of the field theory. While that symmetry breaks down to a compact subgroup in the metallic phase, it spontaneously breaks down to a nilpotent subgroup at criticality. We describe the symmetry-breaking extremal states approximately, by applying the self-consistent theory of localization due to Abou-Chacra, Anderson, and Thouless. The thrust of our work is to provide a field-theoretical foundation for disorder phenomena associated with singular continuous spectra.
\end{abstract}

\maketitle

\section{Introduction}\label{sect:intro}

Disordered electrons undergo a quantum phase transition, known as the Anderson transition, from a metallic state for weak disorder to an insulating state at strong disorder. In the present paper, we consider the situation of electrons treated in the non-interacting (or mean-field) approximation. Following an influential paper \cite{GangOfFour}, it has become standard to assume for the Anderson transition a one-parameter scaling hypothesis. Support for that hypothesis came from the renormalization-group treatment of an effective field theory, the nonlinear $\sigma$-model \cite{Weg79a, Efe83}, analyzed by $(2+\epsilon)$-expansion in the weak-coupling regime near two space dimensions. According to established thought, the nonlinear $\sigma$-model continues to apply in high dimensions \cite{MF94}. There is the caveat, however, that the Anderson transition moves into the strong-disorder regime with increasing dimension, and there exists no proof of renormalizability of the nonlinear $\sigma$-model at the corresponding strong coupling. In fact, we are going to argue here that the standard scenario for Anderson transitions in high dimension needs revision.

That scenario posits that there exist two phases only: (i) the metallic state with absolutely continuous energy spectrum and spatially extended eigenfunctions, and (ii) the insulating state with pure point spectrum and localized eigenfunctions. Now some time ago, it was proposed \cite{AGKL1997, KAI2018} that there may exist a third phase, called NEE for non-ergodic extended, where the energy eigenstates are neither metallic nor insulating, but fractal. This proposal is borne out for simple models of random-matrix type \cite{KKCA2015} but has been criticized \cite{TMS2016, BHT2022, Lemarie2022, SLS2022} for the Anderson model on tree-like graphs. Indeed, there is evidence from analytical and numerical work \cite{TMS2016, BHT2022, Lemarie2022,Tik19} on the Anderson model on random graphs that all of the delocalized phase is asymptotically ergodic: while non-ergodic extended behavior is seen below the scale set by the rapidly diverging correlation volume, ergodicity is recovered beyond that scale. The critical regime is characterized by slow dynamics and strong finite-size effects \cite{MF94,Mirl21}. Two distinct types of critical behavior have been discussed in this context: (i) logarithmic multifractality and (ii) critical localization \cite{Chen24}, with related manifestations found on small-world random graphs \cite{Chen26, Mata26} and random regular graphs \cite{Tik19}. The interpretation of these phenomena in field-theoretical terms and especially a comparison to the one-parameter scaling theory of localization remain under discussion \cite{Mata26}. There does exist recent evidence that the localized phase is described by two length scales instead of just one \cite{Lemarie2022}. Within the relevant scaling regimes, however, a given observable can be governed predominantly by one of these scales. Another recent scaling analysis of the Anderson model on random regular graphs \cite{VAKS2024} found that two scaling parameters are involved, although for large system sizes one-parameter scaling is recovered.

These developments motivate a closer examination of the field-theory states underlying the transition. The present work constructs a novel family of critical states with symmetry properties distinct from those in the localized and extended phases. Their construction is compatible with single-parameter scaling on the metallic side of the transition: the critical broadening of the extremal field-theory states on the way to criticality, and the rescaling of their mixture, are governed by a single scale given by the correlation volume. (We make no claims about the number of distinct length scales on the localized side of the transition, where different observables may probe different characteristic scales.) One consequence of these novel critical states is that the disorder-averaged return probability, energy-projected to a shrinking critical spectral cluster around the mobility edge $E_c$, is predicted to have an algebraic tail $\sim t^{-1/2}$. [Here and throughout this Introduction, non-universal dimensionful prefactors are suppressed.] In contrast, the critical-localization scenario of Ref.~\cite{Tik19} predicts a logarithmically slow approach to a nonzero plateau.

Beyond the motivation from the numerical and analytical results cited above, there are two additional reasons to reconsider the critical field theory of high-dimensional Anderson transitions:

(i) It is known that a random Schr\"odinger operator (or any self-adjoint Hamiltonian for disordered electrons) in the infinite-volume limit may support three (not just two) types of spectrum: apart from absolutely continuous (ac) and pure point (pp) spectrum, there is also the possibility of singular continuous (sc) spectrum \cite{ReedSimon, DJMS94}. Now, a rather natural idea is to link the characteristic features of fractal eigenstates with those of sc spectrum and, indeed, this link has been pursued in a recent article \cite{AK2023}. One goal of our work is to throw further light on the existence or absence of singular continuous spectrum in models of Anderson type.

(ii) The original motivation for the present work was as follows. A few years ago, the senior author of this paper proposed \cite{CFT-IQHT} a conformal field theory for the scaling limit at the critical point between plateaus of the integer quantum Hall (IQH) effect. The key hypothesis underlying the conformal field theory is a novel scenario of target-space reconstruction and partial symmetry breaking. That scenario need not be restricted to the IQH transition or even two dimensions; we expect close relatives thereof to occur for a variety of Anderson transitions at strong coupling. So, by inquiring into a higher-dimensional version of that symmetry-breaking scenario we wish to provide evidence for the existence of a third type of field-theory state (sc, different from ac and pp). This state is naturally connected to fractal eigenfunctions and singular continuous spectrum of the Hamiltonian.

The plan here is to investigate a variant of the standard Anderson model, namely the Wegner $N$-orbital model \cite{Weg79b} of symmetry type $A$ in the Tenfold Way \cite{HHZ}. While previous treatments were mostly concerned with the metallic limit for large $N$, our focus is on the model for $N = 1$, i.e.\ with a single orbital per site, in the regime of strong disorder and high mobility. We shall solve that model in the approximation afforded by the self-consistency equation of Abou-Chacra, Anderson, and Thouless (AAT) \cite{AAT1973}. Exact on the Bethe lattice (an infinite regular tree), the AAT approximation is expected to capture in high dimensions the relevant physics for local observables if not for correlation functions. 

In the remainder of this introduction to a long paper, we outline the backbone of our reasoning. We start from the observation that the physical field-theory state (in the Wegner-Efetov field-theory representation) is ``mixed'', i.e., it decomposes into a family of so-called extremal Gibbs states, whenever spontaneous symmetry breaking occurs. In the random-matrix and disordered-electron community, this kind of decomposition is actually well known under a different name: it is the celebrated ``zero-mode integral'' that delivers Wigner-Dyson random-matrix universality up to the Thouless scale in the metallic phase. Technically speaking, when a global symmetry group $G$ is spontaneously broken to a subgroup $K$ (and the field-theory action contains no terms that break the $G$-symmetry explicitly), the formalism stipulates an integral, the said zero-mode integral, over the quotient space $G/K$, which is also known as the ``Goldstone manifold'' or ``saddle manifold''. This integral is the outermost or last integral to be performed in the field integral. For a fixed coset $g_0 K \in G/K$ (or in notation commonly used, a fixed matrix $Q_0$ on the Goldstone manifold), the non-constant part of the field fluctuates around $g_0 K$ (or $Q_0$) as its mean value. In a high enough space dimension and in the metallic phase, these fluctuations are infrared finite and hence do not restore the broken symmetry. By integrating over them while keeping $Q_0$ fixed, one obtains what is called an \emph{extremal} Gibbs state in the language of statistical mechanics. Key to our reasoning is the proposal that AAT self-consistency, as an equation of mean-field nonlinear type, makes sense as an approximation scheme to be used for \emph{individual} extremal Gibbs states (albeit \underline{not} for the mixed physical state which is assembled by zero-mode integration over the extremal Gibbs states).

Once this proposal is accepted, we can develop a workable scheme of controlling the critical point; the main ideas are as follows. While the special coset $eK \in G/K$ (the ``standard saddle'' $Q_0 = \Sigma_3$) is invariant under the action of $K$, a general coset $gK \in G/K$ is invariant under the deformed group $K^\prime = g K g^{-1}$. The limit of $K^\prime$ under asymptotic Lorentz boosts $g \in G$ is a nilpotent group $N$. [A basic model for such asymptotics is
\begin{equation*}
    K = \exp(\mathbb{R} \cdot \mathrm{i}\sigma_2) \to \lim_{t \to \infty} \mathrm{e}^{t \sigma_3} K \mathrm{e}^{-t \sigma_3} = \exp(\mathbb{R} \cdot \sigma_+) = N \quad \textrm{(Pauli matrices } \sigma_j), 
\end{equation*}
with $\sigma_+ = \frac{1}{2}(\sigma_1 + \mathrm{i}\sigma_2)$ as nilpotent generator.] When the Anderson-critical point is approached from the metallic phase, the extremal Gibbs states expand: they ``thicken'' as the field fluctuations increase. Upon arrival in the insulating phase on the other side of the critical point, they have all merged into a single extremal Gibbs state with unbroken $G$-symmetry, which coincides with the physical state. On the metallic side of the transition, the extremal Gibbs states that dominate the field integral have isotropy groups $K^\prime \cong K$ (cf.\ above). In order for the non-compact symmetry $G$ to be restored across the Anderson transition, statistical weight must move out to infinity on the non-compact manifold $G/K$. This means that on approaching the critical point, the isotropy groups $K^\prime$ of the statistically dominant extremal Gibbs states are pushed towards their nilpotent limits; based on the AAT approximation, we are going to argue that, right at the Anderson-critical point with suitable boundary conditions imposed, they in fact become nilpotent, $N^\prime \cong N$. We expect this mechanism to be robust and not an artifact of the AAT scheme. 

Let us emphasize that the critical extremal Gibbs states individually break a $\mathrm{U}(1)$-symmetry (cf.\ the title of the paper), whereas their incoherent mixture, the physical state, does \underline{not}. The latter state is a zero-mode integral over the $\mathrm{U}(1)$-family of deformed Goldstone manifolds $G/N$. This novel symmetry breaking pattern $G/N$ (as opposed to $G/K$) is proposed to be the universal hallmark of singular continuous (as opposed to absolutely continuous) spectrum for the Wegner model and random Schr\"odinger operators in general.

After several years of revision, our final conclusion differs from that in an early preprint: for the Wegner one-orbital model in symmetry class $A$, and assuming the AAT scheme, our field-theory scenario for singular continuous spectrum is borne out \emph{only at the critical point}. Nonetheless, we still believe that singular continuous spectrum may exist as a true thermodynamic phase for a suitable model class (possibly involving random magnetic background fields).

The organization of this paper is as follows. Sect.~\ref{sect:Wegner} introduces the basic objects: the Wegner model on a lattice $\mathbb{G}$, the retarded and advanced local Green's functions $G_{n,n}(E \pm \mathrm{i} \varepsilon)$, and the Fourier-Laplace transform $Y_{\rm FL}$ of the distribution of the reciprocal local Green's function, which is the focus of our analysis. In Sect.~\ref{Decomposition}, we sharpen the meaning of the ``physical state'' $Y_{\rm FL}$ as an incoherent mixture of extremal states. Sect.~\ref{sect:spec} introduces the retarded-advanced correlator $K_{n,\varepsilon}(E)$ as the disorder-averaged product of $G_{n,n}(E + \mathrm{i}\varepsilon)$ and $G_{n,n}(E - \mathrm{i} \varepsilon)$. By relating $K_{n,\varepsilon} = \int Y_{\rm FL}$ to the energy-projected return probability, we argue that different spectral types are distinguished by the small-$\varepsilon$ behavior of this correlator: $K_{n,\varepsilon} \sim \varepsilon^{-1}$ for pure point (pp) spectrum, $K_{n,\varepsilon} \sim 1$ for absolutely continuous (ac) spectrum, and $K_{n,\varepsilon} \sim \varepsilon^{-\alpha}$, $0 < \alpha < 1$, in the singular continuous (sc) scenario. Using super-calculus, we derive in Sect.~\ref{SubSec:AAT self-consistency} the AAT self-consistency equation, which serves as a tool to approximate the extremal states in high dimension. Sect.~\ref{sect:Gibbs} gives some field-theory background and explains how different symmetry-broken extremal states are selected by boundary conditions. In Sect.~\ref{sect:reduction} we reduce the supersymmetric (SUSY) AAT equation to a purely ``bosonic'' equation. Sect.~\ref{sect:analysis} takes a look at the geometry of the target space and its symmetry-group orbits. Imposing $\mathrm{U}(1)$-invariance, we examine the pp-type state in the insulating phase and the $\mathrm{U}(1)$-invariant metallic extremal state. We also study the full $G/K$ family of metallic extremal Gibbs states generated by Lorentz boosts. In Sect.~\ref{Harmonic analysis} we locate the pp-ac phase boundary by applying harmonic analysis to the AAT equation linearized about the pp solution. The eigenfunctions of the linear transfer operator are then used in Sect.~\ref{U(1)crit} to elucidate how the metallic extremal Gibbs states approach criticality. This analysis yields the inverse correlation-volume scale $C$ and the scaling of the Lorentz boost required to retain nontrivial critical field fluctuations in the infinite-volume limit. Sect.~\ref{sc crit} implements this scaling and constructs the critical extremal states $Y_{\rm sc}$, whose Goldstone manifold is $G/N$ with $N \subset G$ a nilpotent subgroup. Finally, in Sect.~\ref{K sc} we calculate $K_{n,\varepsilon}$ in these critical extremal Gibbs states and reconstruct the result in the corresponding physical state by performing the zero-mode integral. We thus obtain the critical scaling
\begin{equation*}
    K_{n,\varepsilon} \sim \varepsilon^{-1/2}
\end{equation*}
and an algebraic long-time tail $t^{-1/2}$ for the disorder-averaged return probability. In conjunction with the spectral analysis of Sect.~\ref{sect:spec}, this characterizes the critical local spectral measure as singular continuous. Sect.~\ref{Conclusion} summarizes our results and discusses their possible validity beyond the AAT approximation.

\section{Model and technique used}
\label{sect:Wegner}

The object studied in this paper is a random Hamiltonian, $H$, of Schr\"odinger type in the discrete setting of an infinite graph, $\mathbb{G}$, for example $\mathbb{G} = \mathbb{Z}^d$. The Hamiltonian is written as a sum
\begin{equation}\label{eq:Wegner1}
    H = \sum h_{n,n^\prime} \, c_{n^{\vphantom{\prime}}}^\dagger c_{n^\prime}^{\vphantom{\dagger}}
\end{equation}
over the sites $n$ of $\mathbb{G}$. Although $H$ is presented here in basis-independent form using particle creation and annihilation operators, we will analyze it as an operator ($h$) on the single-particle Hilbert space $\ell^2 (\mathbb{G})$. All Hamiltonian matrix elements are taken to be complex Gaussian independent random variables (subject to the Hermiticity condition $\overline{h_{n,n^\prime}} = h_{n^\prime,n}$) with zero mean and variance
\begin{equation}\label{eq:problaW}
    \mathbb{E} \left( h_{n,n^\prime} h_{l,l^\prime} \right) = \delta_{n l^\prime} \delta_{n^\prime l} \left( \delta_{n n^\prime} w_V^2 + A_{n n^\prime} w_T^2 \right) ,
\end{equation}
where $A_{n n^\prime} = A_{n n^\prime} (1 - \delta_{n n^\prime}) = A_{n^\prime n} \in \{ 0 , 1 \}$ is the adjacency matrix of the graph $\mathbb{G}$. The parameter $w_V^2$ sets the scale of variation of the on-site potentials $h_{n,n} \equiv V_n$, while $w_T^2$ sets the kinetic-energy scale for hopping between nearest-neighbor pairs. Officially known as the Wegner $N$-orbital model \cite{Weg79b} of symmetry class $A$ with $N = 1$ (i.e.\ a single orbital per site), the model so defined will be referred to simply as the ``Wegner model''. Our interest here is in the regime of strong diagonal disorder, $w_V^2 \gg w_T^2$, on a regular graph $\mathbb{G}$ with high coordination number $K+1$. 

Developing conceptual ideas and mathematical tools, the initial sections of the paper are qualitative, and we take the liberty of writing relations such as $K_{n,\varepsilon} \sim \varepsilon^{-\alpha}$ to indicate the leading singularity in the regularization ($\varepsilon \to 0$) while omitting non-universal dimensionful prefactors built from $w_T$, $w_V$, $E$, and $K$. We switch to a more precise style in the final sections, where we derive our main results for the critical behavior at the Anderson transition.

By the probability law (\ref{eq:problaW}), the ensemble of Hamiltonians (\ref{eq:Wegner1}) is invariant under local $\mathrm{U}(1)$ phase rotations
\begin{equation}
    c_n^\dagger \mapsto   c_n^\dagger \mathrm{e}^{\mathrm{i} \alpha_n} , \quad
    c_{n^\prime} \mapsto \mathrm{e}^{- \mathrm{i} \alpha_{n^\prime}}  c_{n^\prime}  .
\end{equation}
This property, while not shared by the Anderson model (where the hopping matrix elements are non-random and constant), is expected to emerge \cite{PS82} under renormalization on large length scales in the metallic regime (and, probably, beyond). Its attractive feature is that it facilitates the introduction of matrix variables.

\subsection{Local Green's function}

The analysis starts from the retarded Green's function, $G \equiv G^+$, which for any real energy $E$ is defined by
\begin{equation}\label{eq:defG+}
    \sum_n (E + \mathrm{i} \varepsilon - h)_{n_1 , n} \, {G}_{n , n_2} = \delta_{n_1 , n_2} \,,
\end{equation}
with $\varepsilon$ a positive real number shifting the Green's function poles into the lower half of the complex energy plane. In the sequel, we focus on the \emph{local} Green's function, i.e.\ the diagonal matrix element $G_n \equiv G_{n,n}$.

The goal is to find the probability distribution of the complex variable $G_n$ (for a fixed energy $E$ and coordination number $K+1$) from the given distribution of the matrix elements $h_{n,n^\prime}$, by assuming Eq.~(\ref{eq:AAT}). Of course, with enough computer power this goal could be achieved directly by numerical simulation. Our intention, however, is to develop an analytical understanding. The approach used here is to compute moments by means of the characteristic function; see Sect.~\ref{sect:FLT}. Now the Anderson transition cannot be detected by looking at moments of the retarded Green's function $G_n$ alone. Rather, critical behavior is to be found in the probability law for the \emph{imaginary} part of $1/G_n$, which makes a transition from singular (of Dirac-delta type) for localized states to smooth for extended states. In order to extract that imaginary part from disorder averages, both branches of the Green's function, retarded and advanced, must be brought into play. Let us therefore consider the matrix-valued Green's function (with $\varepsilon > 0$),
\begin{equation}\label{eq for mathcal G}
    \mathcal{G} = \begin{pmatrix} E + \mathrm{i} \varepsilon - h  & 0 \cr 0 &E - \mathrm{i}\varepsilon - h \end{pmatrix}^{-1}.
\end{equation}
We still focus on the local Green's function, $\mathcal{G}_n \equiv \mathcal{G}_{n,n}$, which is now a $2 \times 2$ matrix, albeit a rather trivial one. (We will make more substantial use of the matrix structure when we discuss the consequences of Wegner's hyperbolic symmetry \cite{Weg79a} and its spontaneous breaking.) The inverse matrix, $\mathcal{G}_n^{-1}$, has 2 real degrees of freedom, $A_n$ and $D_n$:
\begin{equation}\label{eq:strucG}
    (\mathcal{G}_{n,n})^{-1} \equiv \mathcal{G}_n^{-1} = \begin{pmatrix} A_n + \mathrm{i} D_n & 0 \cr 0 &A_n - \mathrm{i} D_n \end{pmatrix}, \quad
    A_n = \frac{\mathrm{Re}(G_n(E-\mathrm{i}\varepsilon))} {|G_n(E + \mathrm{i} \varepsilon)|^2}, \quad D_n = \frac{\mathrm{Im}(G_n(E - \mathrm{i} \varepsilon))}{|G_n(E  + \mathrm{i} \varepsilon)|^2}.
\end{equation}
Note that, while $A_n$ may take any real value, the real number $D_n$ is always positive: $D_n > 0$.

\subsection{Fourier-Laplace transform}\label{sect:FLT}

We now want the probability law of $\mathcal{G}_n^{-1}$ or, equivalently, of the random variables $A_n$ and $D_n$. As is often the case, their joint distribution is easier to handle analytically via the characteristic function or Fourier-Laplace transform. Hence we introduce the transform
\begin{equation}\label{eq:def-Y}
    Y_{\rm FL}(Q_n) := \mathbb{E} \big( \mathrm{e}^{ \mathrm{i} \mathrm{Tr}\, Q_n \mathcal{G}_n^{-1}} \big) ,
\end{equation}
where $\mathbb{E}(\bullet)$ means the expectation value with respect to the distribution of the random variables in the matrix $\mathcal{G}_n^{-1}$, and the Fourier-conjugate variable $Q_n$ is a $2 \times 2$ matrix
\begin{equation}\label{eq:rankone}
    Q_n = \begin{pmatrix} u_n \bar{u}_n &- u_n \bar{v}_n \cr v_n \bar{u}_n &- v_n \bar{v}_n \end{pmatrix} = \begin{pmatrix} u_n \cr v_n \end{pmatrix} \otimes \begin{pmatrix} \bar{u}_n &-\bar{v}_n \end{pmatrix} \qquad (u_n , v_n \in \mathbb{C})
\end{equation}
with the properties $Q_n = \sigma_3 Q_n^\dagger \sigma_3$ and $\mathrm{Det}(Q_n) = 0$. 
In combination with $D_n > 0$, the insertion of a minus sign in Eq.\ (\ref{eq:rankone}) (namely, $-\bar{v}_n$ for $\bar{v}_n$) makes the real part of $\mathrm{i} \mathrm{Tr}\, Q_n \mathcal{G}_n^{-1}$ negative, thereby ensuring that the Fourier-Laplace transform $Y_{\rm FL}$ vanishes exponentially as $|u_n|^2 + |v_n|^2 \to \infty$. We note the special value
\begin{equation}\label{eq:Yat0}
	Y_{\rm FL}(Q=0) = \mathbb{E}(\mathrm{e}^0) = 1
\end{equation}
due to normalization of the distribution for $\mathcal{G}_n^{-1}$. [In view of the simple expression
\begin{equation}\label{eq:quad-form}
    \mathrm{i} \mathrm{Tr}\, Q_n \mathcal{G}_n^{-1} = ( \mathrm{i} A_n - D_n) \vert u_n \vert^2 - (\mathrm{i} A_n + D_n) \vert v_n \vert^2,
\end{equation}
some of our readers might wonder why we introduce the non-diagonal matrix $Q_n$. The rationale for this will become transparent in the following subsection (\ref{Decomposition}) and with further details in Sects.\ \ref{SubSec:AAT self-consistency}, \ref{sect:Gibbs}, and \ref{types}.]

Next, consider the group $\mathrm{GL}(2,\mathbb{C})$ of complex linear transformations
\begin{equation}
    \begin{pmatrix} u_n \cr v_n \end{pmatrix} \mapsto g \begin{pmatrix} u_n \cr v_n \end{pmatrix} .
\end{equation}
The real subgroup that preserves the relation $Q_n = \sigma_3 Q_n^\dagger \sigma_3$ is a non-compact group of pseudo-unitary transformations:
\begin{equation}
    \mathrm{U}(1,1) = \{g \in  \mathrm{GL}(2,\mathbb{C}) \mid g^\dagger \sigma_3 g = \sigma_3 \}.
\end{equation}
The $\mathrm{U}(1)$-subgroup generated by the unit matrix acts trivially on $Q_n$ and can therefore be divided out. So, the relevant group, acting on the matrices $Q_n$ by conjugation $Q_n \mapsto g Q_n g^{-1}$, is $\mathrm{U}(1,1) / \mathrm{U}(1) \cong \mathrm{SU}(1,1) / \mathbb{Z}_2$. [We will not pay much attention to the central factor $\mathbb{Z}_2$ and will usually speak simply of $\mathrm{SU}(1,1)$.]

How does the Fourier-Laplace transform $Y_{\rm FL}$ behave for the three different types of spectral measure of the Hamiltonian $h$ (namely, pure point, absolutely continuous, or singular continuous)? To answer this question, we assume an infinite graph $\mathbb{G}$ with translation invariance, so that the distribution of the random variables $A_n$, $D_n$ becomes independent of the site index $n$. We then simplify the notation: $A_n \equiv A$ and $D_n \equiv D$. Coarsely speaking, one encounters two alternatives for the distribution of $D$ to respond to the regulator $\varepsilon$. (The distribution of $A$ carries information about the local density of states but is non-critical and not specific to the spectral type.)

(i) In the limit of $\varepsilon \to 0$, the distribution of $D$ may remain smooth with finite support on the positive real axis $D > 0$. Then the spectrum of $h$ is absolutely continuous, its eigenfunctions are extended, and the function $Y_{\rm FL}$ is invariant under symmetry transformations $Q \mapsto g Q g^{-1}$ only for $g$ restricted to the compact subgroup $\mathrm{U}(1) \subset \mathrm{SU}(1,1)$ generated by $\mathrm{i} \sigma_3$. We will always refer to this subgroup as $\mathrm{U}(1)_{\mathrm{i} \sigma_3}$.

(ii) In the same limit of $\varepsilon \to 0$, the distribution of $D$ may shrink to a Dirac-delta distribution supported at $D = 0$. In that case, the function $Y_{\rm FL}$ acquires a large symmetry,
\begin{equation}\label{eq:sym-rel-Y}
    Y_{\rm FL}(Q) = Y_{\rm FL}(g Q g^{-1}), \quad g \in \mathrm{SU}(1,1),
\end{equation}
pointwise in $Q$ for $\varepsilon\to 0$. The energy spectrum here is singular, but Eq.~(\ref{eq:sym-rel-Y}) alone does not determine the type of singularity. To distinguish between pure point and singular continuous spectrum, one has to take a closer look at the behavior of $Y_\mathrm{FL}$ (or the distribution of $D$) as a function of the regulator $\varepsilon$. We will do so in detail in Sect.~\ref{sect:spec}. 

\subsection{Decomposition of $Y_{\rm FL}$ into extremal states (rough sketch)} \label{Decomposition}

We now come to a conceptual point which is crucial to our analysis. (Meant to be expository, the present subsection is going to be somewhat cavalier about the calculus with Grassmann variables and supersymmetry. We will be clear about these important details later.)

To begin, we note that disorder averages of products of retarded and advanced local Green's functions are obtained by integrating $Y_{\rm FL}(Q_n) \, d^2 u_n \, d^2 v_n$ against polynomials (or more general functions) of $u_n$ and $v_n$. (From these averages, one extracts the spectral and local wave function observables. See Sect.~\ref{sect:spec} for details.) Hence we may regard $Y_{\rm FL}(Q_n)$ as the density function representing a distribution $\big\langle Y_{\rm FL} , \bullet \big\rangle$ which is to be paired (in the Schwartz sense) with functions of $u_n$ and $v_n$. In anticipation of the field-theory interpretation in Sect.~\ref{sect:Gibbs}, we call this distribution the physical state. We note that, by the definition (\ref{eq:def-Y}), the physical state $\big\langle Y_{\rm FL} , \bullet \big\rangle$ is always invariant under $\mathrm{U}(1)_{ \mathrm{i} \sigma_3}$.

For a disordered insulator (i.e.\ unbroken symmetry and pp spectrum), $Y_{\rm FL}$ is structureless and needs no further discussion here. In contrast, a major refinement comes into play when spontaneous symmetry breaking occurs: the infinite-volume state $\big\langle Y_{\rm FL} , \bullet \big\rangle$ then decomposes as an \emph{incoherent mixture of extremal Gibbs states}. For the metallic phase, this decomposition is well known from the effective description by a non-linear $\sigma$-model \cite{EM2008}: the symmetry group $G$ of the effective field theory for the disordered metal [where $G =\mathrm{U}(1,1|2)$ for models in symmetry class $A$] is broken to a compact subgroup $K \subset G$. The order-parameter space (in the sense of Landau) is the coset space $G/K \equiv \mathcal{X}$, and the physical state is an integral (or ``incoherent mixture'') of so-called extremal Gibbs states $Y_x$:
\begin{equation}\label{eq:YFL-Yx}
    \big\langle Y_{\rm FL} , \bullet \big\rangle = \int_\mathcal{X} \!\! dx \; \mathrm{e}^{- \varepsilon r(x)} \big\langle Y_x , \bullet \big\rangle \,,
\end{equation}
with a regulator function $r(x) \geq 0$ and regulator $\varepsilon > 0$. (Regularization is indispensable here because $G/K$ is non-compact.) Specified by one point (or super-point) $x$ on the Goldstone manifold $\mathcal{X} = G/K$, the extremal Gibbs state $\big\langle Y_x , \bullet \big\rangle$ is the ``macroscopic'' result of integrating out the (infrared finite) fluctuations of the field around $x$. If we had to summarize the present paper in a single sentence, we might say that it is an investigation into the structure of the decomposition (\ref{eq:YFL-Yx}) and its consequences beyond the metallic regime.

Now the value of the regulator function increases linearly with the system volume. Therefore, if the infinite-volume limit is taken for a fixed value of $\varepsilon$, only the state $Y_{x_0}$ with minimal $r(x_0)$ survives in the limit. However, one is often interested in the more informative limit of $\varepsilon \sim \mathrm{vol}^{-1} \to 0$. In that limit, \underline{all} extremal Gibbs states contribute to the physical state [and (\ref{eq:YFL-Yx}) becomes the celebrated zero-mode integral that delivers Wigner-Dyson random-matrix universality up to frequency scales of order $\mathrm{vol}^{1-2/d}$ in units of the energy-level spacing \cite{Efe83}].

There exist several ways to select the extremal Gibbs state $Y_x$. For one, we may turn on a constant background that favors the point $x \in G/K$. For another, we may impose boundary conditions that force the field on the spatial boundary to be centered about $x$. In either way, the decomposition (\ref{eq:YFL-Yx}) offers the following option. Instead of calculating the expectation value of an observable $A \equiv A(Q)$ directly as $\big\langle Y_{\rm FL} , A \big\rangle$, we may adopt an equivalent two-step procedure:
\begin{enumerate}
\item First, compute the expectation value $\big\langle Y_x , A \big\rangle$ in the extremal state given by $Y_x$, by carrying out the integral against the corresponding function $A(Q_n)$:
\begin{equation}
    \big\langle Y_x , A \big\rangle = \int dQ_n Y_x (Q_n) A(Q_n) , \quad dQ_n \propto d^2 u_n d^2 v_n .
\end{equation}
\item Second, take the average of $\big\langle Y_x , A \big\rangle$ with respect to the measure $\mathrm{e}^{- \varepsilon r(x)} dx$ (or a similar measure tailored to the observable $A$ of interest). 
\end{enumerate}

In the following section, \ref{sect:spec}, we will explain how we can detect the three different spectral types, especially singular continuous spectrum, directly from $\big\langle Y_{\rm FL} , A \big\rangle$ for a suitable observable $A$. Afterwards, we will employ the decomposition (\ref{eq:YFL-Yx}) (or a suitable modification thereof) to do the calculation indirectly, by way of $\big\langle Y_x , A \big\rangle$ for the extremal states $Y_x$.

\section{Spectral types and how to detect them} \label{sect:spec}

From the spectral theory of self-adjoint operators, we know that singular continuous (sc) spectrum is a feature that random discrete Schr\"odinger operators (such as the Wegner model of Sect.~\ref{sect:Wegner}) may have in the infinite-volume limit. In simple physics language, the integrated local density of states in the case of sc spectrum is a ``devil's staircase'': it is a continuous function (i.e., jumps are absent) but its derivative is singular (does not exist as a function).

Now we recall from the Introduction (Sect.~\ref{sect:intro}) a major question: sc spectrum for random Schr\"odinger operators is commonly believed to occur at critical or isolated points (if not elsewhere) in parameter space \cite{Prange83, Aizenman07, Sadel15}. In view of that hypothesis, how can we detect sc spectrum from our Fourier-Laplace transform $Y_{\rm FL}$? 

\subsection{Spectral types}\label{sect:III-A}

Let us begin by recalling a few basic facts from spectral theory. Given a diagonal matrix element $G_{n,n}$ of the Green's function, one defines the integrated local density of states (using physics parlance) by
\begin{equation}\label{eq:IlDoS}
    I_n(E) = \lim_{\varepsilon \to 0} \int_{-\infty}^E \frac{dE^\prime} {2\pi\mathrm{i}} \big( G_{n,n}(E^\prime - \mathrm{i} \varepsilon) - G_{n,n}(E^\prime + \mathrm{i} \varepsilon) \big) ,
\end{equation}
and to that function $\mathbb{R} \ni E \mapsto I_n(E)$ one associates a spectral measure $d\nu_n$ by
\begin{equation}
    I_n(E) = \int_{-\infty}^E \!\!\!\! d\nu_n \,.
\end{equation}
The functions $I_n(E)$ so defined (or equivalently, the measures $d\nu_n$) come in three different types.

\begin{enumerate}
\item {\bf ac spectrum}. $I_n(E)$ may  be absolutely continuous (ac). If so, the spectral measure $d\nu_n$ possesses a density function, $\rho_n$:
    \begin{equation}\label{eq:lDoS}
    \rho_n(E) = \frac{1}{2\pi\mathrm{i}} \lim_{\varepsilon \to 0} \big( G_{n,n}(E - \mathrm{i} \varepsilon) - G_{n,n}(E + \mathrm{i} \varepsilon) \big), \quad d\nu_n = \rho_n(E)\, dE .
    \end{equation}
    Since $I_n(E)$ is absolutely continuous, the second fundamental theorem of calculus guarantees that its derivative, $\rho_n$, is Lebesgue-integrable, $\rho_n\in L^1$. In most, if not all, physically relevant cases, $\rho_n$ has better regularity than mere $L^1$-integrability; it is often $L^2$, bounded, or even smooth. 
\item {\bf pp spectrum}. If $I_n(E)$ has discontinuities, jumping between intervals of local constancy, the spectral measure $d\nu_n$ is said to be of pure-point (pp) type. In that case, the $\rho_n$ of Eq.~(\ref{eq:lDoS}) does not exist as a function but is a sum of weighted Dirac-delta ``functions'' instead.
\item {\bf sc spectrum}. It may happen in principle that $I_n(E)$ is continuous but not absolutely continuous. The spectral measure $\mathrm{d}\nu_n$ then is concentrated on a Lebesgue zero set without charging single points (i.e., it has no ``atoms''). Here again, $\rho_n(E)$ exists only in the distributional sense. $I_n(E)$ is non-constant even though its derivative fulfills $I_n^\prime(E) = 0$ for Lebesgue-almost every $E$. An example for an $I_n$ of sc-type is the Cantor function. 

\begin{figure}[ht]
\centering
\begin{minipage}[c]{0.6\textwidth}
\includegraphics[width=0.8\textwidth]{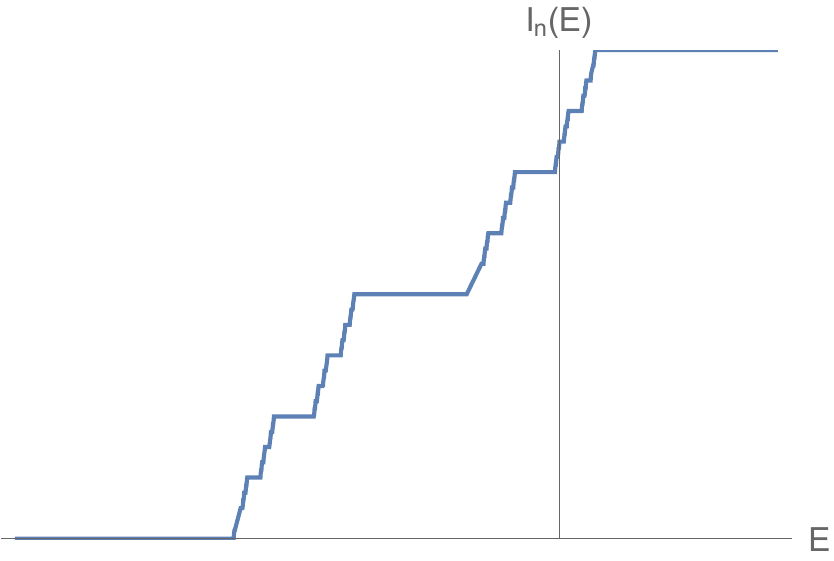}
\end{minipage}
\begin{minipage}{0.35\textwidth}
\caption{The Cantor function -- an example of an integrated local density of states of singular continuous type.}
\end{minipage}
\label{Pic Cantor}
\end{figure}
\end{enumerate}

\subsection{Spectral type from $\mathbb{E}(G^+ G^-)$} \label{sect:III-B}

The Fourier-Laplace transform $Y_{\rm FL}$ gives access to \emph{moments}, or disorder averages of products of local Green's functions. How can this information be used to detect the spectral type? A natural idea is to fix a finite value of $\varepsilon > 0$ and consider the $\varepsilon$-regularized local density of states:
\begin{equation}\label{eq:lDoS-reg}
    \rho_{n,\varepsilon} (E) = \frac{1}{2\pi\mathrm{i}} \big( G_{n,n}(E - \mathrm{i} \varepsilon) - G_{n,n}(E + \mathrm{i} \varepsilon) \big) .
\end{equation}
In order to determine the spectral type, we take the disorder average of the square of $\rho_{n, \varepsilon}$ and inspect its behavior in the limit of vanishing regularization ($\varepsilon \to 0$), as follows.

By the analyticity of the Green's function, the terms $\mathbb{E}(G^+ G^+)$ and $\mathbb{E}(G^- G^-)$ in $\mathbb{E}(\rho_{n, \varepsilon}^2)$ cannot develop singular behavior for $\varepsilon \to 0$. So we drop them and focus instead on the informative quantity
\begin{equation}
    K_{n,\varepsilon}(E) = \mathbb{E} \big( G_{n,n}(E + \mathrm{i} \varepsilon) \, G_{n,n}(E - \mathrm{i} \varepsilon) \big) \qquad (\varepsilon > 0) .
\end{equation}
The diagonal Green's function $G_{n,n}$ is related to the return amplitude $a_n(t) = \langle n | \mathrm{e}^{-\mathrm{i} t h} | n \rangle$ by
\begin{equation}
    G_{n,n} (E + \mathrm{i} \varepsilon) = - \mathrm{i} \int_0^\infty \!\!\!\!\! dt\; \mathrm{e}^{\mathrm{i} (E + \mathrm{i} \varepsilon) t} \, a_n(t) .
\end{equation}
Taking the expectation value of the square, we get
\begin{equation}\label{eq:Kn}
    K_{n,\varepsilon}(E) = \int_0^\infty  \!\!\!\!\! dt \int_0^\infty  \!\!\!\!\! dt^\prime \, \mathrm{e}^{- \varepsilon(t + t^\prime)} \, \mathrm{e}^{\mathrm{i} E(t - t^\prime)} \, \mathbb{E} \left(a_n(t) \, \overline{a_n(t^\prime)} \right) .
\end{equation}
Now in the insulating and metallic phases, where the spectral type is pure point and absolutely continuous, respectively, we can average expression (\ref{eq:Kn}) over macroscopic energy intervals without changing the $\varepsilon$-scaling of $K_{n, \varepsilon}$. Since the typical time scale in Eq.~(\ref{eq:Kn}) diverges as $t , t^\prime\sim\varepsilon^{-1}$, while the characteristic scale for energy dependence remains finite for $\varepsilon \rightarrow 0$, the energy average effectively enforces the equality of times ($t = t^\prime$).

The situation is more challenging in the case of sc spectrum. If sc-type behavior of the local spectral measure occurs only locally at the mobility edge $E_c$, a macroscopic energy average is obviously inappropriate. Here we are helped by the following observation. For the scenario of a fractal singular continuous spectrum at a fixed energy $E = E_c$ considered here, the spectral measure of the interval $[E_c - \varepsilon , E_c + \varepsilon]$ scales with a fractal exponent:
\begin{equation}
   \nu_n([E_c - \varepsilon , E_c + \varepsilon]) \sim \varepsilon^{1  - \alpha} , \quad \alpha \in (0,1).
\end{equation}
A simple consideration then shows that the effective width of the corresponding $\varepsilon$-regularized spectral cluster scales for $\varepsilon \to 0$ as $\eta \sim \varepsilon^\alpha \gg \varepsilon$. In view of that, we shrink the macroscopic energy average in (\ref{eq:Kn}) to an interval $[E_c - \eta , E_c + \eta]$. Averaging $\mathrm{e}^{\mathrm{i} E(t - t^\prime)}$ over that energy interval we get decay in $|t-t^\prime|$ outside a window of size $\eta^{-1} \sim \varepsilon^{-\alpha}$, much smaller than the typical time scale $\varepsilon^{-1}$. So, the energy average effectively filters out $t \approx t^\prime$ in the present case too. 

In summary, for all spectral types (pp, ac, and sc) we have a qualitatively correct relation,
\begin{equation}\label{eq:K and P}
    \frac{\Delta E}{2\pi} K_{n, \varepsilon}(E) \approx \int_0^\infty \!\!\!\!\! dt \; \mathrm{e}^{- 2 \varepsilon t} \,  \mathbb{E} \left( \mathrm{Prob}_{n \rightarrow n}^{E,\Delta E} (t) \right) ,\quad \mathrm{Prob}_{n\to n}^{E,\Delta E} (t) = \Big\vert \left\langle n \mid \mathrm{e}^{- \mathrm{i} t h} \chi_{E,\Delta E}(h) \mid n \right\rangle \Big\vert^2 , 
\end{equation}
with $\chi_{E,\Delta E}$ the characteristic function of the interval of width $\Delta E$ centered at $E$. This expression reveals the dynamical meaning of $K_{n, \varepsilon}$: for a wave packet initialized at the site $n$, $K_{n, \varepsilon}$ is the time-integrated and disorder-averaged return probability (modulo the issue of projecting onto and averaging over an energy interval), regularized by a death rate of $2 \varepsilon $. Informed by formula (\ref{eq:K and P}), we proceed to discuss the different behaviors of $K_{n, \varepsilon}$ for small $\varepsilon$.
\begin{enumerate}
    \item In the metallic phase, the spectral type is ac, and the eigenfunctions of $h$ are extended. The return probability decreases (due to diffusive motion away from the starting point $n$) fast enough in order for the time integral in (\ref{eq:K and P}) to converge. Correspondingly, $K_{n, \varepsilon}$ stays finite in the limit of $\varepsilon \to 0$. In more mathematical terms, we recall that  the local density of states for the case of ac spectrum is (guaranteed to be) an $L^1$-function. According to the Riemann-Lebesgue lemma, its Fourier transform vanishes at infinity. By invoking the spectral representation, 
    \begin{equation}
        a_n(t) \equiv \langle n \mid \mathrm{e}^{-\mathrm{i} t h} \chi_{E,\Delta E}(h) \mid n \rangle = \int_{\mathbb{R}}dE^\prime \rho_n(E^\prime) \, \mathrm{e}^{ -\mathrm{i} t E^\prime} \chi_{E,\Delta E} (E^\prime) ,
    \end{equation}
    we see that the Fourier transform coincides with the return amplitude $a_n(t)$. It then follows that also its squared absolute value, the return probability, vanishes in the large time limit:
    \begin{equation}
        \lim_{t \rightarrow \infty} \, \mathrm{Prob}_{n\rightarrow n}^{E,\Delta E} (t) = 0 .
    \end{equation}
    Furthermore, on physical grounds for the metallic phase, one expects the local density of states to be an $L^2$-function (actually, a smooth bounded function). Then, the Plancherel theorem guarantees that also the Fourier transform, $a_n(t)$, is an $L^2$-function, which means that the time-integrated return probability stays finite for infinite times:
    \begin{equation}\label{typical return prob ac}
        \lim_{T \rightarrow \infty} \int_0^T \!\!\!\!\! dt \; \mathrm{Prob}_{n \rightarrow n}^{E, \Delta E} (t) = \lim_{T \rightarrow \infty} \int_0^T \!\!\!\!\! dt \; | a_n(t) |^2 < \infty.
    \end{equation}
    Since the metallic local density of states consists of broad peaks extending over finite energy intervals, a typical realization of the underlying random Hamiltonian has a finite, nonzero local density of states at the observation energy $E\in\Delta E$. Extreme outliers, where the local density of states is zero or atypically large, are negligibly rare. Disorder averages over expressions such as the time-integrated return probability in Eq.~(\ref{typical return prob ac}) are therefore governed by the typical events where the ac local density of states is nonzero for most of the energies $E\in \Delta E$. For these events, we have just shown that the time-integrated return probability is finite. It follows that
    \begin{equation}
        \lim_{T \rightarrow \infty} \int_0^T \!\!\!\!\! dt \; \mathbb{E} \left( \mathrm{Prob}_{n \rightarrow n}^{E,\Delta E}(t) \right)\sim \lim_{T \rightarrow \infty} \int_0^T \!\!\!\!\! dt \; \big( \mathrm{Prob}_{n \rightarrow n}^{E, \Delta E} (t) \big)_\mathrm{typical} < \infty .
    \end{equation}
    \item In the case of pp spectrum, the eigenfunctions of $h$ are localized and the local density of states is a sum of weighted Dirac-delta functions. Consequently, the return probability at time $t$ coincides with the almost-periodic sum over the atoms of the local spectral measure:
    \begin{equation}\label{pp prob}
        \mathrm{Prob}_{n\rightarrow n}^{E,\Delta E} (t) = \left| \sum_k \nu_n(\{E_k\})\, \mathrm{e}^{-\mathrm{i} t E_k} \right|^2,
    \end{equation}
    where $E_k$ are the discrete energy eigenvalues in the window $\Delta E$. Indeed, the spectral representation lets us express the return amplitude $a_n(t)$ in terms of the pp-type spectral measure 
    \begin{equation}
        d\nu_n = \left( \sum_k \nu_n(\{E_k\}) \, \delta(E-E_k) \right) dE .
    \end{equation}
    {}From this expression we obtain
    \begin{equation}
        a_n(t) = \sum_k \nu_n(\{E_k\}) \, \mathrm{e}^{ - \mathrm{i} t E_k} ,
    \end{equation}
    and end up with the result (\ref{pp prob}) for the return probability. While the return probability may hit zero at certain points in time, its time average is bounded from below by a positive constant:
    \begin{equation}\label{t inf prob pp}
        \lim_{T \rightarrow \infty} \frac{1}{T} \int_0^T \!\!\!\!\! dt \; \mathrm{Prob}_{n \rightarrow n}^{E, \Delta E} (t) = \sum_k \big| \nu_n(\{E_k\}) \big|^2 > 0 .
    \end{equation}
    {}From Eq.~(\ref{t inf prob pp}), one quickly deduces that the $\varepsilon$-regularized time-integrated return probability diverges as $1/\varepsilon$,
    \begin{equation}\label{pp scaling}
        \int_0^\infty \!\!\!\!\! dt \; \mathrm{e}^{-2\varepsilon t} \, \mathrm{Prob}_{n \rightarrow n}^{E,\Delta E} (t) \sim \varepsilon^{-1} ,
    \end{equation}
    for $\varepsilon \to 0$. This scaling holds almost surely for any realization of the disorder. Finally, to arrive at an expression for the disorder average $K_{n, \varepsilon}(E)$, we observe the following. At a fixed observation energy $E$, the disorder average is supported by the rare realizations for which an atomic peak of the local density of states falls into the $\varepsilon$-resolution window centered at $E$. When we take the average over the energy interval of width $\Delta E$, almost every realization features a large number of such peaks. Hence the $\varepsilon$-scaling of $K_{n, \varepsilon}(E)$ evaluated at an energy $E$ in the pp component of the spectrum, coincides with that of (\ref{pp scaling}): $K_{n,\varepsilon}(E) \sim \varepsilon^{-1}$.
    \item In the sc scenario considered here, the (generalized) eigenfunctions are neither localized nor space-filling, but fractal. The temporal dynamics depopulates the initial site $n$, but fails to do so quickly enough for the time integral (\ref{eq:K and P}) to remain convergent in the limit of $\varepsilon \to 0$. The ensuing singular behavior of $K_{n,\varepsilon}$ is expected to be intermediate between $K_{n, \varepsilon} \sim \mathrm{const}$ (ac spectrum) and $K_{n, \varepsilon} \sim \varepsilon^{-1}$ (pp spectrum). In more quantitative terms, when the spectral type in the energy window $\Delta E$ is sc, a plausible scenario for the disorder average $K_{n,\varepsilon}$ is that it diverges with a fractal exponent: $K_{n, \varepsilon} \sim \varepsilon^{ -\alpha}$, $\alpha\in (0,1)$. [Of course, there exist many other ways of interpolating between $K_{n, \varepsilon} \sim \mathrm{const}$ and $K_{n, \varepsilon} \sim \varepsilon^{-1}$, but in the present work we will actually find a fractal scaling $K_{n,\varepsilon} \sim \varepsilon^{-1/2}$ at criticality.] 
\end{enumerate}

Can we be sure that a critical scaling such as $K_{n,\varepsilon} \sim \varepsilon^{-\alpha}$ ($0 < \alpha < 1$) is incompatible with the pp spectrum of an Anderson insulator and the ac spectrum of a disordered metal? Yes, we can! Although one might object that an ac density with no more than $L^1$-regularity could mimic the same small $\varepsilon$-scaling, this alternative is excluded by continuity across the Anderson transition: as the critical point is approached from the metallic side, the distribution of the imaginary part $D \geq 0$ of the reciprocal Green's function for $\varepsilon\to 0$ becomes progressively narrower and concentrates at zero. [Recall from Sect.~\ref{sect:FLT} that the distribution of $D$ in the insulating phase (pp spectrum) is a Dirac-delta distribution at zero, $\delta_0$.] If the Anderson transition is continuous (as we shall assume), the limit distribution of $D$ must collapse to $\delta_0$ on approaching criticality, whereas an ac-critical state would need the remnant of a finite component $D > 0$. Therefore, an intermediate scaling $K_{n,\varepsilon} \sim \varepsilon^{-\alpha}$ is to be identified with singular continuous spectrum. 

Assuming the fractal scaling $K_{n, \varepsilon} \sim \varepsilon^{-\alpha}$ ($0 < \alpha < 1$) and switching back to the time domain by Eq.~(\ref{eq:K and P}), we see that the time- and disorder-averaged return probability for large times $T \sim \varepsilon^{-1}$ (the typical time scale set by the death rate) scales as follows:
\begin{equation}\label{sc time avrg}
   \frac{2\pi}{T \Delta E} \int_0^T \!\!\!\!\! dt \; \mathbb{E} \left( \mathrm{Prob}_{n \rightarrow n}^{E, \Delta E}(t) \right) \sim  T^{\alpha-1}.
\end{equation}
The disorder-averaged return probability itself then has an algebraic long-time tail with fractal exponent:
\begin{equation}\label{eq:frac-tail}
    \frac{2\pi}{\Delta E}\,\mathbb{E} \left( \mathrm{Prob}_{n \rightarrow n}^{E, \Delta E} (t) \right) \sim  t^{\alpha - 1} \quad (\eta^{-1} \ll t \sim \varepsilon^{-1} \rightarrow \infty).
\end{equation}
The difference to the pp scenario is now very transparent: the disorder-averaged return probability in the pp case has a plateau: $(2\pi/\Delta E)\,\mathbb{E} \left( \mathrm{Prob}_{n \rightarrow n}^{E, \Delta E}(t) \right) \sim {\rm const}$ at large times $t\sim \varepsilon^{-1}$, before it decays to zero exponentially in the presence of the death rate. In contrast, in the sc case it has the fractal long-time tail (\ref{eq:frac-tail}) until the death rate kicks in. [The difference to the metallic ac case is equally clear: there, the disorder-averaged return probability has already decayed to zero at large times.]

In the final part \ref{sect:III-C} of this section on spectral theory, we relate the disorder-averaged return probability to the Fourier-Laplace transform $Y_{\rm FL}$. This will deliver a rather clear picture of how $Y_{\rm FL}$ behaves for the three spectral types, and especially of the difference between the pp and the sc cases --- an aspect that remained elusive in the coarse characterization of Sect.~\ref{sect:FLT}.

\subsection{Relating the return probability to $Y_{\rm FL}(Q)$} \label{sect:III-C}

Recalling Eq.~(\ref{eq:quad-form}) and the definition (\ref{eq:def-Y}) of $Y_{\rm FL}(Q)$, we see that 
\begin{equation}\label{eq:K-from-Y}
    K_{n,\varepsilon} = \int dQ \; Y_{\rm FL}(Q), \quad dQ = \pi^{-2} d^2u \, d^2v ,
\end{equation}
by straightforward Gaussian integration. Thus, the disorder average $K_{n, \varepsilon}(E) = \mathbb{E} \big( G_{n,n} (E + \mathrm{i} \varepsilon)\, G_{n,n}(E - \mathrm{i} \varepsilon) \big)$ is simply the normalization integral of the Fourier-Laplace transform $Y_{\rm FL}(Q)$. 

Given the dynamical picture (\ref{eq:K and P}) of $K_{n, \varepsilon}$, the formula (\ref{eq:K-from-Y}) suggests a physical interpretation for $Q$ and $Y_{\rm FL}(Q)$: the matrix $Q$ is something like a positive time [in fact, converting energies $E$ to frequencies $\Omega$ by $E = \hbar \Omega$, we see that the matrix elements of $Q$ carry the physical dimension of time by their appearance in Eq.~(\ref{eq:def-Y})], and $Y_{\rm FL}(Q)$ (or rather, a certain marginal thereof) is the probability for return ($n \to n$) of the time-evolved wave packet after a lapse of ``time'' $Q$. (We note in passing that this interpretation via time dynamics is perfectly understood \cite{ST15,BHS21} for the vertex-reinforced jump process and its field-theory equivalent, the nonlinear $\sigma$-model with target space $\mathrm{H}^{2 \vert 2}$). Let us explain this interpretation in more detail. By again invoking the definition (\ref{eq:def-Y}), we see that $Y_{\rm FL}(Q)$ depends on $Q$ only through the combinations $x_0 = (|u|^2 - |v|^2)/2$ and $x_3 = (|u|^2 + |v|^2) /2$. Defining the ``radial'' marginal
\begin{equation}
    dx_3\, M_{\rm FL}(x_3) = dx_3 \int dQ\, Y_{\rm FL}(Q)\, \delta \left( x_3 - \frac{1}{2} (|u|^2 + |v|^2) \right) ,
\end{equation}
we obtain
\begin{equation}\label{eq: M and P}
    K_{n,\varepsilon} = \int_0^\infty \!\!\!\!\!\! dx_3\, M_{\rm FL}(x_3) \approx \frac{2\pi}{\Delta E} \int_0^\infty \!\!\!\!\!\! dt \; \mathrm{e}^{-2 \varepsilon t}\, \mathbb{E} \left( \mathrm{Prob}_{n\rightarrow n}^{E,\Delta E}(t)\right).
\end{equation}
It should be noted that the reduction from $Y_{\rm FL}(Q)$ to the marginal $M_{\rm FL}(x_3)$ leaves the scaling behavior unchanged, as the integral over the variable $x_0$ is controlled by the finite and non-critical variance of the random variable $A_n$.

The second integral representation of $K_{n, \varepsilon}$ in Eq.~(\ref{eq: M and P}) is governed, for $\varepsilon\to 0$, by the divergent typical scale $t \sim \varepsilon^{-1}$. Since the approximate equality to the first integral representation forces an identification $x_3 \sim t$ in the asymptotic regime of large $t \sim \varepsilon^{-1} \to \infty$, it follows that the asymptotic scaling of the radial marginal of $Y_{\rm FL}(Q) \, dQ$ agrees with that of the disorder-averaged return probability. In other words, everything that we learned in Sect.~\ref{sect:III-B} about spectral types and the distinctive large-time behavior of the disorder-averaged return probability, carries over to the asymptotic behavior ($x_3\sim \varepsilon^{-1}$) of the Fourier-Laplace transform $Y_{\rm FL}$. 

Let us then summarize and further develop the expected consequences for $Y_{\rm FL}$ and $M_{\rm FL}$. First of all, we recall from Eq.~(\ref{eq:Yat0}) that we always have $Y_{\rm FL}{\rm }(Q = 0) = 1$, as befits a return probability when no time has elapsed. Now, as we go away from the time origin $Q = 0$, we encounter three distinct cases:
\begin{enumerate}
\item In the metallic (ac) phase, the return probability, and hence the marginal $M_{\rm FL}(x_3)$, decreases to zero by diffusion away from the initial point $n$. If the space dimension exceeds two, diffusive decay leads to a finite integrated return probability, which is in keeping with the appearance of a finite $D_n = \mathrm{Im} \big( G_n (E + \mathrm{i} \varepsilon)^{-1} \big) > 0$ in the limit of $\varepsilon \to 0$, and is reflected in a spontaneous breaking of the non-compact symmetry $\mathrm{SU}(1,1)$ for $Y_{\rm FL}(Q)$.
\item In the insulating (pp) phase, the disorder-averaged return probability has a plateau of constancy all the way up to the divergent time scale set by the death rate, $t \sim \varepsilon^{-1} \to \infty$. The marginal $M_{\rm FL}(x_3)$ likewise is constant up to the scale $x_3 \sim \varepsilon^{-1}$. The function $Y_{\rm FL}(Q)$ restores the symmetry (\ref{eq:sym-rel-Y}) for $\varepsilon\to 0$, and $\int dQ\; Y_{\rm FL}(Q) \sim \varepsilon^{-1}$. 
\item Turning to the case of sc spectrum, we assume the scaling $K_{n,\varepsilon} \sim \varepsilon^{-\alpha}$ in the frequency domain. (Recall that this is a hypothesis that has to be tested in concrete cases.) The disorder-averaged return probability then has a fractal asymptotic tail $\sim t^{\alpha-1}$ cut off at $t \sim \varepsilon^{-1}$ by the $\varepsilon$-regularization; cf.\ (\ref{eq:frac-tail}). Correspondingly, the large-$x_3$ asymptotics of $M_{\rm FL}(x_3)$ inherits the same fractal power law close to the regulator scale $x_3 \sim \varepsilon^{-1}$. What happens well below that scale? By combining the definition of $Y_{\rm FL}(Q)$ with the expected behavior of $D_n$ (Sect.~\ref{sect:FLT}), we do know that restoration of the symmetry (\ref{eq:sym-rel-Y}) occurs up to a scale given by the inverse characteristic width of the $\varepsilon$-regularized distribution of $D_n$, but how does this width scale with $\varepsilon$? To answer this question, we start from the assumed scaling $K_{n, \varepsilon} \sim \varepsilon^{-\alpha}$ and observe that
\begin{equation}\label{K and A,D}
    K_{n,\varepsilon} = \mathbb{E}\big( |G_n(E + \mathrm{i} \varepsilon)|^2 \big) = \mathbb{E}\left(\frac{1}{A_n^2 + D_n^2}\right).
\end{equation}
Because the distribution of $A_n$ is non-critical and broad while that of $D_n$ is critical and narrow, we can integrate out $A_n$ to obtain
\begin{equation}\label{K and D}
  \varepsilon^{-\alpha} \sim K_{n,\varepsilon} \sim \mathbb{E} \left( \frac{1}{A_n^2 + D_n^2}\right) \sim \mathbb{E} \big( D_n^{-1} \big) .
\end{equation}
Deviating for once from our qualitative style, let us remark here that a more quantitative version of the last relation replaces $\mathbb{E} (D_n^{-1})$ by $\overline{\rho_n(E)}\, \mathbb{E}(D_n^{-1})$ with $\overline{\rho_n(E)}$ the mean local density of states. This is not specific to a singular spectral type but applies whenever the characteristic scale of $D_n$ is small compared to the scale of variation of the $A_n$-distribution. In particular, it applies to metallic states approaching criticality, where $D_n$ remains positive but becomes distributed very narrowly.

Now since the distribution of $D_n > 0$ is dominated by small values, we may assume that $\mathbb{E}(D_n^{-1}) \approx \mathbb{E}(D_n)^{-1}$. Thus, $d_\varepsilon \equiv \mathbb{E}(D_n)$ scales as $d_\varepsilon \sim \varepsilon^\alpha$. Finally, since $D_n$ enters $Y_{\rm FL}$ through the combination $x_3 D_n$, the characteristic width $d_\varepsilon \sim \varepsilon^\alpha$ of the $D_n$-distribution Laplace-transforms to the $x_3$-scale $x_3 \sim \varepsilon^{- \alpha}$. By symmetry restoration (as $d_\varepsilon \sim \varepsilon^\alpha \to 0$), it follows that $M_{\rm FL}(x_3)$ becomes approximately constant for $x_3 \ll \varepsilon^{-\alpha}$. Around the scale $x_3 \sim \varepsilon^{- \alpha}$, crossover to the asymptotic fractal tail sets in. By a short computation, one verifies the consistency of this scenario with the assumed $\varepsilon$-scaling for the integral of $M_{\rm FL}(x_3)$:
\begin{equation}
   \int_0^\infty \!\!\!\!\!\! d x_3 \, M_{\rm FL}(x_3) \sim 
   \int_0^{\varepsilon^{-\alpha}} \!\!\!\!\!\!\!\!\!\!  dx_3 + 
   \int_{\varepsilon^{-\alpha}}^{\varepsilon^{-1}} \!\!\!\!\!\!\!\!\!\! 
   dx_3 \; x_3^{\alpha-1}\sim  \varepsilon^{-\alpha}.
\end{equation}
The two scales have a direct interpretation: on short time scales $x_3 \ll \varepsilon^{-\alpha}$, one only sees the gross structure of the $D_n$-distribution collapsing to $\delta_0$, whereas longer time scales $x_3 \gtrsim \varepsilon^{-\alpha}$ resolve the fine structure near $D_n = 0$, revealing the fractal tail before it ultimately gets cut off by the regulator.
\end{enumerate}

In summary, we can now complete the classification of Sect.~\ref{sect:FLT} by pinpointing the difference between $Y_{\rm FL, pp}$ and $Y_{\rm FL, sc}$. The $\mathrm{SU}(1,1)$-symmetry is restored (pointwise in $Q$) for vanishing death rate $\varepsilon\to 0$ in both cases, but the radial marginal $M_{\rm FL, pp}(x_3)$ has a plateau all the way up to the typical survival time $x_3 \sim \varepsilon^{-1} \to \infty$, whereas $M_{\rm FL, sc}(x_3)$ restores the symmetry by constancy only up to the scale $x_3 \sim \varepsilon^{-\alpha}$ and features a fractal tail from there to $x_3 \sim \varepsilon^{-1}$.

This finishes the present section on spectral theory. We have explained why the observable $K_{n,\varepsilon}$ is suitable as a probe to detect the spectral types. So, our goal from here on is to calculate $K_{n,\varepsilon}$ in different parameter regimes. We will do so indirectly by invoking a decomposition of the kind of (\ref{eq:YFL-Yx}). The corresponding properties of $Y_{\rm FL}$ then follow from the discussion in the present section. As a consistency check, we will verify the presence of the fractal tail in the disorder-averaged return probability for the sc case at the end of Sect.~\ref{K sc}.
 
\section{AAT self-consistency equation for $Y$}\label{SubSec:AAT self-consistency}

Our strategy now is to compute $K_{n,\varepsilon}(E) = \mathbb{E} \big( G_{n,n}(E + \mathrm{i} \varepsilon) \, G_{n,n}(E - \mathrm{i} \varepsilon) \big)$ from the extremal states $Y_x$ of Eq.\ (\ref{eq:YFL-Yx}), for the Wegner model in various parameter regimes and especially at criticality. Ideally, we would do so by a full-fledged field-theory calculation. Leaving that for future work, we shall simplify our task here by assuming an approximation scheme due to Abou-Chacra, Anderson and Thouless (AAT) \cite{AAT1973}. Exact on a tree-like graph, the AAT scheme for a $d$-dimensional lattice such as $\mathbb{Z}^d$ makes the approximation of neglecting (electron) loops. Because loop effects become weaker with increasing dimension $d$, the scheme can be expected to capture the qualitative physics in high dimension. (Here a word of caution is in order. By lacking convexity, our field-theory target space $C^3$, Sect.~\ref{sect:cone-C3}, is unstable under renormalization by coarse graining; yet, the presence of loops does impact the renormalization group flow. The AAT omission of loops may therefore have a non-negligible influence even in high dimension. However, since our symmetry-breaking scenario is based on a robust mechanism mostly due to the non-compactness of the symmetry group, we expect these changes to be more quantitative than qualitative.)

To keep the notation as simple as possible, we will abbreviate $Y_x \equiv Y$ (omitting the subscript for the point $x$ in the order-parameter space $X$), with the understanding that $Y$ always means an extremal state. We must also warn the reader that the same symbol $Y$ will sometimes stand for the exact state and sometimes for its AAT approximation; in instances where the meaning may not be clear from the context, we will specify it parenthetically.

We start from the retarded Green's function $G\equiv G^+$. By basic linear algebra, the reciprocal of the diagonal matrix element ${G}_{n,n} \equiv {G}_n$ can be expressed as
\begin{equation}
    \frac{1}{{G}_n} = E + \mathrm{i}\varepsilon - h_{n,n} - \sum_{n_1 , n_2} h_{n, n_1} {G}_{n_1 , n_2}^\prime h_{n_2 , n} \,,
\end{equation}
where ${G}^\prime$ is the retarded Green's function of $h$ for the graph $\mathbb{G}$ with the distinguished site $n$ omitted.

We now adopt the AAT approximation scheme \cite{AAT1973}: all off-diagonal terms $n_1 \not= n_2$ (and also one of the terms $n_1 = n_2$) are omitted from the sum on the right-hand side, and one assumes that the law of the random variables ${G}_{n_1 = n_2}^\prime$ in the resulting expression can be approximated by that of ${G}_n$. The result of these approximations is the AAT self-consistency equation:
\begin{equation}\label{eq:AAT}
    \frac{1}{{G}_n}\stackrel{!}{=} E + \mathrm{i}\varepsilon - h_{n,n} - \sum_{n^\prime (\not= n)} |h_{n, n^\prime}|^2 {G}_{n^\prime}.
\end{equation}
The notation $\stackrel{!}{=}$ is to say that both sides, as random variables, obey the same law in probability. The sum over $n^\prime$ is restricted to run over $K$ terms (for graph coordination number $K+1$) as part of the approximation scheme.

The AAT self-consistency approximation (\ref{eq:AAT}) naturally extends to the matrix-valued Green's function $\mathcal{G}$ as
\begin{equation}\label{eq:AATprime}
    \mathcal{G}_n^{-1} \stackrel{!}{=} \begin{pmatrix} E + \mathrm{i}\varepsilon - h_{n,n} & 0 \cr 0 &E - \mathrm{i}\varepsilon - h_{n,n} \end{pmatrix} - \sum_{n^\prime (\not= n)} |h_{n,n^\prime}|^2 \mathcal{G}_{n^\prime} .
\end{equation}
The steps leading from here to the reformulation as an equivalent equation for the (AAT-approximated) extremal states $Y$ are spelled out in full detail in the Appendix \ref{deriv: AAT}; the outcome is stated in Eq.\ (\ref{eq:eqn4Y}) below. Next, we sketch the key steps and tools used.

In order to derive the AAT equation for $Y$, we find it useful to pass through a super-analytic continuation $\mathcal{Y}$ of $Y$:
\begin{equation}\label{eq:SUSY-Y}
	\mathcal{Y}(\mathcal{Q}) := Y(Q_{\rm BB} - Q_{\rm FF}) ,
\end{equation}
where $\text{BB}$ stands for ``Bose-Bose'' and $\text{FF}$ for ``Fermi-Fermi''; the supermatrix $\mathcal{Q}$ is defined below. The matrix $Q_\text{BB}$ is the one in (\ref{eq:rankone}) with index $n$ omitted, while $Q_\text{FF}$ is similar but composed of Grassmann variables $\zeta, \bar\zeta, \omega, \bar\omega$:
\begin{equation}\label{eq:Q-BB-FF}
    Q_\text{BB} = \begin{pmatrix} u \bar{u} &- u \bar{v} \cr v \bar{u} &- v \bar{v} \end{pmatrix}, \quad
    Q_\text{FF} = \begin{pmatrix} \zeta \bar{\zeta} & \zeta \bar{\omega} \cr \omega \bar{\zeta} & \omega \bar{\omega} \end{pmatrix}.
\end{equation}
Note that the superfunction $\mathcal{Y}(\mathcal{Q})$ 
inherits from $Y(Q)$ the property
\begin{equation}
    \mathcal{Y} \big\vert_{Q_{\rm BB} = 0 = Q_{\rm FF}} = Y(0) = 1 .
\end{equation}
As a superfunction, $\mathcal{Y}(\mathcal{Q})$ is to be integrated against some kind of Berezin integration form, which we denote by $\mathcal{D} \mathcal{Q}$. The form with the translation invariance appropriate to the present context is
\begin{equation}
\label{eq:Berezin-int}
    \mathcal{D Q} = \frac{d^2 u}{\pi} \frac{d^2 v}{\pi}\times \bigg(- \frac{\partial^2} {\partial\zeta \partial \bar\zeta} \frac{\partial^2}{\partial\omega \partial \bar\omega} \bigg) .
\end{equation}
We see that $\mathcal{D} \mathcal{Q}$ is proportional to the usual Lebesgue measure for $(u,v) \in \mathbb{C}^2$ multiplied by the product of partial derivatives for all the Grassmann variables at hand. 

For present use, we mention a characteristic property of the superfunction $\mathcal{Y} (\mathcal{Q})$. By its definition (\ref{eq:SUSY-Y}), $\mathcal{Y}(\mathcal{Q})$ is expressed in terms of the matrix entries of $Q_{\rm BB} - Q_{\rm FF}$, which are $u \bar{u} - \zeta \bar\zeta$ and three analogous combinations. All of these are invariant under two transformations with odd parameter $\epsilon$, given explicitly as
\begin{equation}\label{eq:deltas}
    \begin{split}
    &\delta_\epsilon^{(1)} u = \epsilon \zeta , \quad \delta_\epsilon^{(1)} \bar\zeta = - \epsilon \bar{u} , \quad \delta_\epsilon^{(1)} v = \epsilon \omega , \quad \delta_\epsilon^{(1)} \bar\omega = \epsilon \bar{v} , \cr
    &\delta_\epsilon^{(2)} \bar{u} = \epsilon \bar\zeta , \quad \delta_\epsilon^{(2)} \zeta = \epsilon u , \quad \delta_\epsilon^{(2)} \bar{v} = - \epsilon \bar\omega , \quad \delta_\epsilon^{(2)} \omega = \epsilon v .
    \end{split}
\end{equation}
Thus $\mathcal{Y}$ has two supersymmetries, $\delta^{(i)}$ ($i = 1, 2$). Moreover, these transformations correspond to Killing vector fields for the Berezin integration form (\ref{eq:Berezin-int}). As a result, a localization principle goes into effect: the integral of $\mathcal{Y}$ against $\mathcal{D Q}$ localizes at $Q_{\rm BB} = Q_{\rm FF} = 0$, the (super-)point distinguished as the sole fixed point of the transformations (\ref{eq:deltas}):
\begin{equation}\label{eq:SUSY-loc}
    \int \mathcal{D Q} \; \mathcal{Y} = \mathcal{Y} \big\vert_{Q_{\rm BB} = Q_{\rm FF} = 0} = 1 .
\end{equation}

We finally outline the derivation of the AAT integral equation for $\mathcal{Y}(\mathcal{Q})$. (This will be sketchy. Please be referred to Appendix \ref{deriv: AAT} for the full derivation.) We begin by inserting into the definition of $\mathcal{Y}(\mathcal{Q})$ the expression for $\mathcal{G}_n^{-1}$ from the AAT self-consistency equation (\ref{eq:AATprime}). We then perform the average over the random potential $h_{n,n}$. Next, to do the average over the random hopping $h_{n,n'}$ ($n \not= n^\prime$), we write each of the $K$ factors originating from the term $\sum |h_{n,n^\prime}|^2 \mathcal{G}_{n^\prime}$ as a Fourier integral against auxiliary variables ($u^\prime, v^\prime$, etc.), while inverting $\mathcal{G}_{n^\prime} \to 1 / \mathcal{G}_{n^\prime}$. This does not produce any $\mathcal{G}_{n^\prime}$-dependent normalization factors, as the integral is a Fourier-Berezin integral with normalized Berezin integration form (\ref{eq:Berezin-int}), see Eq.\  (\ref{eq:SUSY-loc}), for the primed variables. When the final average over the randomness in $1 / \mathcal{G}_{n^\prime}$ is taken, each of the said $K$ factors yields one copy of $\mathcal{Y}(\mathcal{Q})$ as a function of the corresponding primed variables. In this way, we arrive at a closed integral equation, the ``supersymmetric'' (SUSY) AAT equation, for $\mathcal{Y}(\mathcal{Q})$:
\begin{equation}\label{eq:eqn4Y}
    \mathcal{Y}(\mathcal{Q}) = \mathrm{e}^{ \mathrm{i} \, \mathrm{STr}\, \mathcal{Q}(E + \mathrm{i} \varepsilon \Sigma_3) - \frac{1}{2} w_V^2 \mathrm{STr}\, \mathcal{Q}^2} \left( \int \mathcal{D Q}^\prime \, \mathrm{e}^{- w_T^2 \mathrm{STr}\, \mathcal{Q Q}^\prime} \mathcal{Y}(\mathcal{Q}^\prime) \right)^K ,
\end{equation}
where $\mathrm{STr} = \mathrm{Tr}_{\rm B} - \mathrm{Tr}_{\rm F}$ is the supertrace, $\Sigma_3 \equiv \sigma_3 \otimes \mathbf{1}$ is the third Pauli matrix tensored with unity in superspace, and in order to write the integral kernel $\mathrm{e}^{- w_T^2 \mathrm{STr}\, \mathcal{Q Q}^\prime}$ in concise form we introduced a $4 \times 4$ supermatrix
\begin{equation}\label{eq:super-Q}
    \mathcal{Q} := \rowfour{u}{v}{\zeta}{\omega} \otimes \columnfour{\bar{u}} {-\bar{v}} {\bar\zeta} {\bar\omega} \equiv \begin{pmatrix} Q_{\rm BB} &Q_{\rm BF} \cr Q_{\rm FB} &Q_{\rm FF} \end{pmatrix} ,
\end{equation}
which coincides with $Q_{\rm BB}$ in the Bose-Bose sector and $Q_{\rm FF}$ in the Fermi-Fermi sector. Note that $\mathcal{Y}(\mathcal{Q})$ remains a function of $Q_{\rm BB} - Q_{\rm FF}$, the difference of the diagonal blocks. In Sect.\ \ref{sect:reduction}, we will eliminate the Grassmann variables from Eq.\ (\ref{eq:eqn4Y}) to arrive at an equivalent formulation involving only the Bose-Bose matrix $Q_{\rm BB}$. (This step turns out to be somewhat delicate as a generic extremal state $\mathcal{Y}$ is a state of low symmetry.)

\section{Field-theoretical interpretation: $Y_x$ as extremal Gibbs states}\label{sect:Gibbs}

How do we use the AAT scheme (\ref{eq:eqn4Y}) to find an approximation to the extremal states $\mathcal{Y} (\mathcal{Q})$ that go into the decomposition (\ref{eq:YFL-Yx})? The strategy is to re-interpret (\ref{eq:eqn4Y}) as an iterative scheme $\mathcal{Y}_k (\mathcal{Q}) \mapsto \mathcal{Y}_{k+1} (\mathcal{Q})$ with a fixed point reached for $k \to \infty$. To initialize the iteration, we prescribe an initial function $\mathcal{Y}_0 (\mathcal{Q})$ (in the sense of imposing a consistent boundary condition at infinity; see below). This function is centered about a (super-)point $x$ in the order-parameter space $\mathcal{X}$ (a supermanifold) determined by symmetry breaking. We then insert $\mathcal{Y}_0 (\mathcal{Q})$ into the right-hand side of the AAT equation and let the result be $\mathcal{Y}_1 (\mathcal{Q})$. By iterating the procedure we obtain a sequence $\{ \mathcal{Y}_k (\mathcal{Q})\}_{k \geq 0}$. If the sequence converges (as we would expect), the fixed point $\mathcal{Y}_\infty (\mathcal{Q})$ is a solution of the AAT equation (\ref{eq:eqn4Y}). Repeating the whole process for all $x \in \mathcal{X}$ we obtain a continuum of solutions $\mathcal{Y}_\infty \equiv \mathcal{Y}_x$ labeled by the initial data $x$. To construct the physical state $Y_{\rm FL}$, we implement (the analog of) Eq.\ (\ref{eq:YFL-Yx}), which instructs us to take the average of this continuum with respect to a $\mathrm{U}(1)_{\mathrm{i} \sigma_3}$-invariant Berezin integration measure on $\mathcal{X}$.

\begin{figure}[h]
\centering
\includegraphics[width=0.8\textwidth]{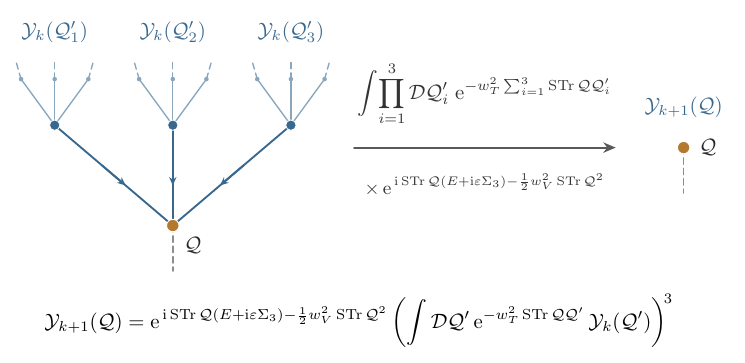}
\caption{Illustration of one step $\mathcal{Y}_k\to \mathcal{Y}_{k+1}$ of the AAT iteration toward an approximate extremal Gibbs state $\mathcal{Y}_\infty$. For illustration, the branching number is chosen as $K=3$. Dashed lines indicate continuation of the lattice.}
\label{Pic aat}
\end{figure}

In the sequel, we will explain the rationale behind this strategy in more detail: the iteration above is interpreted as the recursive integration of a field-theory statistical weight starting from the boundary at spatial infinity (with the boundary field $\mathcal{Q}$ constrained to fluctuate about a fixed value $x \in \mathcal{X}$ of the order parameter breaking the symmetry) and ending at the site $n \in \mathbb{G}$ of the field operator insertion. Thus, the continuum of symmetry-breaking solutions $\mathcal{Y}_x$ of the AAT integral equation (\ref{eq:eqn4Y}) is related to the notion of extremal Gibbs states of a statistical field theory (the field variables being the supermatrices $\mathcal{Q}$ or, after elimination of the Grassmann variables, their Bose-Bose parts $Q_{{\rm BB}}$.)

We recall that the observable $K_{n,\varepsilon}$ has an expression as the normalization integral of the (exact) Fourier-Laplace transform $Y_{\rm FL}(Q)$ defined in Eq.~(\ref{eq:def-Y}):
\begin{equation}
    K_{n,\varepsilon} = \int d Q_n \, Y_{\rm FL}(Q_n), \quad dQ_n = \pi^{-2} d^2 u_n\,d^2 v_n .
\end{equation}
Alternatively, by referring to Eq.~(\ref{eq:SUSY-Y}) or anticipating Eq.~(\ref{eq:expand-Y}) below, we may write $K_{n,\varepsilon}$ as a superintegral with four-fermion insertion:
\begin{equation}
    K_{n,\varepsilon} = \int\mathcal{D}\mathcal{Q}_n \, (-{\bar\zeta}_n \zeta_n {\bar\omega}_n\omega_n)\,\mathcal{Y}_{\rm FL} (\mathcal{Q}_n).
\end{equation}

Next we observe that the AAT self-consistency equation is not always a good approximation scheme to use directly for the physical state $\mathcal{Y}_{\rm FL}$. Indeed, as a \emph{nonlinear} equation with \emph{mean-field} character, the AAT approximation \underline{cannot} be expected to encapsulate the continuum of terms (building the physical state by convex linear combination) when spontaneous symmetry breaking occurs. Rather, we should employ the AAT self-consistency scheme only as a simple model for the field-theory states that emerge when the field variables (in our case, supermatrices) are constrained to fluctuate about a collective value (or mean field) $x \in \mathcal{X}$; these field-theory states are referred to as \emph{extremal} (Gibbs) states. One way to define them is by imposing the collective value $x$ for the fields on the boundary of $\mathbb{G}$ at spatial infinity. (By the long-range order which accompanies the spontaneously broken symmetry, the collective value imposed on the boundary persists into the bulk of the system.) The physical state $\mathcal{Y}_{\rm FL}$ is then built up by superposition of all possible boundary conditions: it is a mixture (i.e., an incoherent sum or integral) of extremal states weighted by a suitable measure for the distribution of boundary conditions.

Let us write this out in more detail. Taking the notation from Eq.~(\ref{eq:YFL-Yx}), we think of $\mathcal{Y}_{\rm FL}$ as a mixture:
\begin{equation}\label{FLmixture}
    \big\langle\mathcal{Y}_{\rm FL}, \bullet \big\rangle = \int_\mathcal{X} \!\! dx \, P(x) \, \big\langle\mathcal{Y}_x , \bullet\big\rangle .
\end{equation}
By evaluation on unity (the trivial observable) we infer the normalization
\begin{equation}
    1 =  \big\langle\mathcal{Y}_{\rm FL}, 1 \big\rangle = \int_\mathcal{X} dx \, P(x) \, \big\langle\mathcal{Y}_x , 1 \big\rangle = \int_\mathcal{X} \!\! dx \, P(x) ,
\end{equation}
as $\langle\mathcal{Y}_x , 1 \rangle$ cannot depend on the variable $x$ parameterizing the Goldstone (super-)manifold $\mathcal{X}$. Since $\mathcal{Y}_{\rm FL}$ is $\mathrm{U}(1)_{\mathrm{i} \sigma_3}$-invariant for all $\varepsilon > 0$, the boundary weight function $P(x)$ needs to have the same property. [In Sect.~\ref{K sc} we regularize the non-compact mixture by a $\mathrm{U}(1)_{\mathrm{i} \sigma_3}$-invariant family $P_\gamma(x)$ with parameter $\gamma > 0$ and take a scaling limit $\gamma \to 0$ as the inverse correlation volume goes to zero at criticality.] For our observable of interest, we obtain
\begin{equation}\label{KFLTandExtremal}
    K_{n,\varepsilon} = \big\langle\mathcal{Y}_{\rm FL}, - {\bar\zeta}_n \zeta_n {\bar\omega}_n \omega_n \big\rangle = \int_\mathcal{X} \!\! dx \, P(x) \int \mathcal{D}\mathcal{Q}_n \, (- {\bar\zeta}_n \zeta_n {\bar\omega}_n \omega_n) \, \mathcal{Y}_x (\mathcal{Q}_n).
\end{equation}

Next, we relate the observable $K_{n,\varepsilon}$ to an expectation value computed from a supersymmetric statistical field theory. This is a standard exercise in applying the Wegner-Efetov supersymmetry method; the calculation is shown in Appendix \ref{App.Efetov SUSY}:
\begin{equation}
    K_{n,\varepsilon} = \int\prod_{j \in \mathbb{G}} \mathcal{D}\mathcal{Q}_j \, (-{\bar\zeta}_n \zeta_n {\bar\omega}_n\omega_n)\, \mathrm{e}^{-S}.
\end{equation}
The action functional $S$ reads:
\begin{align}\label{eq:SUSY-S1}
    \begin{split}
    &S = {S_{\rm reg}} +{S_V} + {S_T} , \quad
    {S_{\rm reg}} = \varepsilon \sum_{j} \mathrm{STr}\, \Sigma_3 \mathcal{Q}_j \,, \cr &{S_V} = \sum_{j} \mathrm{STr} \left( - \mathrm{i} E \mathcal{Q}_j + {\textstyle{\frac{1}{2}}} w_V^2 \mathcal{Q}_j^2 \right) , \quad {S_T} = \frac{1}{2} w_T^2 \sum_{j j^\prime} A_{j j^\prime} \, \mathrm{STr}\, \mathcal{Q}_j \mathcal{Q}_{j^\prime}.
    \end{split}
\end{align}

Let the field-theory state, i.e.\ the proper physical (mixed) state by which to calculate the observables of interest, be denoted by $\langle \bullet_n \rangle$. (The index $n$ is to indicate that we are considering only field observables localized at $n \in \mathbb{G}$. To avoid confusing the reader, let it be clear that there is no difference between $\langle \mathcal{Y}_{\rm FL} , \bullet \rangle$ and $\langle \bullet_n \rangle$. The change of notation is to emphasize that we now think of the former as a quantity to be computed in a specific scheme, namely the Wegner-Efetov supersymmetric field-theory formalism.) This state decomposes into extremal states $\langle \bullet_n \rangle_x$ (when spontaneous symmetry breaking occurs):
\begin{equation}
    \langle \bullet_n \rangle = \int_\mathcal{X} \!\! dx\, P(x) \langle\, \bullet_n \rangle_x .
\end{equation}
In the language of statistical field theory, the latter are called extremal Gibbs states. They are obtained by performing the field integral with a collective boundary condition constraining the field at spatial infinity to fluctuate about $x \in \mathcal{X}$.

By comparing with Eq.~(\ref{FLmixture}) we see that the distribution $\langle \bullet_n \rangle$ coincides with $\big\langle \mathcal{Y}_{\rm FL} , \bullet \big\rangle$, and the distribution $\langle \, \bullet_n \rangle_x$ coincides with $\big\langle \mathcal{Y}_x , \bullet \big\rangle$. The density function $\mathcal{Y}_x$ representing the distribution $\langle \, \bullet_n \rangle_x$ is the result of performing the field integral for all sites but the one with operator insertion ($n \in \mathbb{G}$):
\begin{equation}\label{eq:identYfieldintegral}
    \mathcal{Y}_x (\mathcal{Q}_n) = \int \prod_{j \in \mathbb{G} \setminus \{n \}} \!\!\!\!\!\!
    \mathcal{D}\mathcal{Q}_j \; \mathrm{e}^{- S_x},
\end{equation}
where $S_x$ means that the field at the boundary of $\mathbb{G}$ is constrained to fluctuate about $x$. The idea now is that the AAT scheme can be used to approximate the field integral in Eq.~(\ref{eq:identYfieldintegral}): we imagine some process of integrating out the statistical weight sequentially, starting from infinity and proceeding toward the final site $n$. When loops on the underlying lattice are neglected, this recursive procedure amounts to iterating the AAT integral equation (\ref{eq:eqn4Y}), whose stable fixed points are (AAT approximations to) the functions $\mathcal{Y}_x (\mathcal{Q})$.

The discussion above clarifies why we have been speaking of the solutions $\mathcal{Y}_x$ of the AAT equation as extremal Gibbs states. For a brief overview of the general theory of extremal Gibbs states and the connection to spontaneous symmetry breaking, we refer to Appendix \ref{Interlude}.

To conclude this part, we record the essence of Eq.~(\ref{KFLTandExtremal}): when the derivatives with respect to the Grassmann variables are performed in the rightmost Berezin integral, we are left with the integral over the ``bosonic'' reduction:
\begin{equation}\label{spectral integral}
    K_{n,\varepsilon} = \int_\mathcal{X} \!\! dx\, P(x) \int d Q_n\, Y_x (Q_n).
\end{equation}
Thus, we need only the ``bosonic'' part $Y_x(Q)$ of the solution of the supersymmetric AAT equation, and a natural step is to reduce the supersymmetric equation (\ref{eq:eqn4Y}) to a ``bosonic'' equation. This is done in the following section, \ref{sect:reduction}. [In that section, we return to the abbreviated notation of writing $\mathcal{Y} (\mathcal{Q})$ or $Y(Q)$ for the solutions of the AAT integral equation, i.e., we omit the subscript $x$, it being understood that the solutions correspond to the extremal states].

\section{Reduction to a ``bosonic'' equation}
\label{sect:reduction}

We now wish to convert Eq.~(\ref{eq:eqn4Y}) to a reduced equation no longer featuring the Grassmann variables $\zeta, \bar\zeta, \omega, \bar\omega$ of $Q_{\rm FF}$. The naive initial step in this quest is to simply set $Q_{\rm FF} \equiv 0$ on the left-hand side, thereby restricting $\mathcal{Y} (\mathcal{Q})$ to $Y(Q_{\rm BB} \equiv Q)$. In this way, however, one does not arrive at a closed equation for the restricted function $Y(Q)$, as the integration (actually, differentiation) w.r.t.\ the Grassmann variables on the right-hand side introduces derivatives of $Y(Q)$. Some effort has to be invested to account for those derivatives while retaining the minimal domain for $Y(Q)$.

\subsection{Three-dimensional cone $C^3$}
\label{sect:cone-C3}

We begin with an investigation of the geometric structure underlying the supermatrix $\mathcal{Q}$. Let $P$ be a positive Hermitian $2 \times 2$ matrix. The space, $\mathrm{Herm}^+(2)$, of such matrices is an open convex cone of real dimension 4 and is in bijection with the symmetric space $\mathrm{GL}(2,\mathbb{C}) / \mathrm{U}(2)$ by the mapping $g \cdot \mathrm{U}(2) \mapsto g g^\dagger = P$. Now in view of the key role played by our (total) symmetry group $\mathrm{U}(1,1)$, it is advantageous to base the analysis on a closely related 4-manifold, namely the set, $C_+^4$, of matrices $P^\prime = P \sigma_3$ with $P \in \mathrm{Herm}^+(2)$. Because such matrices satisfy $P^{\prime \, \dagger} = \sigma_3 P^\prime \sigma_3$, they form a subspace of $\sqrt{-1} \times \mathrm{Lie}\, \mathrm{U} (1,1)$ with $\mathrm{Lie}\, \mathrm{U}(1,1)$ the solution set of the equation $A^\dagger = - \sigma_3 A \sigma_3$ for complex $2\times 2$ matrices $A$.

Our interest is in the 3-dimensional boundary, $C^3$, of the convex cone $C_+^4$:
\begin{equation}
    C^3 = \partial C_+^4 := \{ Q \in \overline{C_+^4} \mid \mathrm{Det}\, Q = 0 \} ,
\end{equation}
where at least one of the two positive eigenvalues of the Hermitian matrix $P^\prime \sigma_3$ has gone to zero. (Unlike $C_+^4$, the space $C^3$ is \underline{not} a smooth manifold). $C^3$ is relevant here because our matrices $Q$ parameterized according to Eq.~(\ref{eq:rankone}) by $(u,v) \in \mathbb{C}^2$, constitute $C^3$. In the sequel, we will make good use of the inclusion mapping
\begin{equation}\label{eq:i-map}
    i : \; C^3 \to \overline{C_+^4} ,
\end{equation}
interpreting the matrices $Q$ as points in the closure $\overline{C_+^4}$ of the open set $C_+^4$.

Why is the inclusion map (\ref{eq:i-map}) useful? The short answer is that the Fourier-Laplace transform $Y_{\rm FL}$, by its formal definition in Eq.~(\ref{eq:def-Y}), extends as a smooth function from $C^3$ to $\overline{C_+^4}$. The same is true for the extremal states $\mathcal{Y} = \mathcal{Y}_x$ (and their bosonic restrictions $Y = Y_x$): the field-integral representation (\ref{eq:identYfieldintegral}) remains well defined when the range of $Q_n$ is extended to $\overline{C_+^4}$. As a matter of fact, if we were analyzing the Wegner model with $N \geq 2$ orbitals, we would simply replace the parametrization (\ref{eq:rankone}) of $Q$ by a sum over orbitals:
\begin{equation}\label{eq:N-geq-2}
    Q = \sum_{\alpha = 1}^N  \begin{pmatrix} u_\alpha \bar{u}_\alpha & - u_\alpha \bar{v}_\alpha \cr
    v_\alpha \bar{u}_\alpha & - v_\alpha \bar{v}_\alpha  \end{pmatrix} \in \overline{C_+^4} \qquad (u_\alpha , v_\alpha \in \mathbb{C}) .
\end{equation}

Next, we recall that our superfunction $\mathcal{Y}$ depends only on $Q_\text{BB} - Q_\text{FF}$. We express that difference as
\begin{equation}\label{eq:Q-from-xmu}
    \mathrm{i}(Q_{\rm BB} - Q_{\rm FF}) = \hat{x}_0 \mathrm{i} \mathbf{1} + \hat{x}_3\mathrm{i} \sigma_3 + \hat{x}_1 \sigma_1 + \hat{x}_2 \sigma_2
\end{equation}
in terms of basic superfunctions $\hat{x}_\mu$ ($\mu = 0, 1, 2, 3$). By inverting this expression, we get (with $\sigma_0 \equiv \mathbf{1}$)
\begin{equation}
    \hat{x}_j = \frac{1}{2} \mathrm{Tr}\, (Q_{\rm BB} - Q_{\rm FF})\, \sigma_j \quad (j = 0, 3),  \quad \hat{x}_k = \frac{\mathrm{i}}{2} \mathrm{Tr}\, (Q_{\rm BB} - Q_{\rm FF}) \, \sigma_k \quad (k = 1, 2).
\end{equation}

We now want to insert (\ref{eq:Q-from-xmu}) into $\mathcal{Y}$ and expand at all points $Q_{\rm BB} \equiv Q$ (with $Q \in C^3$) with respect to the nilpotent parts (i.e.\ the Grassmann variables) in the $\hat{x}_\mu$. This plan meets with a complication: since the number parts of the four functions $\hat{x}_\mu$ ($\mu = 0, 1, 2, 3$) cannot be linearly independent as functions on the 3-dimensional space $C^3$, the operation of taking partial derivatives with respect to them is not well-defined. Nonetheless, given the preparations made above, there exists a natural fix: taking the cue from (\ref{eq:Q-from-xmu}), we introduce four real-valued functions $x_\mu$ (without hat) by
\begin{equation}\label{eq:P-from-xmu}
    \mathrm{i} P^\prime = x_0 \mathrm{i} \mathbf{1} + x_3 \mathrm{i} \sigma_3 + x_1 \sigma_1 + x_2 \sigma_2 \in \mathrm{Lie}\, \mathrm{U}(1,1). 
\end{equation}
These may serve as coordinate functions for the 4-dimensional space $C_+^4$. As such, they are subject to the conditions
\begin{equation}
    x_3 > 0, \quad x_3^2 > x_0^2 + x_1^2 + x_2^2 \,.
\end{equation}
The boundary $C^3 \subset \overline{C_+^4}$ is reached by allowing (and demanding, respectively) equality:
\begin{equation}
    C^3 : \quad x_3 \geq 0 , \quad x_3^2 = x_0^2 + x_1^2 + x_2^2 \,.
\end{equation}
Indeed, the last equality is the coordinate expression for the equation $\mathrm{Det}\, Q = 0$ singling out the subspace $C^3$. We note that under pullback $i^\ast x_\mu \equiv x_\mu \circ i$ by the inclusion map (\ref{eq:i-map}) we have
\begin{equation}\label{eq:first coordinates}
    i^\ast x_0 = \frac{1}{2} (\bar{u}u -\bar{v}v) , \quad i^\ast x_1 = \frac{\mathrm{i}}{2} (\bar{u}v -\bar{v}u),\quad i^\ast x_2 = \frac{1}{2} (\bar{u} v +\bar{v} u),\quad i^* x_3 = \frac{1}{2}(\bar{u} u + \bar{v} v).
\end{equation}
Now by invoking the second of Eqs.~(\ref{eq:Q-BB-FF}), we extend the superfunctions $\hat{x}_\mu$ to the 4-dimensional space $\overline{C_+^4}$ as
\begin{equation}\label{eq:extend-xmu}
    \begin{split}
    &\hat{x}_0 = x_0 + \frac{1}{2} (\bar{\zeta}\zeta + \bar\omega \omega), \quad \hat{x}_1 = x_1 + \frac{\ii}{2} (\bar\zeta \omega + \bar\omega \zeta) , \cr
    &\hat{x}_2 = x_2 + \frac{1}{2} (\bar\zeta \omega - \bar\omega \zeta) , \quad \hat{x}_3 = x_3 + \frac{1}{2} (\bar\zeta \zeta - \bar\omega \omega) .
    \end{split}
\end{equation}

\subsection{Expanding the superfunction $\mathcal{Y}$: enter the Laplacian}

We are finally in a position to execute our plan of expanding the superfunction $\mathcal{Y}$ with respect to the nilpotents:
\begin{align}\label{eq:expand-Y}
    \begin{split} \mathcal{Y}(\mathcal{Q}) &= Y(Q) + \frac{1}{2} (\bar\zeta \zeta + \bar\omega \omega) \frac{\partial Y} {\partial x_0} (Q) + \frac{1}{2} (\bar\zeta \zeta - \bar\omega \omega) \frac{\partial Y}{\partial x_3}(Q) \cr &+ \frac{\ii}{2} (\bar\zeta \omega + \bar\omega \zeta) \frac{\partial Y}{\partial x_1} (Q) + \frac{1}{2} (\bar\zeta \omega - \bar\omega \zeta) \frac{\partial Y}{\partial x_2}(Q) - \bar\zeta \zeta \bar\omega \omega \Delta Y(Q), \qquad Q\in C^3 .
\end{split}
\end{align}
The second-order differential operator $\Delta$ in the last summand is of Laplace-type:
\begin{equation}\label{eq:def-Delta}
    \Delta = \frac{1}{4} \bigg( \frac{\partial^2} {\partial x_3^2} - \frac{\partial^2} {\partial x_0^2} - \frac{\partial^2} {\partial x_1^2} - \frac{\partial^2} {\partial x_2^2} \bigg).
\end{equation}
We recall that $Y(Q)$ extends smoothly from $C^3$ to $\overline{C_+^4}$. Thus $\Delta Y$ and all partial derivatives $\partial Y / \partial x_\mu$ are well-defined. They still exist as one-sided derivatives in the limit of approaching the boundary points $Q \in C^3 \subset \overline{C_+^4}$.

We now set $Q_{\rm FF} = 0$ in Eq.~(\ref{eq:eqn4Y}) and observe that
\begin{equation}
    \mathrm{e}^{-w_T^2 \mathrm{STr} \, \mathcal{Q} \mathcal{Q}^\prime} \Big\vert_{ Q_{\rm FF} \equiv 0} = \mathrm{e}^{-w_T^2 \mathrm{Tr} \, Q_{\rm BB}  Q_{\rm BB}^\prime} ,
\end{equation}
which tells us (in the case of $Q_{\rm FF} = 0$) that all dependence on the Grassmann variables of $Q_{\rm FF}^\prime$ in the integrand on the right-hand side of Eq.~(\ref{eq:eqn4Y}) resides in $\mathcal{Y}(\mathcal{Q}^\prime)$. Consequently, integration over those Grassmann variables picks up nothing but the last term on the r.h.s.\ of (\ref{eq:expand-Y}). Thus we obtain for $Y(Q) = \mathcal{Y}(\mathcal{Q}) \big\vert_{Q_{\rm FF} = 0}$ the equation
\begin{equation}\label{eq:eqn4YBB}
    Y(Q) = \mathrm{e}^{\mathrm{i} \, \mathrm{Tr}\, Q (E + \mathrm{i} \varepsilon \sigma_3 ) - \frac{1}{2} w_V^2 \mathrm{Tr}\, Q^2} \left( \int d\mu(Q^\prime) \, \mathrm{e}^{-w_T^2 \mathrm{Tr}\, Q Q^\prime} (\Delta Y) (Q^\prime) \right)^K ,
\end{equation}
where both $Q$ and $Q^\prime$ run through $C^3$. For the measure of integration $d\mu(Q)$ we may take $\pi^{-2} d^2 u \, d^2 v$ from Eq.~(\ref{eq:Berezin-int}) or, even better, the push forward of this measure to $C^3$. Note that for the rank-one matrix $Q$ (in $C^3$) one has the identity $(\mathrm{Tr} \, Q)^2 = \mathrm{Tr}\, (Q^2)$.

As it stands, Eq.~(\ref{eq:eqn4YBB}) fails to be  closed for $Y$ as a function on $C^3$. The reason is that $\Delta Y$ is \emph{not} determined by the values that $Y$ takes on $C^3$; some information on its directional derivatives \emph{normal} to $C^3$ is needed. What is the minimal information that has to be added? To answer this question, consider the first-order differential operators
\begin{align}\label{def:K_j}
\begin{split}
    &K_1 = x_2\frac{\partial}{\partial x_3} + x_3 \frac{\partial}{\partial x_2}, \quad K_2 = x_3 \frac{\partial}{\partial x_1} + x_1 \frac{\partial}{\partial x_3},\cr
    &K_3 = x_1 \frac{\partial}{\partial x_2} - x_2 \frac{\partial}{\partial x_1}, \quad
    D = \sum_{\mu=0}^3 x_\mu \frac{\partial} {\partial x_\mu} \,.
\end{split}
\end{align}
All of these annihilate the function $\mathrm{Det}\, Q = x_0^2 + x_1^2 + x_2^2 - x_3^2$ (in the case of $D$ this holds after restriction to $C^3$, which is good enough for our purposes). Since $C^3$ is defined by $\mathrm{Det}\, Q = 0$, this means that each of $D$ and $K_j$ ($j = 1, 2, 3$) corresponds to a vector field \emph{tangential} to $C^3$. Armed with this information, one does a straightforward calculation [Appendix \ref{Laplace manipulation}] to show that $\Delta$ operating on functions on $C_+^4$ has the expression
\begin{equation}\label{eq:Delta-VF}
    \Delta = \frac{1}{4} (x_3^2 - x_2^2 - x_1^2)^{-1} \bigg(\Delta^\flat - 2 D\circ x_0 \frac{\partial}{\partial x_0} + (x_0^2 + x_1^2 + x_2^2 - x_3^2) \frac{\partial^2} {\partial x_0^2}\bigg) ,
\end{equation}
where $\Delta^\flat$ is a second-order differential operator that will appear repeatedly in the following:
\begin{equation}\label{eq:abbrev}
    \Delta^\flat = K_3^2 - K_2^2 - K_1^2 + D^2 + D .
\end{equation}
We will need $(\Delta Y)(Q)$ only in the limit of taking $Q$ to $C^3$, in which case the expression (\ref{eq:Delta-VF}) simplifies to
\begin{equation}\label{eq:Lapl-restr}
    (\Delta Y)(Q) = \frac{1}{4 x_0^2} (\Delta^\flat Y)(Q) - \frac{1}{2 x_0} (D+1) \frac{\partial Y} {\partial x_0}(Q) ,
\end{equation}
where we have used $i^*(x_0^2 + x_1^2 + x_2^2 - x_3^2) = 0$ and the commutation relation $D \circ x_0 = x_0 (D+1)$. It is clear from the definition (\ref{eq:def-Delta}) that the operator $\Delta$ commutes with the group action $Q \mapsto g Q g^{-1}$ by $g \in \mathrm{SU}(1,1)$. As a check on the $C^3$-restricted Laplacian (\ref{eq:Lapl-restr}), one verifies that this commutation law still holds.

\subsection{Closing the system of equations}
\label{sect:closeq}

What do we learn from the formula (\ref{eq:Lapl-restr})? The first summand on the right-hand side is expressed in terms of the $C^3$-tangential operators $D$ and $K_j$, which restrict to first-order differential operators on $C^3$. If this were the only summand, the equation for $Y(Q)$ would already be closed. However, the second summand involves the partial derivative $\partial Y / \partial x_0$, which is not tangential to $C^3$. Therefore, in order to close the equation, we must carry along $\partial Y / \partial x_0$ and augment the equation for $Y(Q)$ by an equation for $(\partial Y / \partial x_0) (Q)$.

To obtain an equation for $\partial Y / \partial x_0$, we return to Eq.~(\ref{eq:eqn4Y}) and apply $\partial / \partial x_0$ on both sides. Having done so, we proceed as before: we set $Q_{\rm FF} = 0$ and carry out the Grassmann integration on the right-hand side. The resulting equation is somewhat cumbersome; to present it in a manageable form, we introduce the abbreviation
\begin{equation}\label{eq:def-X(Q)}
    X(Q) \equiv \int d\mu(Q^\prime) \, \mathrm{e}^{-w_T^2 \mathrm{Tr}\, Q Q^\prime} \left( \frac{1}{4 {x^\prime_0}^2} (\Delta^\flat Y)(Q^\prime) - \frac{1}{2 x^\prime_0} (D+1) \frac{\partial Y} {\partial x_0}(Q^\prime) \right) .
\end{equation}
The equation for $\partial Y / \partial x_0$ then reads
\begin{align}\label{eq:derY-x0}
    \begin{split}
    \frac{\partial Y}{\partial x_0} (Q) &= \mathrm{e}^{\mathrm{i} \, \mathrm{Tr}\, Q (E + \mathrm{i} \varepsilon \sigma_3  ) - \frac{1}{2} w_V^2 \mathrm{Tr}\, Q^2} \bigg( 2 (\mathrm{i} E - w_V^2 \mathrm{Tr}\, Q)\, X(Q)^K \cr
    &+ K X(Q)^{K-1} \int d\mu(Q^\prime) \, \mathrm{e}^{- w_T^2 \mathrm{Tr}\, Q Q^\prime} w_T^2 (1 - w_T^2 \mathrm{Tr}\, Q Q^\prime)
    \frac{\partial Y}{\partial x_0} (Q^\prime) \bigg).
    \end{split}
\end{align}
We observe that our system of equations is now closed, as no further unknowns enter on the right-hand side. Note that with the abbreviation (\ref{eq:def-X(Q)}), the equation (\ref{eq:eqn4YBB}) for $Y$ takes the shortened form
\begin{equation}\label{eqn4Y-short}
    Y(Q) = \mathrm{e}^{\mathrm{i} \, \mathrm{Tr}\, Q (E + \mathrm{i} \varepsilon \sigma_3 ) - \frac{1}{2} w_V^2 \mathrm{Tr}\, Q^2} X(Q)^K .
\end{equation}

In summary, we have derived a system of two equations, (\ref{eqn4Y-short}) and (\ref{eq:derY-x0}), for two functions, $Y$ and $\partial Y / \partial x_0$. A distinctive  feature of this system (in the limit of vanishing regulator $\varepsilon$) is its $\mathrm{SU} (1,1)$-equivariance: if $Y(Q)$ and $(\partial Y / \partial x_0)(Q)$ form a solution of the system, then so do the transformed functions $Y(g Q g^{-1})$ and $(\partial Y / \partial x_0)(g Q g^{-1})$ for any $g \in \mathrm{SU}(1,1)$. Now, in all calculations done later on, we will use some trick [of restricting the class of solutions $Y$ by breaking $\mathrm{SU}(1,1)$] to avoid the appearance of the auxiliary function $\partial Y / \partial x_0$. Nonetheless, as a matter of principle and for potential use in further work, we wanted to demonstrate here how the system of equations can be closed in a manner that fully respects the symmetry group $\mathrm{SU}(1,1)$.

\subsection{Reduction to ``bosonic'' boundary data} \label{Grassmann in Yx}

We recall that Eq.~(\ref{spectral integral}) expresses our observable of interest, $K_{n, \varepsilon}$, as a mixture of extremal Gibbs states, which we intend to model approximately as solutions $Y_x(Q)$ of the reduced AAT equation (\ref{eqn4Y-short}). The solution parameter $x$ runs through an order-parameter superspace $\mathcal{X}$ yet to be specified. So, what is the concrete meaning of the integral over $x \in \mathcal{X}$ in (\ref{spectral integral}) and, in particular, how should we handle the Grassmann degrees of freedom in $x$? Given the long-wavelength description of the metallic phase by a nonlinear $\sigma$-model \cite{Efe83, EM2008}, the order-parameter space $\mathcal{X}$ is known to be a supermanifold $\widehat{G} / \widehat{K}$ of Goldstone modes for the standard symmetry-breaking scenario where $\widehat{G} = \mathrm{U}(1,1|2)$ and $\widehat{K} = \mathrm{U}(1|1) \times \mathrm{U}(1|1)$. However, in the present paper we work with the microscopic (i.e.\ unrenormalized) field degrees of freedom $\mathcal{Q}$ of the $N = 1$ Wegner model. So we ask: what is the order-parameter space $\mathcal{X}$ appropriate to our setting? What is the family of boundary conditions to be imposed at the boundary of the infinite lattice $\mathbb{G}$? And how do these boundary conditions translate into the initial data that are to be fed into the AAT equation?

Roughly speaking, the answer is as follows. We pick a weight function, say $\mathcal{I}_\gamma$, controlling the values taken by the super\-matrix field $\mathcal{Q}$ at the boundary. To distinguish the boundary field from the bulk field, we denote the former by $\mathcal{Q}_\infty$. Motivated by the physical idea of particle absorption at the boundary, we choose the weight function to be
\begin{equation}
    \mathcal{I}_\gamma(\mathcal{Q}_\infty) = \mathrm{exp}\left( - \gamma w_T \STr \mathcal{Q}_\infty \Sigma_3 \right) ,
\end{equation}
with dimensionless regulator $\gamma > 0$ corresponding to a local absorption energy width (or absorption frequency) $\gamma w_T$. We note that $\mathcal{I}_\gamma$ has the $\widehat{K}$-invariance that comes with the implementation of the absorption process by $\varepsilon \Sigma_3 \to \gamma w_T \Sigma_3$ at the boundary. To detect the fine structure of the energy spectrum (in particular, the difference between ac- and sc-type) it will be necessary to take an informed limit $\gamma \to 0$. Altogether then, a tentative answer to the posed question is that we compute the physical state as a mixture
\begin{equation}\label{microscopic mixture}
    \langle \mathcal{Y}_\mathrm{FL}, \bullet \rangle = \lim_{\gamma \to 0}\int\mathcal{D}\mathcal{Q}_\infty \; \mathcal{I}_\gamma(\mathcal{Q}_\infty)\, \langle \mathcal{Y}_{\mathcal{Q}_\infty} , \bullet \rangle ,
\end{equation}
where $\mathcal{Y}_{\mathcal{Q}_\infty} (\mathcal{Q})$ is the solution of the SUSY-AAT equation (\ref{eq:eqn4Y}) with initial condition $\mathcal{Y}_0 (\mathcal{Q}) = \delta( \mathcal{Q} - \mathcal{Q}_\infty)$; cf.\ Sect.~\ref{sect:Gibbs}.

Now for our observable $K_{n,\varepsilon}$ we benefit from the special (and greatly simplifying) feature that the four-fermion insertion (\ref{KFLTandExtremal}) at the observation site $n$ projects the SUSY solution $\mathcal{Y}_{\mathcal{Q}_\infty}$ onto its bosonic component, the first term in the expansion (\ref{eq:expand-Y}). As we have seen, the SUSY-AAT equation for that component reduces to a closed system of bosonic type (devoid of Grassmann degrees of freedom). By the same token, the initial data given by $\delta( \mathcal{Q} - \mathcal{Q}_\infty)$ get reduced to their bosonic component $\delta(Q - Q_\infty)$, provided that we are calculating $K_{n, \varepsilon}$. In other words, all dependence on the Grassmann variables of $\mathcal{Q}_\infty$ resides in the boundary weight function $\mathcal{I}_\gamma (\mathcal{Q}_\infty)$. By integrating out that dependence in Eq.~(\ref{microscopic mixture}), we obtain a reduced boundary weight function,
\begin{equation}\label{Igam}
    \widetilde{I}_\gamma(Q_\infty) = (\gamma w_T)^2 \mathrm{e}^{- \gamma w_T \mathrm{Tr}\, Q_\infty \sigma_3} ,
\end{equation}
which is normalized with respect to $d Q_\infty :$ $\int dQ_\infty \, \widetilde{I}_\gamma (Q_\infty) = 1$. The expression for $K_{n, \varepsilon}$ now takes the form
\begin{equation}\label{mixt intermediate}
    K_{n,\varepsilon} = \lim_{\gamma \to 0}  \int dQ_\infty \, \widetilde{I}_\gamma(Q_\infty) \, \langle Y_{Q_\infty}, 1 \rangle.
\end{equation}

At this stage, we must concede that the prescription given above does not fully conform to the standard protocol for extremal Gibbs states and spontaneous symmetry breaking. Indeed, by pinning all of the boundary field $\mathcal{Q} \big\vert_{\partial \mathbb{G}}$ to a single collective value $\mathcal{Q}_\infty$, we are over-specifying the data at the boundary. While collective pinning is appropriate for the Goldstone degrees of freedom that undergo symmetry breaking and develop long-range order that persists throughout the bulk and its boundary, it is inappropriate for the massive field degrees of freedom, which fluctuate independently beyond the length scale set by their correlation length.

How does our field $\mathcal{Q}$ decompose into massive versus symmetry degrees of freedom? There exists an evidently massive degree of freedom in $\mathcal{Q}$, namely the $\widehat{G}$-invariant $\mathrm{STr}\, \mathcal{Q}$. Now it is not difficult to see that the Lie supergroup $\widehat{G} = \mathrm{U}(1,1|2)$ of symmetries acts transitively on the codimension-one supermanifolds given as the level sets of $\mathrm{STr}\, \mathcal{Q}$. Thus all remaining degrees of freedom are of symmetry type, and $\mathrm{STr}\, \mathcal{Q}$ is the one and only $\mathcal{Q}$-field degree of freedom that is definitely massive. To execute the proper protocol to perfection, we should take apart the boundary constraint $\delta(\mathcal{Q} - \mathcal{Q}_\infty$) so as to let the boundary values of the massive field component $\mathrm{STr}\, \mathcal{Q}$ fluctuate about the boundary-pinned Goldstone degrees of freedom. Having said so, we shall turn around and keep to the simple prescription formulated above. Our justification is that, within the AAT-approximation scheme used (if not more generally), the massive degree of freedom $\mathrm{STr}\, \mathcal{Q}$ relaxes under the AAT-iteration from the boundary to the observation site $n$. Thus the initial condition for $\mathrm{STr}\, \mathcal{Q}$ gets forgotten and we may as well model the initial data by $\delta(\mathcal{Q} - \mathcal{Q}_\infty)$.

What we take away from this discussion is that it makes good sense to adhere to Eq.~(\ref{mixt intermediate}) but for one modification: we need to eliminate the massive degrees in $Q_\infty$, which are $\mathrm{Tr}\, Q_\infty$ and, depending on the case, some symmetry degrees of freedom that stem from the symmetry restoration of a subgroup $S \subset G$. Given that $Y_{Q_\infty}(Q)$ does not vary with these freedoms in $Q_\infty$, we can push the probability measure $dQ_\infty \, \widetilde{I}_\gamma (Q_\infty)$ forward to a probability measure $dx \, P_\gamma(x)$ on the Goldstone manifold $X = G/S$ of broken symmetries. Thus we arrive at
\begin{equation}
    K_{n,\varepsilon} = \lim_{\gamma \to 0} \int_X \!\! dx \, P_\gamma(x) \int dQ_n \, Y_x(Q_n).
\end{equation} 
It will be the theme of Sect.~\ref{types} to describe the exact details of $X$ and its realization in the various regimes.

Whenever the $\gamma$-regularization is not essential to the discussion, we suppress the subscript $\gamma$ and the corresponding limit and simply write $dx\, P(x)$. We postpone the explicit construction of this induced distribution until it is needed in Sect.~\ref{K sc}.

\section{Extremal Gibbs states for the insulating and metallic phases}
\label{sect:analysis}

With the ``bosonic'' reduction (\ref{eq:eqn4YBB}) in hand, we are in a position to determine the families of extremal Gibbs states $Y_{x}$ in different parameter regions. As a preparation, we take a closer look at the geometry of the space of matrices $Q$. 

\subsection{Target space geometry}\label{geom}

A characteristic feature of the field theory for the Anderson transition is the non-compactness of the symmetry group $G = \mathrm{SU}(1,1)$, which acts on the bosonic target space $C^3$ by Lorentz transformations. (In what follows, we assume the material of Sect.~\ref{sect:cone-C3} to be understood.) The 3-dimensional space $C^3$ is in bijection with $\mathbb{R}^3$ by
\begin{equation}\label{eq:biject}
    \mathbb{R}^3 \to C^3 , \quad (x_0,x_1,x_2) \mapsto \Big( x_0,x_1,x_2,x_3 = + \sqrt{x_0^2 + x_1^2 + x_2^2} \Big) ,
\end{equation}
where the $x_\mu$ ($\mu = 0, 1, 2, 3$) were introduced in Eq.~(\ref{eq:P-from-xmu}) as quasi-Cartesian coordinates for the 4-dimensional manifold $C_+^4$. [The bijection (\ref{eq:biject}) is a diffeomorphism from $\mathbb{R}^3 \setminus \{0\}$ to $C^3 \setminus \{ Q = 0\}$ but has a cone singularity at the ``tip'' $Q = 0$.] We recall the characterization of $C^3$ as a boundary, $C^3 = \partial\, \overline{C_+^4}$. We also recall that our function $Y(Q)$ extends from $C^3$ into $C_+^4$.

Geometrically (and roughly) speaking, $C^3$ is a (non-convex) cone foliated by two-dimensional hyperboloids, $\rm{H}_{x_0}^2$. Each of the latter is given as the solution set of a quadratic equation,
\begin{equation}
   \mathrm{H}_{x_0}^2 : \;\; x_3^2 - x_1^2 - x_2^2 = x_0^2 \quad (x_3 \geq |x_0|) ,
\end{equation}
for the variables $x_j$ ($j = 1, 2, 3$) with the parameter $x_0 \in \mathbb{R}$ held fixed. (Invoking the language of special relativity, we might say that $x_3$ is a positive energy-like coordinate, $x_1$ and $x_2$ are momentum-like coordinates, and $|x_0|$ is the corresponding invariant mass parameter of a relativistic particle.) The non-compact group $G$ acts transitively on each of the hyperboloids $\mathrm{H}_{x_0}^2$ by conjugation $Q \mapsto g Q g^{-1}$. Since the subgroup $\mathbb{Z}_2 \subset G$ of elements $\pm \mathbf{1}$ acts trivially, we may replace $G = \mathrm{SU}(1,1)$ by the Lorentz group $\mathrm{SU}(1,1) / \mathbb{Z}_2 \cong \mathrm{SO}(1,2) \equiv G^\prime$. Doing so, we may refer to $G$-invariant objects as ``Lorentz invariants''.

\begin{figure}[ht]
\centering
\begin{minipage}[c]{0.55\textwidth}
\includegraphics[width=0.4\textwidth]{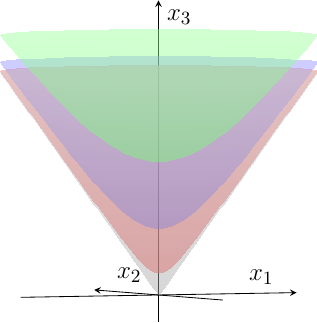}
\end{minipage}
\begin{minipage}{0.4\textwidth}
\caption{Sketch of four hyperboloids in the non-convex cone $C^3$: the special orbit with zero mass $|x_0|=0$ (a.k.a.\ light cone) in gray and three other hyperboloids $\mathrm{H}^2_{x_0 > 0}$.}
\end{minipage}
\label{Pic C3}
\end{figure}

\subsection{Types of order-parameter orbits}\label{types}

Our goal is to determine the extremal states that go into the decomposition (\ref{eq:YFL-Yx}). [Please be reminded that by an ``extremal state'' we mean a solution of the AAT equation (for $\varepsilon = 0$), not a mixture of such solutions.] We have seen in Sect.~\ref{sect:closeq} that the system of two equations, (\ref{eqn4Y-short}) and (\ref{eq:derY-x0}), is $\mathrm{SU}(1,1)$-equivariant for $\varepsilon = 0$. Thus, given an extremal state $Y$, every
\begin{equation}
   {}^g Y(Q) \equiv Y(g^{-1} Q g), \quad g \in G, 
\end{equation}
is again an extremal state (for $\varepsilon = 0$). Let $S\subset G$ denote the stabilizer of $Y$, i.e., ${}^s Y = Y$ for $s \in S$. The family of symmetry-related extremal states is then naturally identified with the homogeneous space
\begin{equation}
    G \cdot Y\cong G/S .
\end{equation}
In Sect.~\ref{Decomposition} we denoted the metallic stabilizer by $K$, where $K$ is a maximal compact subgroup. Here we use the more general notation $S$, since the stabilizer need not be compact. As the system parameters are varied, the stabilizer $S$, and hence the homogeneous space $G/S$, may change. We refer to the orbit $G/S$ realized at fixed parameters as the order-parameter orbit. The collection of extremal states, as the system parameters are varied, thus forms an abstract $G$-space. At fixed system parameters, the order-parameter orbit is the $G$-orbit of an extremal state in this space.

For the present model, the orbit types realized by the extremal states  happen to correspond to the orbit types of the $G$-action on the field target space $C^3$. Thus, the latter provide a simple geometric guide to the possible patterns of spontaneous symmetry breaking. Let us therefore examine the orbit types of the $G$-action on $C^3$.

For $x_0\neq 0$, the $G$-orbits in $C^3$ are the two-dimensional hyperboloids $\mathrm{H}^2_{x_0} \cong \mathrm{H}^2$. Each point of such a hyperboloid is stabilized by some compact one-parameter subgroup $K$ isomorphic to $\mathrm{U}(1)$, so we have $\mathrm{H}_{x_0}^2 \cong G/K \cong \mathrm{SU}(1,1)/ \mathrm{U}(1)$. We will see below that the same orbit type $G/K$ is realized by the metallic extremal Gibbs states. 

Because the $G$-action on $C^3$ is linear, it transforms the zero matrix $Q = 0$ into itself. Thus $Q = 0$ constitutes a zero-dimensional $G$-orbit $G/G$ of point type. It is natural to associate the point-type orbit $G/G$ with the pp-type state, which shares with $Q = 0$ the full $G$-invariance.

Now the orbits $G/K$ of metallic and $G/G$ of insulating type do not exhaust the orbit types found in our target space $C^3$. In the subspace given by $x_0 = 0$, there exists another type, namely the ``light cone''
\begin{equation}
    \mathrm{H}^2_0 : \; x_3^2 - x_1^2 - x_2^2 = x_0^2 = 0 \quad (x_3 > 0) ,
\end{equation}
which is diffeomorphic to a punctured Poincar\'e disk, $\mathrm{D}^2\setminus \{ 0 \}$. Any point on this $G$-orbit is stabilized by some nilpotent one-parameter subgroup $N \subset G$, so the orbit is two-dimensional and of type $G/N$. (Note that the nilpotent subgroups $N$ of $G = \mathrm{SU}(1,1)$ form a one-parameter family, whereas the compact subgroups $K$ are a two-parameter family. This distinction is not reflected in the Table below.)

In summary, motivated by the set of $G$-orbit types found in $C^3$, we anticipate that we are going to encounter 3 types of order-parameter orbits, namely $G/K$, $G/G$, and $G/N$, for extremal Gibbs states of metallic, insulating, and sc-type, respectively. It will be convenient to model these by stratification maps $\Psi_S : \; G/S \to X_S$, where $X_K = \mathbb{R}^2$, $X_G = \{ 0 \}$, and $X_N = \mathbb{R}^2 \setminus \{0\}$. Acting on $G/S$ by left translation [$L_{g_0}: \; gS \mapsto (g_0 g)S$], group elements $g \in G$ act on $X_S$ by $\Psi_S \circ L_g \circ \Psi_S^{-1}$; thus each model space $X_S$ is naturally a $G$-space. In this stratification, adjoining the point-type orbit $G/G$ to the nilpotent orbit $G/N$ amounts to completing
\begin{equation}
    \mathbb{R}^2\setminus\{0\} \to \big( \mathbb{R}^2\setminus\{0\} \big) \cup \{0\} = \mathbb{R}^2,
\end{equation}
with the added point $\{0\}$ fixed by the full group $G$.

In the following subsections, we address the ac- and pp-type extremal Gibbs states and verify their association with the orbit types $G/K$ and $G/G$, respectively. Afterwards, in Sect.~\ref{sc crit}, we return to the geometric picture sketched here and show how it anticipates the emergence of the nilpotent orbit $G/N$ at criticality.

We summarize the anticipated types of order-parameter orbits $G/S$ and their model spaces $X_S$ in a table:
\begin{center}
	\begin{tabular}{||cc||}
	\hline
	orbit type & extremal Gibbs state(s)  \\
	[0.5ex]
	\hline\hline
	$G/K \cong \mathbb{R}^2$ & ac-type   \\
	\hline
	$G/G \cong \{0\}$  & pp-type  \\
	\hline
	$G/N \cong \mathbb{R}^2\setminus\{0\}$ & critical/ sc-type  \\
	\hline
    \end{tabular}
\end{center}

\subsection{$\mathrm{U}(1)_{\mathrm{i}\sigma_3}$-symmetry reduction}\label{sect:U1red}

With the geometric guideline given above, we now start the analysis of the solutions of the ``bosonic'' AAT equation (\ref{eq:eqn4YBB}). A natural first step is to study the extremal Gibbs state with $\mathrm{U}(1)_{\mathrm{i} \sigma_3}$-symmetry. Though technically rather simple, this study already delivers some interesting results, and it has the charm that we can test the whole formalism by computing various expressions explicitly, for parameter values in some limiting regimes. Still keeping the notation short, we here denote the $\mathrm{U}(1)_{\mathrm{i} \sigma_3}$-invariant solution of the AAT equation by $Y(Q)$. By its extension to $\overline{C_+^4} \supset C^3$, this solution depends on two variables only: the Lorentz-invariant function $x_0$ and the $\mathrm{U}(1)_{\mathrm{i} \sigma_3}$-invariant function $x_3$, both defined as quasi-Cartesian coordinates for $C_+^4$ in Eq.~(\ref{eq:P-from-xmu}).

To pull back the analysis from $\overline{C_+^4}$, we introduce two coordinate functions $p, q$ on $C^3$ by
\begin{equation}\label{eq:pullx03}
    p = i^\ast (x_3 + x_0) , \quad q = i^\ast (x_3 - x_0) ,
\end{equation}
and we define their partial derivatives $\partial / \partial p$ and $\partial / \partial q$ by the convention that $i^\ast x_1$ is held fixed. (The special choice of $x_1$ is made because in later sections we will study a class of $\mathrm{U}(1)_{\mathrm{i}\sigma_3}$-symmetry broken solutions $Y(Q)$ which have an additional dependence on $i^\ast x_2$.) Now for functions $Y$ on $\overline{C_+^4}$ that depend on $x_3$, $x_0$ (but not on $x_1$, $x_2$) the result $\Delta Y$ of applying the Laplacian (\ref{eq:def-Delta}) restricts to
\begin{equation}
    i^\ast (\Delta Y) = i^\ast \frac{1}{4} \left( \frac{\partial^2} {\partial x_3^2} - \frac{\partial^2} {\partial x_0^2} \right) Y = \frac{\partial^2}{\partial p \, \partial q} i^\ast Y \equiv \frac{\partial^2} {\partial p \, \partial q} Y .
\end{equation}
[Here and in the following, we abuse the notation by simply writing $Y$ instead of $i^\ast Y$ (as the meaning will always be clear from the context).] We see that $i^\ast \Delta Y$ is now expressed by $C^3$-data alone. Hence the equation (\ref{eq:eqn4YBB}) for $Y$ is already closed, and there is no need to carry along the extra information in $\partial Y / \partial x_0$.

Let us then make the closed equation (\ref{eq:eqn4YBB}) explicit in $p$ and $q$. Another view of (\ref{eq:pullx03}) is $p = |u|^2 \geq 0$ and $q = |v|^2 \geq 0$, where $u,v$ appeared in the early going of Sect.~\ref{sect:FLT}. To complete the coordinate system for $C^3$, we augment $p, q$ by an angular variable, $\phi = \mathrm{arg}(\bar{u} v)$. By push forward of the initial measure $\pi^{-2} d^2 u \, d^2 v$, the integration measure $d\mu(Q)$ on $C^3$ is expressed as
\begin{equation}
    d\mu(Q) = dp \, dq \, \frac{d\phi}{2\pi} ,
\end{equation}
and the exponent of the integral kernel $\mathrm{e}^{- w_T^2 \mathrm{Tr}\, Q Q^\prime}$ becomes
\begin{equation}
    \mathrm{Tr}\, Q Q^\prime = p p^\prime + q q^\prime - 2 \sqrt{p q p^\prime q^\prime} \cos(\phi - \phi^\prime) .
\end{equation}
Since the $\mathrm{U}(1)_{\mathrm{i}\sigma_3}$-invariant function $Y$ does not depend on $\phi$, we may average the kernel over angles to arrive at
\begin{equation}\label{eq:eqn4Y-U1}
    Y(p,q) = \mathrm{e}^{\mathrm{i}E(p-q) - \varepsilon (p+q) - \frac{1}{2} w_V^2 (p-q)^2} \left( \int_{\mathbb{R}_+^2} dp^\prime dq^\prime \, \mathrm{e}^{-w_T^2 (p p^\prime + q q^\prime)} I_0(2w_T^2 \sqrt{p q p^\prime q^\prime}) \frac{\partial^2} {\partial p^\prime \partial q^\prime} Y(p^\prime, q^\prime) \right)^K ,
\end{equation}
with $I_0$ a modified Bessel function of the first kind.

It turns out that some special values of the solutions $Y$ of Eq.~(\ref{eq:eqn4Y-U1}) are already determined by a dimensionally reduced equation -- these are the function values on the two subspaces
\begin{equation}
    H^\pm : \; x_3 \pm x_0 = 0 , \quad x_1 = x_2 = 0 ,
\end{equation}
which are central half-axes in $C^3$. To see how the reduction works, set $q \equiv 0$ in Eq.~(\ref{eq:eqn4Y-U1}). The integral kernel then becomes independent of $q^\prime$, so that the derivative with respect to $q^\prime$ is a total one and can be integrated. Using that $Y(p^\prime,\infty) = 0$, and abbreviating $Y(p,0) \equiv Z_\varepsilon (p)$, we obtain the reduced equation
\begin{equation}\label{eq:eqn4Z}
    Z_\varepsilon (p) = \mathrm{e}^{\mathrm{i} (E + \mathrm{i}\varepsilon) p - \frac{1}{2} w_V^2 p^2} \bigg( - \int_0^\infty \!\!\!\! dp^\prime \mathrm{e}^{- w_T^2 p p^\prime} \frac{\partial}{\partial p^\prime} Z_\varepsilon (p^\prime) \bigg)^K ,
\end{equation}
which determines $Z_\varepsilon (p)$ for $p \geq 0$. To extend $Z_\varepsilon$ to the negative real axis, we vary the blueprint above: we set $p = 0$, and $Y(0,q) = Z_\varepsilon(-q)$, and so on, in order to obtain the variant of Eq.~(\ref{eq:eqn4Z}) which determines the second branch of $Z_\varepsilon$, where its argument is negative. That branch glues continuously to the first one because $Z_\varepsilon (0+) = Z_\varepsilon (0-) = 1$. The finished product of the construction is a smooth function $\mathbb{R} \ni r \mapsto Z_\varepsilon(r)$ with the property $Z_\varepsilon (-r) = \overline{Z_\varepsilon (r)}$.

Eq.~(\ref{eq:eqn4Z}) is relatively easy to analyze. Its solution interpolates between $Z_\varepsilon (r=0) = 1$ and $Z_\varepsilon (r \to \pm \infty) = 0$. For the special energy $E = 0$, the function $Z_\varepsilon (r)$ is real-valued, even, positive, and monotonically decreasing on $\mathbb{R}_+$.

\subsection{Mean local density of states in high dimension}\label{sect:lDoS}

We recall that in the present paper we are mainly interested in the regime of high dimension and strong disorder. Now for a large dimension parameter $K \gg 1$, the analysis of Eq.~(\ref{eq:eqn4Z}) simplifies due to $\lim_{K \to \infty} (1 - y/K)^K = \mathrm{e}^{-y}$. In order to exploit that limit law, we change variables by $y = p \sqrt{K} w_T$, $y^\prime = p^\prime \sqrt{K} w_T$, and we introduce the dimensionless parameters $w^2 / K = w_V^2 / K w_T^2$ and $\eta = (E + \mathrm{i}\varepsilon) / \sqrt{K} w_T$, which are to be kept fixed as $K$ grows large. By partially integrating in Eq.~(\ref{eq:eqn4Z}) we are then led to the scaled equation (let $Z_\varepsilon \equiv Z$ for now)
\begin{equation}\label{eq:eqn4Zs}
    Z(\cdot) = \mathrm{e}^{\mathrm{i}\eta y - \frac{1}{2} w^2 y^2 / K} \bigg( 1 - \frac{y}{K} \int_0^\infty \!\!\!\! dy^\prime Z(\cdot^\prime) \, \mathrm{e}^{- y y^\prime / K} \bigg)^K .
\end{equation}
As will be seen presently, the integral of $Z$ over $\mathbb{R}_+$ remains finite in the proposed scaling limit for large $K$. In anticipation of that, we take $K \to \infty$ in (\ref{eq:eqn4Zs}) to obtain the equation
\begin{equation}
    Z(\cdot) = \exp \left( \mathrm{i}\eta y - \frac{w^2 y^2}{2K} - y \int_0^\infty \!\!\!\! Z(\cdot^\prime) \, dy^\prime \right) \quad (K \to \infty) .
\end{equation}
Its solution for $w^2 = 0$, the limit of vanishing disorder potential (referred to as the ``weak-disorder limit'' for short), is exponential in $y = p \sqrt{K} w_T$:
\begin{equation}
    Z = \mathrm{e}^{- \big( \sqrt{1 - (\eta/2)^2} - \mathrm{i}\eta / 2\big) y} .
\end{equation}
Using $\overline{\rho_n(E)} = \pi^{-1} \mathrm{Im}\, \mathbb{E}(G_n^-) = \mathrm{Re} \int_0^\infty Z(p)\, dp$, we then see that the mean local density of states $\overline{\rho_n}$ obeys Wigner's semicircle law:
\begin{equation}
    \overline{\rho_n(E)} \, dE = \frac{1}{\pi} \sqrt{1 - E^2/(4K w_T^2)} \, \frac{dE}{\sqrt{K} w_T} ,
\end{equation}
in the form of an energy band $- 2 \sqrt{K} w_T \leq E \leq 2 \sqrt{K} w_T$ of width set by the random hopping with strength $w_T$.

Our interest, however, is in the opposite limit of large $w^2/K = w_V^2 / K w_T^2$ (``strong disorder''). In that case, the solution for $Z$ is a Gaussian with offset:
\begin{equation}\label{eq.Z4strongdisorder}
   Z = \mathrm{e}^{\mathrm{i} \eta y - \frac{1}{2} w^2 y^2 / K} .
\end{equation}
By returning to the variable $p$ and integrating as before, we obtain a result that one would expect:
\begin{equation}
    \overline{\rho_n(E)}\, dE = \frac{dE}{\pi}\, \mathrm{Re}\, \int_0^\infty \!\!\!\! Z(p)\, dp = \exp \left( - \frac{E^2}{2 w_V^2} \right) \frac{dE}{ \sqrt{2\pi}\, w_V} ,
\end{equation}
Indeed, by simple reasoning, the mean local density of states in the strong-disorder limit is given by the distribution of the on-site potential disorder (which is Gaussian for the Wegner model under study).

\subsection{Solution of pp-type}\label{sect:pptype}

We now communicate that Eq.~(\ref{eq:eqn4Y-U1}) for $\varepsilon \to 0$ always has a solution, $Y_{\rm pp}$, which possesses the full Lorentz symmetry. (As a disclaimer, we must caution that, while this solution always exists, it is unstable outside the pp-phase.)

Taking Lorentz invariance for granted, the solution $Q \mapsto Y_{\rm pp}(Q)$ is uniquely determined as follows. First, assume $\mathrm{Tr}\, \sigma_3 Q > 0$. In that case, there exists a Lorentz transformation $Q \mapsto g Q g^{-1}$ that conjugates $Q$ to $\mathrm{diag}(p,0)$ with $p > 0$, and we must have $Y_{\rm pp}(Q) = Z_{\varepsilon = 0} (p)$  as explained at the end of Sect.~\ref{sect:U1red}. Second, if $\mathrm{Tr}\, \sigma_3 Q < 0$, then we Lorentz-conjugate to $\mathrm{diag}(0,-q)$ (still with $q > 0$) and set $Y_{\rm pp}(Q) = Z_0 (-q)$. Third, the values of $Y_{\rm pp}(Q)$ for $\mathrm{Tr}\, \sigma_3 Q = 0$ (in which case $Q$ typically fails to be diagonalizable) are given by continuous extension. The outcome of the whole procedure can be succinctly summarized as
\begin{equation}\label{eq.Ypp}
    Y_{\rm pp}(Q) = Z_0 (\mathrm{Tr}\, Q) .
\end{equation}
By construction, the function $Y_{\rm pp}$ so defined is the unique Lorentz-invariant function that satisfies the requirement of taking the boundary values prescribed by Eq.~(\ref{eq:eqn4Z}) on the two half-axes $H^\pm \subset C^3$. Now, on physical and mathematical grounds, the pp-phase certainly exists and is thermodynamically stable for strong enough disorder. Therefore, the solution $Y_{\rm pp}$ is expected to be stable with respect to generic perturbations in the strong-disorder regime ($w_V^2 \gg K w_T^2$). 

\begin{figure}[ht]
\centering
\begin{minipage}[c]{0.6\textwidth}
\includegraphics[width=1\textwidth]{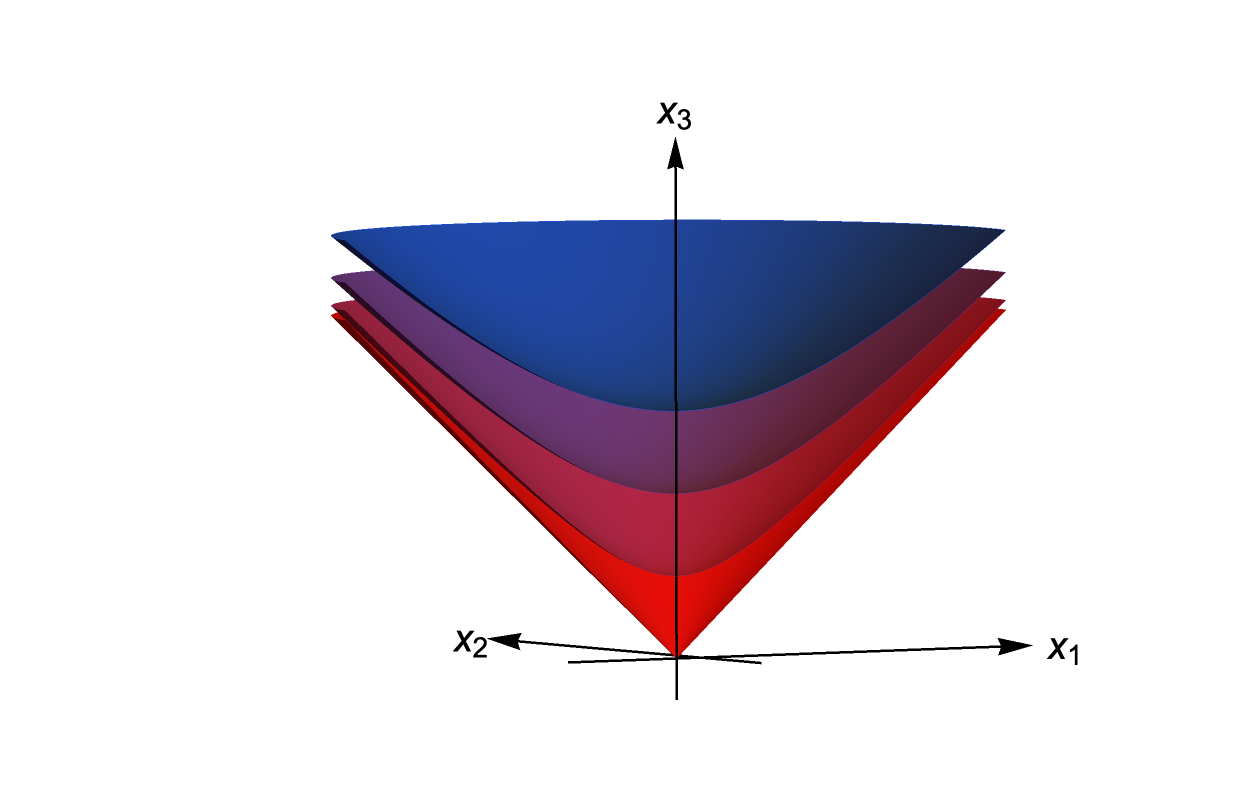}
\end{minipage}
\begin{minipage}{0.35\textwidth}
\caption{Sketch of the solution $Y_{\rm pp}$ for $x_0\geq 0$. It is constant on every hyperboloid $\mathrm{H}^2_{x_0}$ but decays in the direction of the massive variable $x_0$. The decay is indicated by a change of color from red to blue. For better visibility the hyperboloids are cut in the middle.}
\end{minipage}
\label{Pic pp sol}
\end{figure}

Having constructed the pp-solution for $\varepsilon = 0$, we now argue that it does correspond to the spectral type pp. For that, we need the solution for small but finite $\varepsilon > 0$. We first observe that the AAT equation has a unique stable solution in the pp phase. In other words, the order-parameter orbit is the trivial space $X = G/G$, the mixture in Eq.~(\ref{spectral integral}) is trivial, and $Y_{\rm pp}^{\varepsilon}$ coincides with the physical state $Y_{\rm FL}$. Deep in the pp-phase, where the correlation length (a.k.a.\ localization length) is small, one expects the simple approximation $Y_{\rm pp}^\varepsilon \approx \mathrm{e}^{- 2\varepsilon x_3} Y_{\rm pp}^{\varepsilon = 0}$ to be qualitatively correct. As the critical point is approached, the dependence on $\varepsilon$ will be somewhat modified, but we still expect a plateau in $x_3$ up to a scale set by $\varepsilon^{-1}$. By consequence, the radial marginal $M_{\rm FL}(x_3)$ of $Y_{\rm FL} = Y_{\rm pp}^\varepsilon$ has a plateau up to the cutoff scale $x_3 \sim \varepsilon^{-1}$ and we obtain
\begin{equation}\label{YFL and YQinf pp phase}
    K_{n,\varepsilon} = \int d\mu(Q_n)\, Y_\mathrm{pp}^\varepsilon (Q_n) = \int dx_3\, M_{\rm FL}(x_3) \sim \overline{\rho_n(E)}\, \varepsilon^{-1} \quad (\varepsilon \to 0). 
\end{equation}
As we have learned in \ref{sect:III-B} and \ref{sect:III-C}, this behavior is characteristic of a pp-type spectrum.

\subsection{Solutions of ac-type}\label{sect:ac}

In the opposite limit of weak disorder ($w_V^2 \ll w_T^2$), $K$ large and energy $E$ not too far from the band center, the solution $Y_{\rm pp}$ is certain to be unstable with respect to generic perturbations. [In other words, the recursion scheme explained at the end of Sect.~\ref{SubSec:AAT self-consistency} will converge to fixed points different from $Y_\mathrm{pp}$.] Guided by the established theory of weakly disordered metals \cite{Efe83,Weg79a}, we expect a different type, $Y_{\rm ac}$, of stable solutions describing metallic behavior. The $\mathrm{U}(1)_{\mathrm{i} \sigma_3}$-invariant solution $Y_{\rm ac}$ exhibits fast decay in the variables $p$ and $q$, thereby breaking the non-compact symmetry of the one-parameter subgroups generated by $\sigma_1$ and $\sigma_2$ in the symmetry group $\mathrm{SO} (1,2) \cong \mathrm{SU}(1,1) / \mathbb{Z}_2$.  
As a demonstration of principle, let us present the solution $Y_{\rm ac}$ in the deep-metallic limit of $w_V^2 = 0$ and $K \to \infty$. Invoking the large-$K$ scaling of Sect.~\ref{sect:lDoS}, we see that the Bessel function $I_0$ in Eq.~(\ref{eq:eqn4Y-U1}) can be approximated as $I_0 \approx 1$ in that limit. The integral over $p^\prime, q^\prime$ then separates, and so does the solution:
\begin{equation}\label{eq:Yac}
  Y_{\rm ac}(p,q) = Z_0(p)\, \overline{Z_0(q)} = \mathrm{e}^{- \sqrt{K w_T^2 - (E/2)^2} (p+q) + \mathrm{i} (E/2) (p-q)} \qquad (w_V^2 = 0 , \quad K \gg 1) .
\end{equation}

For the related case of the Anderson model, a solution of the type of Eq.~(\ref{eq:Yac}) was exhibited in the work of \cite{MF1992} using methods similar to ours. 

\subsection{Metallic extremal Gibbs states and ac-type behavior}\label{sect:ac boost}

We now look more closely into the geometry of the mechanism of spontaneous symmetry breaking in the metallic phase. The results of the present subsection will pave the way for a full understanding of the critical regime and especially the construction of the critical sc-type states. The latter will be the subject of the final sections.

To begin, notice that the weak-disorder solution (\ref{eq:Yac}) inside the energy band ($|E| < 2 \sqrt{K} w_T$) has finite integral norm:
\begin{equation}\label{eq:fin-norm}
    \int_{C^3} \!\! d\mu(Q) \, Y_{\rm ac}(Q) < \infty .
\end{equation}
This is a general feature (not restricted to the weak-disorder limit) of metallic solutions $Y_{\rm ac}$. A hallmark of spontaneous symmetry breaking, the finite norm results from the fast decay of $Y_{\rm ac}$ (which persists for a vanishing regulator $\varepsilon$) in the non-compact directions traced out by the symmetry group $G = \mathrm{SU}(1,1)$.

We wish to translate (\ref{eq:fin-norm}) into $K_{n,\varepsilon}$ being finite. To that end, first let $\varepsilon = 0$, in which case the AAT equation has the full $G$-symmetry. Given the solution $Y_{\rm ac}$ (augmented by $\partial Y_{\rm ac} / \partial x_0$), we obtain a related solution ${}^g Y_{\rm ac}$ [augmented by $\partial ({}^g Y_{\rm ac}) / \partial x_0$] by making a Lorentz transformation:
\begin{equation}\label{Yac boost}
    {}^g Y_{\rm ac}(Q) = Y_{\rm ac}(g^{-1} Q g) \qquad (g \in G).
\end{equation}
If we transform by $k \in K = \mathrm{U}(1)_{\mathrm{i}\sigma_3} \subset G$, the solution stays the same: 
${}^k Y_{\rm ac}(Q) = Y_{\rm ac}(Q)$. So, the metallic order-parameter orbit is $X = G/K$ and the incoherent mixture (\ref{eq:YFL-Yx}) of extremal Gibbs states takes the form
\begin{equation}
    Y_{\rm FL}(Q_n) = \int_{G/K} \!\! dg_K \, P(gK) \; {}^g Y_{\rm ac}(Q_n) ,
\end{equation}
with normalized boundary measure, $\int_{G/K} dg_K \, P(gK) = 1$. Because Lorentz transformations $g \in G$ are isometries of $\big( C^3 , d\mu(Q) \big)$, the integral norm is invariant:
\begin{equation}\label{g isometry}
    \int_{C^3} d\mu(Q) \, {}^{g} Y_{\rm ac}(Q) = \int_{C^3} d\mu(Q) \, Y_{\rm ac}(Q) \qquad (g \in G) .
\end{equation}
Hence the integral over the order-parameter orbit factors out and we obtain
\begin{equation}
    \int d\mu(Q_n) \, Y_{\rm FL}(Q_n) = \int d\mu(Q_n) \, Y_{\rm ac}(Q_n) < \infty .
\end{equation}

While all this was for $\varepsilon = 0$, we can easily infer a statement about $K_{n,\varepsilon}$ for $\varepsilon > 0$. As a time-integrated return probability with death rate $2\varepsilon$, the function $\varepsilon \mapsto K_{n,\varepsilon}$ is monotonically decreasing:
\begin{equation}
    \lim_{\varepsilon \to 0} K_{n,\varepsilon} \geq K_{n,\varepsilon} . 
\end{equation}
Now by dominated convergence we may take the limit inside the integral to obtain the desired inequality:
\begin{equation}
    K_{n,\varepsilon} \leq \lim_{\varepsilon \to 0} \int d\mu(Q_n) \, Y_{\rm FL}^\varepsilon (Q_n) = \int d\mu(Q_n) \, Y_{\rm FL}^{\varepsilon = 0} < \infty .
\end{equation}
By the material of Sect.~\ref{sect:spec} it follows that the solutions ${}^g Y_{\mathrm{ac}}$ and their order-parameter orbit $G/K \cong \mathbb{R}^2$ are associated with an ac spectrum of the random Hamiltonian $h$ in Eq.~(\ref{eq:Wegner1}).

In the remainder of this section, we inquire into the fate of ${}^g Y_{\rm ac}$ under extreme Lorentz boosts $g \in G$. Here our first observation is that the non-compact space $G/K$ has an asymptotic boundary at infinity -- in the Poincar\'e disk model $\mathrm{D} = \{ z \in \mathbb{C} \mid |z| < 1 \}$ of $\mathrm{H}^2$ this is the circle $\mathrm{S}^1 = \{ z \in \mathbb{C} \mid |z| = 1 \}$, and that circle is a $G$-orbit. The existence of such a boundary distinguishes the present setting from the ``classical'' one with a compact and closed order-parameter space, say the two-sphere $\mathrm{S}^2$ for a model of interacting vectors with compact symmetry group $\mathrm{O}(3)$. 

What happens when ${}^g Y_\mathrm{ac}$ is Lorentz-boosted toward the boundary circle? For simplicity, we look at the band center $E = 0$ and consider Lorentz transformations generated by the vector field $K_{23} \equiv x_3 \, \partial / \partial x_2 + x_2 \, \partial / \partial x_3$ (on $C^3$). Such transformations take an especially simple form when expressed in terms of horospherical $\mathbb{R}$-valued coordinate functions $\varphi$, $b$, and $m$ on $C^3 \setminus \{ Q = 0 \} \cong \mathbb{R}^3 \setminus \{0\}$:
\begin{equation}\label{eq:horoscoord}
    i^\ast x_0 = w_T^{-1} m, \quad i^\ast x_1 = - w_T^{-1} b, \quad i^\ast x_2 = \frac{1}{2} w_T^{-1} \big( \mathrm{e}^\varphi - (m^2 + b^2) \, \mathrm{e}^{-\varphi} \big), \quad i^\ast x_3 = \frac{1}{2} w_T^{-1} \big( \mathrm{e}^\varphi + (m^2 + b^2) \, \mathrm{e}^{-\varphi} \big),
\end{equation}
(The tip $Q = 0$ of $C^3$ is reached by setting $m = b = 0$ and sending $\varphi \to -\infty$.)
The function $m = \frac{1}{2} w_T \mathrm{Tr}\, Q$ is Lorentz-invariant. The coordinate lines of $b$ are horocycles, and $\varphi$ is the coordinate associated with Lorentz-boost geodesics in the $x_2 x_3$-plane. The variables $b$ and $\varphi$ are symmetry degrees of freedom. In these coordinates, the vector field $K_{23}$ simply generates translations of the coordinate $\varphi$.

It can be checked that for functions on $\overline{C_+^4}$ that depend on $x_3$, $x_0$, $x_2$ (but not on $x_1$) the result $\Delta Y$ of applying the Laplacian (\ref{eq:Lapl-restr}) restricts to $C^3$ as
\begin{equation}\label{Laplace horosph}
    i^\ast(\Delta Y) = i^\ast\frac{1}{4} \left(\frac{\partial^2}{\partial x_3^2} - \frac{\partial^2}{\partial x_0^2} - \frac{\partial^2}{\partial x_2^2}\right)Y = w_T^2 \left( -\frac{1}{4} \left(\frac{\partial^2}{\partial m^2} + \frac{\partial^2}{\partial b^2}\right) + \frac{1}{2b}\frac{\partial}{\partial b}\circ D\right)i^\ast Y.
\end{equation}
(As of now, we again suppress the pullback symbol: $i^\ast Y\equiv Y$, etc.) Note that the equation (\ref{eq:eqn4YBB}) for $Y$ is already closed, and once more, there is no need to carry along the extra information in $\partial Y / \partial x_0$.

The restricted Laplacian (\ref{Laplace horosph}) is translation-invariant in $\varphi$. Consequently, the restricted AAT equation (\ref{eq:eqn4YBB}) is equivariant under the Lorentz boost $g : \; \varphi \mapsto \varphi - \tau$ ($\tau \in \mathbb{R}$). In the large-$K$ metallic regime of $w_T^2 \ll w_V^2 \ll K w_T^2$, the $\mathrm{U}(1)_{\mathrm{i} \sigma_3}$-invariant solution $Y_\mathrm{ac}$ takes the form
\begin{equation}\label{Gaussian type ac soln}
    Y_\mathrm{ac} \approx \exp\left(-2 w_V^2 x_0^2 - 2 \sqrt{K} w_T x_3\right) ,
\end{equation}
by straightforward modification of (\ref{eq:Yac}). When expressed in the horospherical coordinate system, this becomes Gaussian in $m$ and $b:$
\begin{equation}
    Y_\mathrm{ac} = \exp\left( -2 w^2 m^2 - \sqrt{K} \left(\mathrm{e}^\varphi + (m^2 + b^2)\, \mathrm{e}^{-\varphi} \right) \right) ,
\end{equation}
and the solution is boosted ($g : \; \varphi \mapsto \varphi - \tau$) to
\begin{equation}\label{boosted Gaussian}
    {}^g Y_\mathrm{ac} = \exp\left(-2 w^2 m^2 - \sqrt{K} \left(\mathrm{e}^{\varphi - \tau} + (m^2 + b^2)\, \mathrm{e}^{-\varphi + \tau} \right)\right).
\end{equation}
We see that large boosts ($\tau \gg 1$) strongly suppress ${}^g Y_\mathrm{ac}$, unless $m^2 + b^2 = 0$. In the latter case, the dependence on $\varphi$ due to $\exp\big( - \sqrt{K} \, \mathrm{e}^{\varphi - \tau} \big)$ is weakened as $\tau$ grows. Thus, large boosts elongate the support of ${}^g Y_\mathrm{ac}$ in a neighborhood of the ``light ray''
\begin{equation}\label{boost ray}
    L_0 = \{(m,b) = (0,0),\ \varphi \in \mathbb{R} \} ,
\end{equation}
and the support of ${}^g Y_\mathrm{ac}$ concentrates on $L_0$ in the limit of $\tau \to \infty$.

Another description of the light ray $L_0$ is as the solution set of $x_0 = x_1 = 0$ and $x_2 = x_3 > 0$. This ray aims at the point $z := (x_1 + \mathrm{i} x_2) / x_3 = \mathrm{i}$ of the asymptotic boundary $\mathrm{S}^1$ of the Poincar\'e disk $\mathrm{D}$. Clearly, we could $\mathrm{U}(1)_{\mathrm{i} \sigma_3}$-rotate the ray $L_0$ to make it end at a different point of the boundary circle $\mathrm{S}^1$; in this way, we would obtain another large-boost limit of the metallic solution. The technical analysis would have to be adapted to a different class of functions (depending on $x_3$, $x_0$ and $x_1 \cos\phi + x_2 \sin\phi$ for some angle $\phi$), but there would be no essential change. Hence, keeping the general situation in mind, we will continue doing the analysis for functions $Y$ on the 3-dimensional space $C^3$ which extend to the 4-dimensional space $\overline{C_+^4}$ as functions that are independent of $x_1$.

\begin{figure}[H]
\centering
\includegraphics[width=0.8\textwidth]{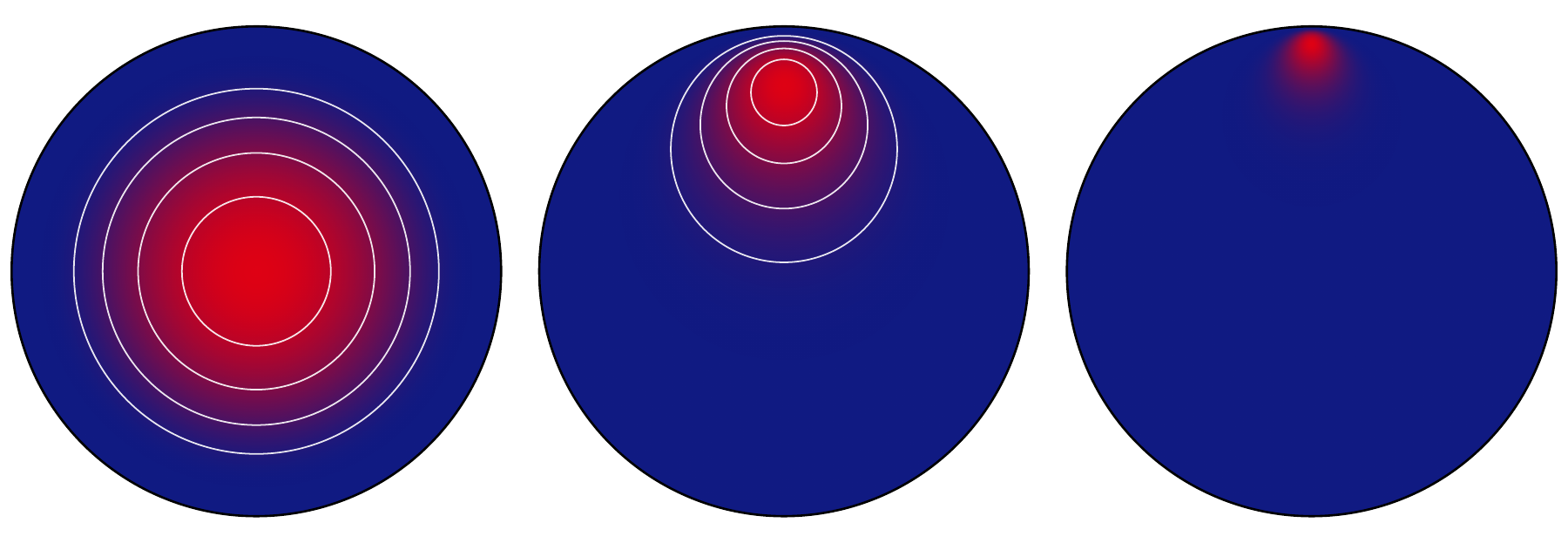}
\caption{Sketch of three $G$-equivalent metallic solutions on $\mathrm{H}^2_{x_0}$ for fixed $x_0 = w_T^{-1} m > 0$: the $\mathrm{U}(1)_{\mathrm{i} \sigma_3}$-invariant solution $Y_{\rm ac}$ on the left, a moderately boosted version ${}^g Y_{\rm ac}$ in the middle, and a strongly boosted one on the right. Each solution is invariant under some (boosted) compact subgroup $K \subset G = \mathrm{SU}(1,1)$ and hence constant on hyperbolic circles around its respective center. As before, the decay of the solution is indicated by a change of color from red to blue. Under large boosts, the support of ${}^g Y_{\rm ac}$ is pushed to a boundary point of the Poincar\'e disk model of the standard two-hyperboloid. That point corresponds to a light-like geodesic on $\mathrm{H}^2_{0}$.}
\label{Pic ac}
\end{figure}

We will now start our investigation of the critical regime, where the metallic extremal Gibbs states ${}^g Y_{\rm ac}$ restore the full $G$-symmetry.

\section{Harmonic analysis and location of critical manifold}\label{Harmonic analysis}

In this section, we derive an expression for the critical manifold which is asymptotically exact in the limit of large branching number $K\rightarrow\infty$:
\begin{equation}\label{Eq. strong coupling line of instability}
    K = \frac{1}{\sqrt{2}} \frac{w_c}{\ln w_c}\exp\left({\frac{E^2}{2w_V^2}}\right).
\end{equation}
Note that, apart from a logarithmic correction, the critical disorder strength $w = w_c$ is proportional to $K$. [By the ``critical manifold'' we mean the set of points in the $(K , w = w_V / w_T , E / w_V)$ parameter space where the pp solution (\ref{eq.Ypp}) becomes linearly unstable against perturbations $\delta Y$.] 

To derive Eq.\ (\ref{Eq. strong coupling line of instability}), it suffices to consider perturbations invariant under $\mathrm{U}(1)_{\mathrm{i}\sigma_3}$, since the Lorentz equivariance of the AAT equation (\ref{eq:eqn4YBB}) lets us transfer the criterion of (in-)stability to Lorentz-transformed perturbations. Linearization of the AAT equation yields a linear integro-differential operator, whose eigenfunctions we compute. This is possible because the pair $Y_\mathrm{pp},(\partial Y/\partial x_0)_\mathrm{pp}$ has the full Lorentz symmetry, opening the door for harmonic analysis on the target space $C^3$ with isometry group $\mathrm{SU}(1,1)$. Having found the dependence of the eigenfunctions on the symmetry degrees of freedom $g \in \mathrm{SU}(1,1)$, we can eliminate $g$ to obtain a linear operator in the Lorentz-invariant mass parameter $w_T^{-1} m$. It then remains to calculate the eigenvalues of this reduced operator. When the maximal eigenvalue reaches magnitude one, the point of linear instability of the pp solution is reached. In the strong-coupling limit $K \sim w \gg 1$, we are able to determine the maximal eigenvalue and the corresponding eigenfunction analytically. This will yield the formula for the critical manifold given in (\ref{Eq. strong coupling line of instability}). 

\subsection{Linearization}

To further our understanding of the full AAT equation and its solutions, we now address the question of linear stability of the pp solution $Y_\mathrm{pp},(\partial Y/\partial x_0)_\mathrm{pp}$: we perturb $Y_\mathrm{pp}$ by a function $\delta Y$ on $C^3$,
\begin{equation}\label{eq: pert4Ypp}
	Y = Y_\mathrm{pp} + s\, \delta Y \qquad (s \in \mathbb{R}) ,
\end{equation}
we insert this ansatz on both sides of the AAT equation, Eq.~(\ref{eq:eqn4YBB}), and we differentiate with respect to $s$ at $s=0$. The result is a linear mapping transforming the perturbation $\delta Y$ into a new perturbation $\delta Y^{\rm new} = \mathcal{L}\, \delta Y$ given by
\begin{equation}\label{eq:eqn4deltaYBB}
	\begin{aligned}
		&\delta X^{\text{new}}(Q) = \int d\mu(Q^\prime)\, \mathrm{e}^{-w_T^2\Tr QQ^\prime}\Delta\,\delta Y(Q^\prime),\\
		&\delta Y^\text{new}(Q) = K \, \frac{Y_\text{pp}(Q)}{X_{\text{pp}}(Q)} \, \delta X^{\text{new}}(Q) \equiv (\mathcal{L}\, \delta Y)(Q) .
	\end{aligned}
\end{equation}
(We have simplified the expression for $\delta Y^{\rm new}$ by using that $Y_{\rm pp}$ solves the AAT equation.) In the remainder of this section, we will be concerned with the spectral analysis of the integro-differential operator $\mathcal{L}$.

\subsection{Extracting the eigenfunctions}

To perform the ensuing calculations, we find it convenient to introduce hyperbolic polar coordinates:
\begin{equation}\label{eq.hyperbolicpolar}
i^\ast x_0 = w_T^{-1} m, \quad i^\ast x_3 = w_T^{-1} |m| \cosh\theta, \quad i^\ast x_1 = w_T^{-1} |m| \sinh\theta \, \cos\phi, \quad i^\ast x_2 = w_T^{-1} |m| \sinh\theta \, \sin\phi .
\end{equation}
Here $w_T^{-1} |m|$ is the Lorentz-invariant ``mass'' (of a ``relativistic particle'' with energy-like coordinate $i^\ast x_3$ and momentum-like coordinates $i^\ast x_1$ and $i^\ast x_2$), and the coordinates $\theta$ and $\phi$ are the hyperbolic radius and polar angle, respectively. The latter parameterize the symmetry orbits $\mathrm{H}^2_m \equiv \mathrm{H}^2_{x_0 = m/w_T}$ generated by the group action $Q \mapsto g Q g^{-1}$ on a base point $q_0 \in \mathrm{H}^2_m$:
\begin{equation}\label{Eq.eq4conjugation hyperbolic polar}
    i^\ast Q = g q_0 g^{-1}, \quad q_0 \equiv q_0(m) = \frac{1}{w_T} \big( m \cdot \mathbf{1} + |m|\, \sigma_3 \big) , \quad g = \mathrm{e}^{\mathrm{i} \phi \sigma_3 / 2} \mathrm{e}^{\theta \sigma_1 / 2}.
\end{equation}
When acting on functions that extend to $\overline{C_+^4}$ as functions independent of $x_1$, the Laplace operator (\ref{eq:Lapl-restr}) in these coordinates reduces to:
\begin{equation}\label{Laplace in polar}
    i^\ast \Delta = w_T^2 \bigg\{\frac{1}{4m^2} \left(-\frac{1}{\sinh\theta} \frac{\partial}{\partial\theta}\sinh\theta \frac{\partial}{\partial\theta} - \frac{1}{\sinh^2 \theta} \frac{\partial^2}{\partial\phi^2}\right) - \frac{1}{4} \frac{\partial^2}{\partial m^2} + \frac{1}{2m} \frac{\partial}{\partial m} \left( \frac{\cosh\theta}{\sinh\theta} \frac{\partial}{\partial\theta} - \frac{\tan\phi}{\sinh^2\theta} \frac{\partial}{\partial\phi}\right) \bigg\} \, i^\ast.
\end{equation}
For better readability, we will henceforth omit the pullback symbol $i^\ast$, with the understanding that all functions are restricted to $C^3$.

A natural choice for the eigenfunctions is a product ansatz, separating the dependence on the $G$-orbit coordinates $\theta, \phi$ from the dependence on the mass parameter $m$:
\begin{equation}\label{eq:prodansatz4deltaY}
    \delta Y(Q) = \sum_{n\in\mathbb{Z}} \sum_\lambda \delta y_{\lambda,n}(m) \, \Omega_{\lambda,n}(\theta) \, \mathrm{e}^{\mathrm{i} n \phi} .
\end{equation}
In Eq.\ (\ref{eq:prodansatz4deltaY}) $\lambda$ is a real number specifying the radial part of the eigenfunctions, $\Omega_{\lambda,n}$. As announced above, we restrict our considerations to the $\mathrm{U}(1)_{\mathrm{i} \sigma_3}$-invariant mode $n = 0$:
\begin{equation}
    \delta Y = \sum_{\lambda} \delta y_\lambda(m) \, \Omega_\lambda(\theta).
\end{equation}
Our next step is to find the functions $\Omega_{\lambda}$ and derive a reduced equation for the $m$-dependent perturbations $\delta y_\lambda(m)$. A plausible ansatz is that the $\Omega_{\lambda}$ are given by $\phi$-independent eigenfunctions, so-called zonal spherical functions $\varphi_\lambda (\theta)$, of the Laplace-Beltrami operator, $L_{\mathrm{H}^2}$, on $\mathrm{H}^2\cong \mathrm{SU}(1,1)/\mathrm{U}(1)$. This ansatz is motivated by the fact that the Laplace-Beltrami operator on $\mathrm{H}^2$ appears in our Laplace operator, more concretely in $\Delta^\flat$ defined in Eq.\ (\ref{eq:abbrev}), as
\begin{equation}
    - L_{\mathrm{H}^2} = \Delta^\flat - (D^2+D) = K_3^2 - K_2^2 -K_1^2 .
\end{equation}
Nonetheless, close inspection shows that zonal spherical functions on $\mathrm{H}^2$ are \underline{not} the right choice for $\Omega_\lambda(\theta)$.

One can see that they are not suitable by observing the following: when evaluated on the central half axes, $H^\pm$, which in the above polar coordinate system (\ref{eq.hyperbolicpolar}) are the level sets $\theta = 0$, the zonal spherical functions take the value $\varphi_\lambda (0) = 1$. This is at variance with the boundary condition for the perturbations $\delta Y$. Indeed, the solutions of the $\mathrm{U}(1)_{\mathrm{i} \sigma_3}$ symmetry-reduced equation (\ref{eq:eqn4Y-U1}) are constrained along the two central half axes to agree with the pp solution $Y_\mathrm{pp}$; see the end of Sect.\ \ref{sect:U1red}. Consequently, the functions $\Omega_\lambda$ must vanish along these two half axes in $C^3$. It turns out that the ``correct'' functions to use are connected to (but not identical to) the spherical functions with angular quantum number $\pm 1$. They are defined as follows. Let $\varphi_{\lambda, \pm 1}(\theta) \, \mathrm{e}^{\pm\mathrm{i}\phi}$ be an eigenfunction of $L_{\mathrm{H}^2}$ with eigenvalue $- (\lambda^2 + 1/4)$ ($\lambda\in\mathbb{R}$). Then the two functions $\varphi_{\lambda, \pm 1}(\theta) \equiv \widetilde{\varphi}_\lambda(\theta)$ are the same conical function and have the integral representation
\begin{equation}
    \widetilde{\varphi}_{\lambda}(\theta) = \int_0^{2\pi}\frac{d \phi}{2\pi}\,  \mathrm{e}^{\mp \mathrm{i}\phi} ( \cosh\theta + \sinh\theta \, \cos\phi)^{- \frac{1}{2} - \mathrm{i} \lambda}, \quad \lambda\in\mathbb{R}.
\end{equation}
As we shall see in the next subsection, the linear perturbation $\delta Y$ of the AAT solution $Y_{\rm pp}$ is correctly expanded as
\begin{equation}\label{eq: eigenfunction expansion of fixed mode}
	\delta Y(m,\theta) = \sum_{\lambda}\delta y_{\lambda}(m) \, \Omega_{\lambda}(\theta), \quad \Omega_{\lambda}(\theta) \equiv \widetilde{\varphi}_\lambda(\theta)\, \sinh\theta \, ,
\end{equation}
where $\sum_\lambda$ actually means an \emph{integral} over $\lambda$.

\subsection{Demonstration via partial integration}

In the $\mathrm{U}(1)_{\mathrm{i}\sigma_3}$-invariant sector the Laplace operator reduces to
\begin{equation}\label{eq.Delta U(1)polar}
	\Delta \, \delta Y(m,\theta) = w_T^2 \bigg(-\frac{1}{4m^2}\frac{1}{\sinh\theta}\frac{\partial}{\partial \theta} \sinh\theta \frac{\partial}{\partial \theta} - \frac{1}{4} \frac{\partial^2}{\partial m^2} + \frac{\cosh\theta} {\sinh\theta} \frac{\partial}{\partial\theta} \frac{1}{2m} \frac{\partial}{\partial m} \bigg) \, \delta Y(m,\theta),
\end{equation}
and the first of the two equations in (\ref{eq:eqn4deltaYBB}) becomes
\begin{equation}\label{eq:eq4deltaXn=0}
	\begin{aligned}
    &\delta X^\mathrm{new}(m,\theta) = \int d\mu(Q^\prime)\,  \mathrm{e}^{-2 m m^\prime - 2 |mm^\prime| (\cosh\theta \, \cosh\theta^\prime  - \sinh\theta \, \sinh\theta^\prime \, \cos(\phi-\phi^\prime))} \Delta \, \delta Y(m^\prime,\theta^\prime) \\
    &= 2 w_T^{-2} \int|m^\prime| d m^\prime\int\sinh\theta^\prime d\theta^\prime \mathrm{e}^{-2 mm^\prime - 2 |mm^\prime |\cosh\theta \, \cosh\theta^\prime} I_0(2 |mm^\prime| \sinh\theta \, \sinh\theta^\prime) \, \Delta \, \delta Y(m^\prime,\theta^\prime).
\end{aligned}
\end{equation}
Having ruled out the zonal spherical functions as a system in which to expand the perturbation $\delta Y$, one does not know immediately how to proceed. The following heuristic is offered as an inspiration. While we have made the reduction to an equation of purely ``bosonic'' type, our reduced AAT equation still reflects its supersymmetric origin by the presence of the Laplacian $\Delta$ due to the elimination of the Grassmann variables. Now the latent supersymmetry suggests that the harmonic analysis to be administered isn't that of $\mathrm{H}^2$ but rather that of its super-analog $\mathrm{H}^{2|2}$. [Without getting into technical details \cite{Zirn91} which are unnecessary here, let us just mention that the Laplacian on zonal spherical functions for $\mathrm{H}^{2|2}$ transforms into the Laplacian on the conical functions $\widetilde{\varphi}_\lambda$ for $\mathrm{H}^2$ by conjugation with $\sinh\theta$. This is the heuristic behind the factor $\sinh\theta$ in Eq.\ (\ref{eq: eigenfunction expansion of fixed mode}).] The practical hint we can take from this observation is that we should transfer the supersymmetry-remnant $\Delta$ from acting on $\delta Y$ to acting on the integral kernel in (\ref{eq:eq4deltaXn=0}). We are now going to do so by executing two partial integrations.

For the stated purpose, we recall from Sect.\ \ref{sect:U1red} that the Laplacian $\Delta$ acting on $\mathrm{U}(1)_{\mathrm{i}\sigma_3}$-invariant functions has the particularly simple form $\Delta = \partial^2 / \partial p \partial q$ in the coordinates $p, q$ given by
\begin{equation}
	m = w_T \frac{p-q}{2}, \quad |m|\cosh\theta = w_T \frac{p+q}{2},
\end{equation}
and that Eq.\ (\ref{eq:eq4deltaXn=0}) simplifies to
\begin{equation}\label{eq: linearized eq involving Delta_0 in MF coordinates}
	\delta X^\mathrm{new}(p,q) = \int_{\mathbb{R}^+ \times \mathbb{R}^+} \!\!\!\!\!\!\!\!\!\!\!\!\!\!\!\!\!\!\!
	dp^\prime dq^\prime \mathrm{e}^{- w_T^2 (pp^\prime + qq^\prime)} I_0 \big(2 w_T^2 \sqrt{pqp^\prime q^\prime}\big) \, \Delta \, \delta Y(p^\prime,q^\prime) .
\end{equation}
We emphasize again that the perturbations $\delta Y$ vanish on the central half axes $p = 0$ and $q = 0$. This causes the boundary terms of the intended partial integration with respect to $p^\prime$ and $q^\prime$ to vanish on that locus. The boundary terms at infinity require a more careful analysis, which is provided in Appendix \ref{App. boundary}.

The Laplacian acting on the integral kernel (after partial integration w.r.t.\ $p^\prime$ and $q^\prime$) takes the convenient form
\begin{equation}\label{Delta on kernel}
	\begin{aligned}
		&\frac{\partial^2}{\partial p^\prime\partial q^\prime}\bigg(\e^{-w_T^2(pp^\prime+qq^\prime)}I_0 \big( 2w_T^2\sqrt{pqp^\prime q^\prime} \big) \bigg) = w_T^2\sqrt{\frac{pq}{p^\prime q^\prime}}\left(1+w_T^2\frac{\partial}{\partial w_T^2}\right)\e^{-w_T^2(pp^\prime+qq^\prime)} I_{1} \big( 2w_T^2\sqrt{pqp^\prime q^\prime} \big) ,
	\end{aligned}
\end{equation}
where $\partial/\partial w_T^2$ denotes partial differentiation at fixed $p,q,p',q'$ (equivalently, at fixed $Q,Q'$). This relation is easily verified by using the identities
\begin{equation}
    I_0^\prime(x) = I_1(x), \quad I_1^\prime(x) = I_0(x) - \frac{I_1(x)}{x} .
\end{equation}

Next, we transform the right-hand side of Eq.\ (\ref{Delta on kernel}) back to hyperbolic polar coordinates and restore the polar-angle integral. The result is:
\begin{equation}\label{eq:partialintegratedeq4deltaXn=0}
	\begin{aligned}
		&\delta X^\text{new}(m,\theta) = w_T^2 \int d\mu(Q^\prime)\bigg|\frac{m}{m^\prime}\bigg|\frac{\sinh\theta}{\sinh\theta^\prime} \, \mathrm{e}^{\mathrm{i}(\phi - \phi^\prime)} \left(\left(1 + w_T^2\frac{\partial}{\partial w_T^2}\right) \mathrm{e}^{- w_T^2 \Tr QQ^\prime}\right) \, \delta \, Y(m^\prime,\theta^\prime).
	\end{aligned}
\end{equation}
Note that the r.h.s.\ of Eq.\ (\ref{eq:partialintegratedeq4deltaXn=0}) depends only on the difference $\phi - \phi^\prime$ and thus is independent of $\phi$ after the $\phi^\prime$ integration. Why is the partially integrated version (\ref{eq:partialintegratedeq4deltaXn=0}) useful? In short, the reason is that the integral kernel is separated into a Lorentz-invariant part (the integration measure and the factor in the big brackets) and the rest, which fails to be Lorentz-invariant. The correct ansatz for the eigenfunctions is now apparent:
\begin{equation}\label{eigenfcts Omega}
	 \Omega_{\lambda}(\theta) = \widetilde{\varphi}_{\lambda}(\theta) \, \sinh\theta ,
\end{equation}
since the conical functions $\widetilde{\varphi}_\lambda$ will be reproduced by integration against the Lorentz-invariant kernel in the large round brackets of Eq.~(\ref{eq:partialintegratedeq4deltaXn=0}). Note that the functions $\Omega_\lambda$ do satisfy the correct condition on the two center half axes $H^\pm$:
\begin{equation}
    \lim_{\theta \rightarrow 0} \Omega_\lambda(\theta) = 0 .
\end{equation}
Unlike the zonal spherical functions on $\mathrm{H}^2$, they behave as $\sim \mathrm{e}^{\theta / 2}$ for $\theta \rightarrow \infty$ (which is the asymptotic behavior of the zonal spherical functions for $\mathrm{H}^{2|2}$). In the following subsection, we will see that with the choice (\ref{eq: eigenfunction expansion of fixed mode}), we can reduce the linearized equation (\ref{eq:eqn4deltaYBB}) to a one-dimensional integral equation in the invariant mass parameter $m$.

\subsection{Intermezzo: Integral identity for spherical functions}

Here we inject an important identity from the harmonic analysis on $\mathrm{H}^2$ \cite{HEL}. Let $z \mapsto f(z)$ be a function on $\mathrm{H}^2 \cong \mathrm{SU}(1,1) / \mathrm{U}(1)_{\mathrm{i}\sigma_3}$. If $f$ solves the differential equation $L_{\mathrm{H}^2} f = - (\lambda^2 + \frac{1}{4})f$, then it satisfies the identity
\begin{equation}\label{eq: functional equation}
	 \int_{\mathrm{U}(1)_{\mathrm{i}\sigma_3}} \!\!\!\!\!\!\!\!\!\!\!\!\!\!\!\!
	 dk \, f(gk\cdot z) = f(g\cdot o) \, \varphi_\lambda (z), \quad g \in \mathrm{SU}(1,1) ,
\end{equation}
where $o = k \cdot o$ is the $\mathrm{U}(1)_{\mathrm{i}\sigma_3}$-invariant base point of $\mathrm{H}^2$, $\varphi_\lambda(z)$ denotes the zonal spherical function with eigenvalue $- (\lambda^2 + \frac{1}{4})$, and $dk$ is the unit-mass Haar measure on $\mathrm{U}(1)_{\mathrm{i}\sigma_3}$. The proof of this equation is short: because the invariant differential operator $L_{\mathrm{H}^2}$ commutes with left translations by elements of $\mathrm{SU}(1,1)$, the function on the left-hand side, $z \mapsto \int dk \,  f(gk\cdot z)$, is a $\mathrm{U}(1)_{\mathrm{i}\sigma_3}$-invariant eigenfunction of $L_{\mathrm{H}^2}$ with eigenvalue $ - (\lambda^2 + \frac{1}{4})$. Up to multiplication by constants, there exists only one such function, namely $\varphi_\lambda$. Therefore, the function on the right-hand side must be proportional to $z \mapsto \varphi_\lambda (z)$. The constant of proportionality is determined by setting $z = o$ and using $\varphi_\lambda (o) = 1$ and the normalization of the Haar measure $dk$.

\subsection{Reduction to a one-dimensional integral equation}

To proceed, we insert the ansatz of Eq.\ (\ref{eq: eigenfunction expansion of fixed mode}) into (\ref{eq:partialintegratedeq4deltaXn=0}). By linearity, it suffices to consider one term $\delta y_\lambda(m) \, \Omega_\lambda(\theta)$ in the sum of Eq.\ (\ref{eq: eigenfunction expansion of fixed mode}).  Thus, fixing any value of $\lambda$, we have
\begin{equation}\label{reduction 1}
    \delta X(m,\theta)^\text{new} = \sinh\theta \; \mathrm{e}^{\mathrm{i}\phi} \int d\mu(Q^\prime) \, w_T^2 \bigg|\frac{m}{m^\prime} \bigg| \, \delta y_{\lambda}(m^\prime) \bigg(1 + w_T^2 \frac{\partial}{\partial w_T^2}\bigg)\, \mathrm{e}^{-w_T^2\Tr QQ^\prime} \widetilde{\varphi}_\lambda(\theta^\prime) \, \mathrm{e}^{- \mathrm{i}\phi^\prime} 
\end{equation}	
for the new perturbation after one AAT iteration. The function $\widetilde{\varphi}_\lambda(\theta^\prime) \, \mathrm{e}^{- \mathrm{i}\phi^\prime} \equiv \Phi_\lambda(g^\prime \cdot o)$ on the right-hand side is an eigenfunction of $L_{\mathrm{H}^2}$ with eigenvalue $-(\lambda^2 + 1/4)$. This means that we have the option to use the relation (\ref{eq: functional equation}). 

To prepare that step, we write $Q = g q_0 g^{-1}$ for some $g \in \mathrm{SU}(1,1)$. We then make a substitution $g^\prime \to g g^\prime$, $Q^\prime \to g Q^\prime g^{-1}$ of integration variables and exploit the $\mathrm{SU}(1,1)$-invariance of the integration measure $d\mu(Q^\prime)$. This results in the following change of integrand:
\begin{equation}
    \mathrm{e}^{-w_T^2\Tr QQ^\prime} \Phi_\lambda(g^\prime \cdot o) \to \mathrm{e}^{-w_T^2 \Tr q_0 Q^\prime} \Phi_\lambda(gg^\prime\cdot o).
\end{equation}
Next, we make another substitution of integration variables: $g^\prime \to k g^\prime$, $Q^\prime \to k Q^\prime k^{-1}$ with $k \in \mathrm{U}(1)_{\mathrm{i}\sigma_3}$, and we use that $\mathrm{e}^{- w_T^2 \mathrm{Tr}\, q_0 Q^\prime}$ is invariant under such a substitution. Our integrand then changes as
\begin{equation}
    \Phi_\lambda(g g^\prime\cdot o) \to \Phi_\lambda(g k g^\prime\cdot o) ,
\end{equation}
all other factors remaining the same. Now since $k \in \mathrm{U}(1)_{\mathrm{i}\sigma_3}$ arose by a change of integration variables, we may take the Haar-average over $k$ without changing the expression for $\delta X^{\rm new}$. So, applying relation (\ref{eq: functional equation}) we get 
\begin{equation}
    \int dk \; \Phi_\lambda(g k g^\prime\cdot o) =  \Phi_\lambda(g \cdot o) \, \varphi_\lambda(g^\prime \cdot o) .
\end{equation}
By observing that $\Phi_\lambda(g \cdot o) \, \mathrm{e}^{\mathrm{i}\phi} \sinh\theta = \widetilde{\varphi}_\lambda(\theta) \sinh\theta = \Omega_\lambda(\theta)$, we obtain
\begin{align*}
	&\delta X(m,\theta)^\text{new} = \Omega_{\lambda}(\theta) \, w_T^2 \int d\mu(Q^\prime) \bigg|\frac{m}{m^\prime}\bigg| \, \delta y_{\lambda}(m^\prime) \bigg(1+w_T^2\frac{\partial}{\partial w_T^2}\bigg)\, \mathrm{e}^{-w_T^2\Tr q_0 Q^\prime} \varphi_\lambda (\theta^\prime).
\end{align*}
We see that we have reproduced the function $\Omega_{\lambda}(\theta)$, thereby confirming that $\Omega_\lambda$ is the good eigenfunction to use for the dependence of the perturbation on the hyperbolic radius $\theta$. The remaining degrees of freedom are the functions $\delta y_\lambda(m)$ depending on the mass variable $m$. 

The expression for $\delta X^{\rm new}$ can be simplified further by noting the factored form $d\mu(Q) = w_T^{-2} \pi^{-1} |m|\,  dm\, \sinh\theta \, d\theta \, d\phi$ of the integration measure. By using this information and including the factors taking $\delta X^{\rm new}$ to $\delta Y^{\rm new}$ in Eq.\ (\ref{eq:eqn4deltaYBB}), we finally arrive at the transfer $\delta y_\lambda(m) \to \delta y_\lambda^{\rm new}(m)$ in the form of a single integral over $m^\prime$:
\begin{equation}\label{eq: reduced linear integral equation with orbit integral}
    \delta y_{\lambda}(m)^\text{new} = K \frac{Y_\mathrm{pp}(m)}{X_\mathrm{pp}(m)}\, |m|\, \int_\bbR dm^\prime \, \mathcal{I}_\lambda (m, m^\prime) \, \delta y_{\lambda}(m^\prime) ,
\end{equation}
where the kernel $\mathcal{I}_\lambda (m, m^\prime)$ is an integral over $\theta^\prime$:
\begin{equation}\label{Ilam}
    \mathcal{I}_\lambda (m, m^\prime) = 2 \int_0^\infty \!\!\!\! d\theta^\prime \sinh\theta^\prime \bigg(1 + w_T^2 \frac{\partial}{\partial w_T^2} \bigg) \, \mathrm{e}^{-w_T^2\Tr q_0 Q^\prime} \, \varphi_\lambda (\theta^\prime) .
\end{equation}
This integral is computed in Appendix (\ref{g integral in horospherical coordinates}) and expressed as
\begin{equation}\label{eq: final reduced m equation 1}
    \mathcal{I}_\lambda(m, m^\prime) = \frac{\mathrm{e}^{-2 m m^\prime}}{\sqrt{\pi | m m^\prime |}}
    \bigg( (1-4 mm^\prime) K_{i\lambda}(2 |mm^\prime|) - 2 |mm^\prime| \big( K_{1+i\lambda} + K_{1-i\lambda} \big) (2 |mm^\prime|) \bigg)
\end{equation}
in terms of modified Bessel functions $K_\nu$ of the second kind.

\subsection{Strong-coupling limit and location of critical point}

We now compute the eigenvalue with the maximum magnitude of the transfer operation $\delta y_\lambda \mapsto \delta y_\lambda^{\rm new}$ defined in Eq.\ (\ref{eq: reduced linear integral equation with orbit integral}). (It is easy to see that the maximal eigenvalue occurs for $\lambda = 0$. We therefore simplify our calculation by setting $\lambda$ to that value.) To start, we recall from Eqs.~(\ref{eq.Ypp}) and (\ref{eq.Z4strongdisorder}) that the pp solution for strong disorder ($w^2 = w_V^2 / w_T^2 \gg K > 1$) takes the form
\begin{equation}\label{eq:Ypp-strongdis}
    Y_\mathrm{pp} (m) = \mathrm{e}^{-2 w^2 m^2 + 2\mathrm{i} \frac{E}{w_T} m} , \quad X_\mathrm{pp} (m) = 1 .
\end{equation}
The presence (here, and on the right-hand side of the AAT equation) of the factor $\mathrm{e}^{-2 w^2 m^2}$ sets the scale for $m$ and $m^\prime$ to be $w^{-1}$. Thus $2 w^2 m^2 \sim 1$ and the argument $2 |m m^\prime| \sim w^{-2} \ll 1$ of the modified Bessel functions is typically small. Therefore, we may use the small-argument expansion
\begin{equation}
    K_0(x) \sim -\ln x, \quad K_1(x) \sim \frac{1}{x} .
\end{equation}
For our purposes, it suffices to keep only the term of leading order in $1/w^2$. The kernel $\mathcal{I}_0(m,m^\prime)$ then simplifies to
\begin{equation}\label{strong coupling kernel}
    \mathcal{I}_0(m,m^\prime) \sim - \frac{\ln (2 |m m^\prime|)}{\sqrt{\pi |m m^\prime|}} 
    \sim \frac{\ln (w^2)}{\sqrt{\pi |m m^\prime|}} ,
\end{equation}
and Eq.~(\ref{eq: reduced linear integral equation with orbit integral}) takes the form
\begin{equation}\label{separated integral operator}
    \delta y(m)^\text{new} = 2K \frac{\ln w}{\sqrt{\pi}} \, \mathrm{e}^{-2 w^2 m^2 + 2\mathrm{i} \frac{E}{w_T} m}
    \sqrt{|m|} \int\frac{d m^\prime}{\sqrt{|m^\prime|}} \; \delta y(m^\prime) .
\end{equation}
We see that the strong-coupling limit turns the transfer $\delta y \mapsto \delta y^{\rm new}$ into (a multiple of) the rank-one projection onto the real line $\mathbb{R}\, \delta y$ for the (eigen-)function 
\begin{equation}\label{Target space of strong coupling operator}
    \delta y(m) = \mathrm{e}^{- 2 w^2 m^2 + 2 \mathrm{i} \frac{E}{w_T} m} \sqrt{|m|} .
\end{equation}
As a consistency check, we note that $\delta y \to 0$ for $m \to 0$, as required. The eigenvalue is
\begin{equation}
    \nu_{\rm max} = \nu(\lambda = 0) = 2K \frac{\ln w}{\sqrt{\pi}} \int dm \, \mathrm{e}^{- 2 w^2 m^2 + 2\mathrm{i} \frac{E}{w_T} m} = \sqrt{2} K \frac{\ln w}{w} \exp\left(- \frac{E^2}{2 w_V^2}\right) .
\end{equation}
The critical manifold is determined by the requirement that $\nu_{\rm max}$ reaches magnitude one. Consequently, the critical disorder $w = w_c$, as a function of the branching number $K$ and the dimensionless energy $E / w_V$, is given as the solution of the equation
\begin{equation}\label{crit point HA}
    K = \frac{1}{\sqrt{2}}\frac{w_c}{\ln w_c} \exp\left(\frac{E^2}{2 w_V^2}\right) ,
\end{equation}
in the limit of large $K$.  

In deriving Eq.~(\ref{crit point HA}) we have taken into account only the leading asymptotics. To check the validity of our approximations, we have carried out precise numerical calculations of $\nu$ based on the formula (\ref{strong coupling kernel}) for $\mathcal{I}_0(m , m^\prime)$. The exact values differ from the approximate ones only by a minor amount, unless $K$ is close to unity. Since these deviations will make no difference for our qualitative analysis of the critical behavior, we do not report them here.

\section{$Y_\mathrm{ac}$ near the critical manifold}\label{U(1)crit}

\subsection{Scale of maximal decay $\Lambda$ and equation for $Y_\mathrm{ac}-Y_\mathrm{pp}$}

In previous sections (\ref{sect:ac} and \ref{sect:ac boost}), we used the coordinates $p$, $q$ of Eq.~(\ref{eq:pullx03}), or the horospherical coordinates defined in Eq.~(\ref{eq:horoscoord}), to express the function $Y_\mathrm{ac}$. For the purpose of the present section, we switch to  hyperbolic polar coordinates, Eq.~(\ref{eq.hyperbolicpolar}). [Please be reminded that $Y_\mathrm{ac}$ stands for the AAT-approximate $\mathrm{U}(1)_{\mathrm{i} \sigma_3}$-invariant extremal  Gibbs state; not for the physical mixed state, which is a zero-mode integral over the Goldstone manifold $G/K$.]

The function $Y_{\rm ac}(Q)$ depends on $Q$ only through the hyperbolic radius $\theta = \theta(Q)$ and the (dimensionless) mass parameter $m = m(Q)$. As $w$ approaches $w_c$ from below, $Y_\mathrm{ac}$ converges pointwise to the pp solution $Y_\mathrm{pp}$ in the following way. For $0 < w_c - w \ll 1$, there emerges a large scale $\Lambda$ separating two regimes: for $\theta$ well below $\Lambda$, the function $Y_{\rm ac}$ has come close to reaching $Y_{\rm pp}$, while for $\theta$ well above $\Lambda$, the difference $Y_{\rm pp} - Y_{\rm ac}$ remains large as $Y_{\rm ac}$ is still close to zero. Between these regimes, there is a range of $\theta$-values where $Y_{\rm ac}$ drops off fast. The size of this range is expected to be of order unity and independent of the large and growing scale $\Lambda$, which goes to infinity as $w \nearrow w_c$. This is the scenario for the dependence on the (``massless'') variable $\theta$. In contradistinction, the characteristic scale for the ``massive'' variable $m$ remains finite ($m \sim 1/w$) on approaching criticality.

How does the scale $\Lambda$ depend on the critical parameter $w_c-w$? This question will now be answered by harmonic analysis from the preceding section. By reorganizing the AAT integral equation (\ref{eq:eqn4YBB}), we can formulate an equation satisfied by the difference $Y_\mathrm{ac} - Y_\mathrm{pp} \equiv \delta Y_\mathrm{ac}$:
\begin{equation}\label{Yac-Ypp}
    \delta Y_\mathrm{ac}(Q) = Y_\mathrm{pp}(Q) \left( \int d\mu(Q^\prime)\, \mathrm{e}^{-w_T^2 \mathrm{Tr}\, Q Q^\prime} \Delta \left(Y_\mathrm{pp} (Q^\prime) + \delta Y_\mathrm{ac}(Q^\prime) \right) \right)^K - Y_{\rm pp}(Q) , 
\end{equation}
where we used the strong-disorder expression (\ref{eq:Ypp-strongdis}) for $Y_{\rm pp}$. By the definition of the scale $\Lambda$, the difference $\delta Y_\mathrm{ac}$ is small for $\theta \ll \Lambda$ and Eq.~(\ref{Yac-Ypp}) can be linearized in that regime, using that the kernel $\mathrm{e}^{- w_T^2 \mathrm{Tr}\, Q Q^\prime}$ has essentially finite range. The outcome, 
\begin{equation}\label{Yac-Ypp lin}
    \delta Y_\mathrm{ac}(Q) = K\, Y_\mathrm{pp}(Q) \int d\mu(Q^\prime) \, \mathrm{e}^{- w_T^2 \mathrm{Tr}\, Q Q^\prime} \Delta\,\delta Y_\mathrm{ac}(Q^\prime) \qquad (\theta(Q) \ll \Lambda) ,
\end{equation}
is simply the linear integro-differential equation (\ref{eq:eqn4deltaYBB}) from Sect.~\ref{Harmonic analysis}. We have already found the eigenfunctions of the transfer operator of Eq.~(\ref{Yac-Ypp lin}). The maximal eigenvalue $\nu(\lambda=0)$ equals unity at $w = w_c$. On the metallic side of the transition ($w < w_c$), the maximal eigenvalue $\nu(0)$ has magnitude greater than one, and Eq.~(\ref{Yac-Ypp lin}) is solved by the eigenfunction $\delta Y_{\lambda^\ast}(Q)$ with eigenvalue $\nu(\lambda^\ast) = 1$. The quantum number $\lambda^\ast$ goes to zero as $w \nearrow w_c$. In a later part of this section, we will quantify the dependence of $\lambda^\ast$ on the critical parameter $w_c - w$.

\subsection{Calculation of the scale $\Lambda$}

The eigenfunctions $\delta Y_{\lambda^\ast}$ are given by
\begin{equation}\label{HA eigenfunctions}
    \delta Y_{\lambda^\ast} = C(\lambda^\ast) \, \sinh\theta \; \widetilde{\varphi}_{\lambda^\ast}(\theta)\, \delta y_{\lambda^\ast}(m),
\end{equation}
where the constant $C(\lambda^\ast)$ is arbitrary \emph{a priori}. It will be determined later by the requirement that $Y_\mathrm{ac}$ and $\partial_\theta Y_\mathrm{ac}$ be continuous in the crossover region $\theta \sim \Lambda$. We recall that
\begin{equation}\label{integral phigam}
    \widetilde{\varphi}_{\lambda^\ast}(\theta) = \int_0^{2\pi} \frac{d\phi}{2\pi} \left(\cosh\theta + \sinh\theta \, \cos\phi \right)^{-\frac{1}{2} - \mathrm{i}\lambda^\ast} \cos\phi .
\end{equation}

We are now interested in the asymptotic regime $1 \ll \theta \ll \Lambda$ and $|\lambda^\ast| \ll 1$. From the integral representation (\ref{integral phigam}) one deduces directly that $\widetilde{\varphi}_{\lambda^\ast}$ has the asymptotic behavior
\begin{equation}\label{asymptotics for phigam}
    \widetilde{\varphi}_{\lambda^\ast}(\theta) \sim \, - \frac{1}{\lambda^\ast}\, \mathrm{e}^{-\theta /2}\, \sin \big(\lambda^\ast (\theta - a) \big) ,
\end{equation}
where $a = 2 - 2\ln 2$. Accordingly, the asymptotics of $\delta Y_\mathrm{ac}$ is
\begin{equation}\label{delta Yasy}
    \delta Y_\mathrm{ac} \sim - \frac{C(\lambda^\ast)}{\lambda^\ast} \, \mathrm{e}^{\theta/2} \, \sin \big( \lambda^\ast (\theta - a) \big) \, \delta y_{\lambda^\ast}(m) .
\end{equation}
To determine the coefficient $C(\lambda^\ast)$ and the growth law of the scale $\Lambda$, we exploit the regularity of the ac-type solution: $Y_\mathrm{ac}$ and $\partial_\theta Y_\mathrm{ac}$ must be continuous near $\theta \sim \Lambda$. This requirement is equivalent to the two conditions
\begin{equation}\label{cond on deltaY}
    \frac{\partial}{\partial\theta}\, \delta Y_\mathrm{ac} \bigg|_{\theta = \Lambda} \approx 0 , \quad \delta Y_\mathrm{ac}\bigg|_{\theta = \Lambda} \approx - Y_\mathrm{pp} , 
\end{equation}
which lead to
\begin{equation}\label{cond on delta,C}
    \Lambda \sim \frac{\pi}{|\lambda^\ast|}, \quad C \sim \mathrm{e}^{- \Lambda / 2} .
\end{equation}
We note that the scaling (\ref{cond on delta,C}) is in agreement with the corresponding result obtained for the Anderson model within the AAT approach; see Ref.~\cite{MF1992}. The present derivation has the neat feature that it relies solely on the eigenfunctions $\delta Y_{\lambda^\ast}$ given by harmonic analysis.

\subsection{Formula for $Y_\mathrm{ac}$ near criticality}

We can now get an approximate view of the $\mathrm{U}(1)_{\mathrm{i} \sigma_3}$-invariant metallic solution $Y_\mathrm{ac}$ near criticality. From the construction above, we know that for $\theta \ll \Lambda$ we have
\begin{equation}\label{Yaccrit th ll Lam}
    Y_\mathrm{ac} = Y_\mathrm{pp} + C(\lambda^\ast) \, \sinh\theta\; \widetilde{\varphi}_{\lambda^\ast}(\theta)\, \delta y_0(m), \quad C(\lambda^\ast) \sim \mathrm{e}^{-\frac{1}{2} \pi/|\lambda^\ast|}.
\end{equation}
For $1\ll\theta\ll\Lambda$ we can expand,
\begin{equation}\label{expand}
    \widetilde{\varphi}_{\lambda^\ast}(\theta) \sim -\frac{1}{\lambda^\ast}\, \mathrm{e}^{-\theta/2}\, \sin\big( \lambda^\ast(\theta-a)\big) \sim -\theta\, \mathrm{e}^{-\theta/2} \sim \widetilde{\varphi}_0(\theta) < 0.
\end{equation}
The last relation in (\ref{expand}) states the asymptotics of $\widetilde{\varphi}_0 (\theta)$. It follows that in Eq.~(\ref{Yaccrit th ll Lam}) we can replace $\widetilde{\varphi}_{\lambda^\ast}$ by $\widetilde{\varphi}_0$.

Next, we recall that $\delta Y_\mathrm{ac}$ as constructed above is continuous and differentiable through the crossover region $\theta\sim \Lambda$, and $Y_{\rm ac}$ is strongly suppressed for $\theta \gg \Lambda$. As mentioned at the beginning of the present section, the width of this crossover is expected to remain of order unity as $\Lambda\to\infty$. A quantitative estimate of this width can be had as follows. We introduce the ``front'' coordinate $s = \theta - \Lambda$. Using Eq.~(\ref{delta Yasy}) together with the matching conditions (\ref{cond on deltaY}), we find for fixed $s$ as $\lambda^\ast \to 0$:
\begin{equation}
   R(s) \equiv \frac{Y_{\rm ac}(\Lambda+s,m)}{Y_{\rm pp}(m)} \longrightarrow 1 - \mathrm{e}^{s/2}\, \left(1-\frac{s}{2}\right) \quad (s\leq 0).
\end{equation}
The ratio $R(s)$ is monotonic in $s$ between $R(-\infty) = 1$ and $R(0) = 0$. Therefore, if we choose any $\delta$, $0 < \delta < 1/2$, the width of the interval $[s_1 , s_2]$ defined by $R(s_1) = 1 - \delta$ and $R(s_2) = \delta$ is finite and independent of the diverging scale $\Lambda$. 

The harmonic analysis yields the solution in the inner region $\theta \ll \Lambda$ but does not determine the nonlinear crossover profile. A quantitative description of the crossover region,
\begin{equation}\label{U(1)invt critical solution}
    Y_\mathrm{ac} = Y_\mathrm{pp} \, F\left(C \, \sinh\theta\, \widetilde{\varphi}_0(\theta) \, \sqrt{|m|} \right),
\end{equation}
is by some function $F(u)$ with fast decay for $u \to -\infty$ and the small-$u$ behavior
\begin{equation}
    F(u) = 1 + u + \mathcal{O}(u^2) \quad (-\infty < u \leq 0; \quad u \to 0)
\end{equation}
to reproduce the linearized solution. To get from Eq.~(\ref{Yaccrit th ll Lam}) to (\ref{U(1)invt critical solution}) we used that
\begin{equation}
    \delta y_0 (m) = \sqrt{|m|} \, Y_{\rm pp} \, .
\end{equation}

It remains to ascertain how $\lambda^\ast$ approaches zero as the disorder strength tends to its critical value. That calculation is given in an Appendix, \ref{app:Scaling gamma ast}. For $w_c - w \ll 1$ we find
\begin{equation}
    \lambda^\ast \sim \sqrt{w_c-w} \quad (0 < w_c - w \ll 1) .
\end{equation}
Given in Eq.~(\ref{Yaccrit th ll Lam}), the parameter $C$ vanishes with $w_c - w$ in non-analytic fashion (due to the approximation of $\mathbb{G}$ by a tree):
\begin{equation}\label{scaling tilde C}
    C \sim \exp\left( -\frac{1}{2}\frac{\pi}{\sqrt{w_c-w}} \right) .
\end{equation}
The inverse of $C$ sets the scale for the correlation volume $\mathcal{V}_\xi$ within the AAT approximation (i.e., $C \sim \mathcal{V}_{\xi}^{-1}$, where $\xi \sim \Lambda$ is the corresponding length). On length scales well below $\xi$, the system behaves as if the governing extremal Gibbs state were $Y_{\rm pp}$ (taken at the critical disorder strength $w = w_c$), and only on scales much larger than $\xi$ does it cross over to metallic behavior. In other words, $\mathcal{V}_\xi$ is the volume over which the field fluctuations retain their critical character, interpolating between pp- and ac-type behavior, and $C$ is the corresponding inverse critical scale.

\section{Construction of solutions of sc type}\label{sc crit}

\begin{figure}[H]
\centering
\includegraphics[width=0.8\textwidth]{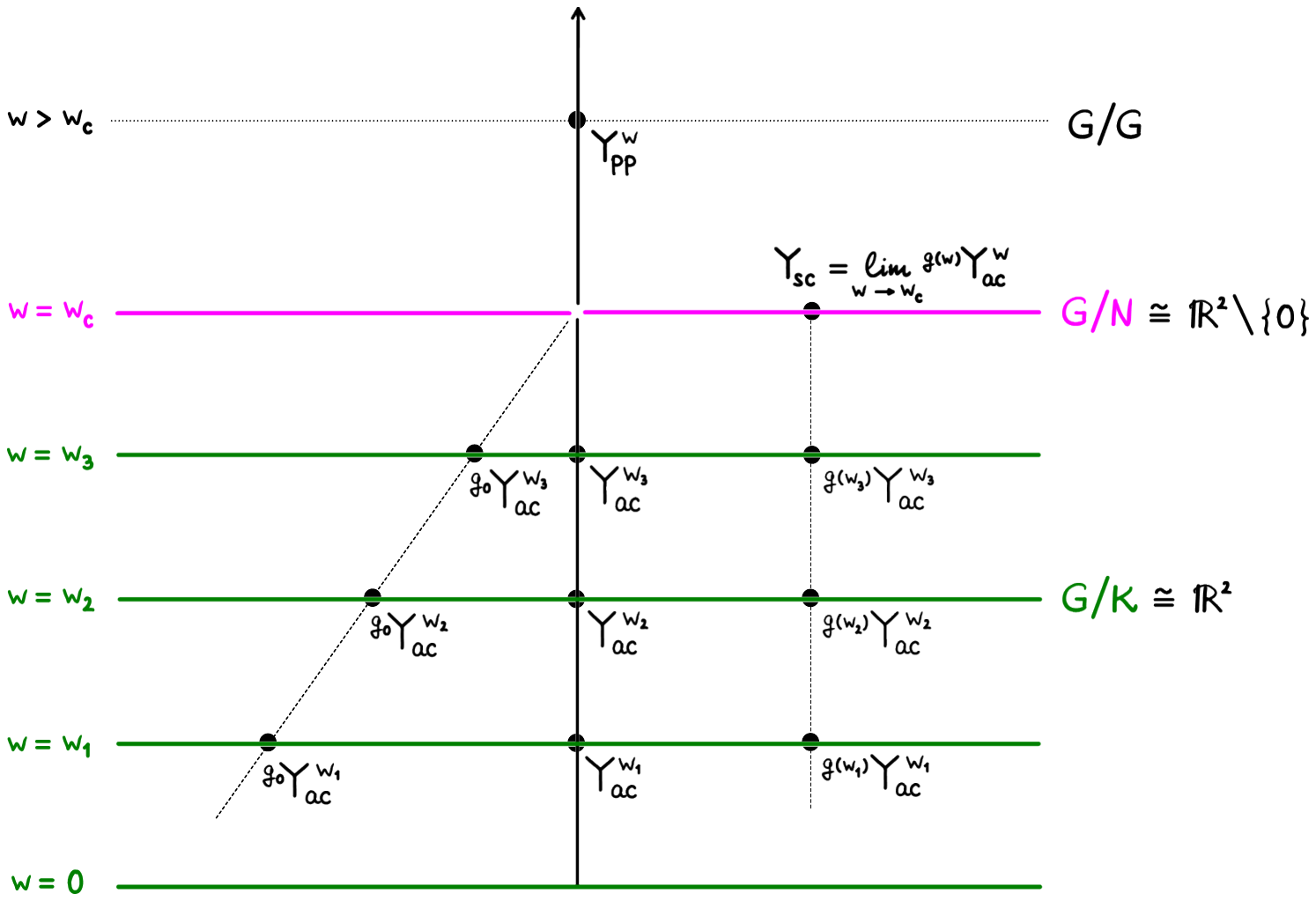}
\caption{Schematic illustration of the critical double-scaling mechanism. As $w$ is sent to $w_c$ from below, every metallic solution ${}^{g_0} Y_{\rm ac}$ (for a Lorentz boost $g_0$ independent of $w$) converges to $Y_{\rm pp} \big\vert_{w_c}$, i.e., the metallic order-parameter orbit $G/K \cong \mathbb{R}^2$ collapses to the trivial orbit $G/G \cong \{0\}$ of the insulating phase. Nonetheless, by tuning a diverging boost $Y_{\rm ac} \to {}^{g(w)} Y_{\rm ac}$ to the shrinking inverse correlation-volume scale $C(w) \to 0$, a nontrivial order-parameter orbit $G/N$ of solutions $Y_{\rm sc}$ arises at the critical point $w = w_c$.}
\label{Fig:Ysc}
\end{figure}

With all the preliminaries from the previous sections in hand, we are finally in a position to construct the critical solutions of sc-type. The basic idea is sketched in Fig.~\ref{Fig:Ysc}. For any fixed symmetry-group element $g\in G$, the metallic extremal state ${}^g Y_{\rm ac}$ approaches the fully symmetric state $Y_{\rm pp}$ of the Anderson insulator in the limit of $w \to w_c$. Put differently, the metallic order-parameter orbit $X = G/K \cong \mathbb{R}^2$ collapses to the trivial orbit $G/G \cong \{0\}$ at criticality.

The crucial observation here is that the collapse ${}^g Y_{\rm ac} \to Y_{\rm pp}$ for $w \to w_c$ fails to be uniform in non-compact directions of $G/K$ and happens only \emph{if $g$ is kept fixed}. By that non-uniformity, the door is open for a third type of solution to exist at $w = w_c$, as indicated by the purple line in Fig.~\ref{Fig:Ysc}. From Sect.~\ref{U(1)crit} we already know that the convergence ${}^{g=e} Y_{\rm ac} \to Y_{\rm pp}$ proceeds pointwise and is governed by the approach to zero of the inverse correlation volume $C$. So we may offset the collapse ${}^g Y_{\rm ac} \to Y_{\rm pp}$ by following the trajectory of a $w$-sequence (e.g., the vertical line in Fig.~\ref{Fig:Ysc}) of metallic solutions
\begin{equation}
    Y_{x(w)} \big\vert_w \equiv {}^{g(w)} Y_{\rm ac}\big\vert_w
\end{equation}
which are Lorentz-boosted by elements $g(w)$ with diverging rapidity for $w \to w_c$. The critical scaling of $g(w)$ is dictated by the vanishing of the critical scale $C(w) \to 0$. Information about the precise dependence of $g(w)$ on $C(w)$ will be provided by the AAT equation below.

The stabilizer of the boosted solution $Y_{x(w)} \big\vert_w$ is the conjugate compact subgroup $\widetilde{K}_w = g(w) K g(w)^{-1}$, whose large-boost limit is a nilpotent subgroup $N \subset G$. We are thus led to expect a critical family of extremal states of orbit-type $G/N$ instead of $G/G$. Geometrically speaking, the process of adjoining the $G/G$-orbit to $G/N \cong \mathbb{R}^2\setminus \{0\}$ amounts to adding the tip $\{0\}$, which represents the fixed-boost limit of the metallic family. We now show that the AAT equation indeed realizes this critical scenario.

\subsection{Construction}\label{construction} 

In what follows, we present the calculation that implements the intuitive picture described above. We first isolate the part in Eq.~(\ref{U(1)invt critical solution}) that depends on the symmetry degrees of freedom:
\begin{equation}\label{deltaY/Y_pp manipulated}
    \frac{\delta Y_0}{Y_\mathrm{pp}} = C \sqrt{w_T} \int_0^{2\pi} \frac{d\phi}{2\pi} \frac{\sqrt{x_1^2+x_2^2}\, \cos\phi} {\big(x_3 + \sqrt{x_1^2 + x_2^2}\, \cos\phi \big)^{1/2}} = C \sqrt{w_T} \int_0^{2\pi} \frac{d\phi}{2\pi} \frac{x_2\cos\phi + x_1\sin\phi}{\left(x_3 + x_2\cos\phi + x_1\sin\phi\right)^{1/2}} .
\end{equation}
Here and in the following, we again omit the pullback symbol $i^\ast$. The $x_\mu$ are the quasi-Cartesian coordinate functions defined in (\ref{eq:P-from-xmu}) and restricted (by $i^\ast$) from $\overline{C_+^4}$ to $C^3$. For the first equality in (\ref{deltaY/Y_pp manipulated}) we used the coordinate transformation (\ref{eq.hyperbolicpolar}). The second equality follows by shifting $\phi$ so as to reduce $x_2 \cos\phi + x_1 \sin\phi$ to $\sqrt{x_1^2 + x_2^2}\, \cos\phi$. 

The average over the angle $\phi$ renders the expression in Eq.~(\ref{deltaY/Y_pp manipulated}) invariant under the action of $\mathrm{U}(1)_{\mathrm{i} \sigma_3}$. In the next step, we use the $G$-equivariance of the AAT equation to construct solutions \emph{without} $\mathrm{U}(1)_{\mathrm{i} \sigma_3}$-symmetry, by applying Lorentz boost transformations taken from $G = \mathrm{SU}(1,1)$. 

What happens to the right-hand side of (\ref{deltaY/Y_pp manipulated}) under a Lorentz boost? For concreteness, we consider the one-parameter group of boosts $g = g(\tau)$ generated by the vector field $x_3 \, \partial / \partial x_2 + x_2 \, \partial / \partial x_3$. The result of the boost action is then succinctly expressed in terms of ``light-cone coordinates'' $x_\pm = x_3 \pm x_2 \geq 0$:
\begin{equation}
    g(\tau) : \; x_\pm \mapsto \mathrm{e}^{\mp \tau} x_\pm \quad (\tau \in \mathbb{R}).
\end{equation}
Thus, the terms in the integrand on the r.h.s.\ of (\ref{deltaY/Y_pp manipulated}) transform as
\begin{equation}
\begin{aligned}
    &x_2 \cos\phi + x_1\sin\phi \mapsto \frac{1}{2} \left(x_+ \, \mathrm{e}^{-\tau} - x_- \, \mathrm{e}^{\tau} \right) \cos\phi + x_1 \sin\phi,\\
    & x_3 + x_2 \cos\phi + x_1 \sin\phi \mapsto x_+ \, \mathrm{e}^{-\tau} \frac{1}{2} \left(1 + \cos\phi\right) + x_- \, \mathrm{e}^{\tau} \frac{1}{2}\left(1-\cos\phi\right) + x_1 \sin\phi.
\end{aligned}
\end{equation}
Note that in the horospherical coordinates (\ref{eq:horoscoord}) we have
\begin{equation}
    x_+ = w_T^{-1} \mathrm{e}^\varphi, \quad x_- = w_T^{-1} (m^2 + b^2) \, \mathrm{e}^{-\varphi}.
\end{equation}

We now expand the boosted version of the r.h.s.\ of (\ref{deltaY/Y_pp manipulated}) to extract the asymptotic behavior in the limit of a large boost rapidity $\tau \to \infty$. Noting that the main contribution to the integral comes from $\phi$ near zero and keeping only the leading terms (and a regulating term in the denominator), we obtain
\begin{equation}
    {}^{g(\tau)} \big( \delta Y_0 / Y_{\rm pp} \big) \approx -C \sqrt{w_T}\, x_- \, \mathrm{e}^{\tau} \int_0^\pi \frac{d\phi}{2\pi} \left( x_+ \, \mathrm{e}^{-\tau} + x_- \, \mathrm{e}^{\tau} \phi^2 / 4 \right)^{-1/2} .
\end{equation}
This integral is brought into standard form by straightforward manipulations and by absorbing some irrelevant constants into $C$:
\begin{equation}
    {}^{g(\tau)} \big( \delta Y_0 / Y_{\rm pp} \big) \approx - C \sqrt{w_T^{\vphantom{\cdot}}\, x_-}\, \mathrm{e}^{\tau /2} \int_0^{\frac{\pi}{2}\sqrt{\frac{x_-}{x_+}} \, \mathrm{e}^{\tau}} \! \!\!\!\! \frac{dx}{\sqrt{1 + x^2}} = - C \sqrt{w_T^{\vphantom{\cdot}}\, x_-}\, \mathrm{e}^{\tau /2} \, \mathrm{arsinh} \left(\frac{\pi}{2} \sqrt{\frac{x_-}{x_+}} \, \mathrm{e}^{\tau}\right).
\end{equation}
Lastly, we use $\mathrm{arsinh}(x/2) \sim \ln x$ for $x\gg 1$ to obtain
\begin{equation}
    {}^{g(\tau)} \big( \delta Y_0 / Y_{\rm pp} \big) \approx
    - C \, \tau \, \mathrm{e}^{\tau /2} \sqrt{w_T^{\vphantom{\cdot}}\, x_-}\, .
\end{equation}
This already finishes the main part of the calculation. In the limit at hand ($\tau \gg 1$) and by passing from the linear approximation to the nonlinear crossover profile $F$ in (\ref{U(1)invt critical solution}), we find that the boosted solutions of the AAT equation turn into
\begin{equation}\label{boosted crit solution}
    {}^{g(\tau)} Y_\mathrm{ac} = Y_\mathrm{pp}\, F\left(- C\, \tau \, \mathrm{e}^{\tau /2} 
    \sqrt{w_T^{\vphantom{\cdot}}\, x_-} \right).
\end{equation}

It is now clear that an extremal state of novel type, to be called $Y_{\rm sc}$, is attained by demanding the product $C\,\tau\,\mathrm{e}^{\tau /2}$ to stay fixed and of order unity as the critical point is approached. For $\beta \in \mathbb{R}$, let a function $w \mapsto \tau_\beta(w)$ be defined by
\begin{equation}
    C(w)\, \tau_\beta(w)\, \mathrm{e}^{\tau_\beta(w) / 2} = \mathrm{e}^{\beta / 2},
\end{equation}
with the inverse correlation volume $C$ scaling as in (\ref{scaling tilde C}). The function $\tau_\beta(w)$ can be expressed by the principal real Lambert W function, defined as the solution of the equation $\mathrm{W}(x)\, \mathrm{e}^{\mathrm{W}(x)} = x$:
\begin{equation}
    \tau_\beta(w) = 2 \mathrm{W} \left( \frac{\mathrm{e}^{\beta/2}}{2 C(w)}\right) .
\end{equation}
By sending $w \nearrow w_c\,$, we then arrive at a family $\{ {}^\beta Y_{\rm sc} \}_{\beta \in \mathbb{R}}$ of critical solutions:
\begin{equation}\label{Ysc}
    \lim_{w \to w_c} {}^{g(\tau_\beta(w))} Y_{\rm ac}\big\vert_{w} \equiv {}^\beta Y_\mathrm{sc} = Y_\mathrm{pp} \, F\left( - \mathrm{e}^{\beta/2} \sqrt{w_T^{\vphantom{\cdot}}\, x_-} \right) = Y_\mathrm{pp}\, F \left( - \sqrt{m^2 + b^2}\; \mathrm{e}^{(\beta -\varphi) / 2} \right) .
\end{equation}
The behavior of $\tau_\beta$ near criticality is
\begin{equation}
    \tau_\beta (w) = \beta + \tau_C(w), \quad \tau_C(w) \sim -2 \ln C(w) \sim \frac{\pi}{\sqrt{w_c - w}} ,
\end{equation}
where $\tau_C$ sets the characteristic scale for boosts $Y_{\rm ac} \to {}^{g(\tau)} Y_{\rm ac}$ to offset the contraction ${}^{g_0} Y_{\rm ac} \to Y_{\rm pp}$ as $C(w) \to 0$.

We note that the function $x_- = x_3 - x_2$ is annihilated by the sum $K_2 + K_3$ of first-order differential operators in Eq.~(\ref{def:K_j}), representing the generator $\sigma_+ \equiv \frac{1}{2}(\sigma_2 + \mathrm{i} \sigma_3)$ of $G = \mathrm{SU}(1,1)$. This generator is nilpotent: $\sigma_+^2 = 0$. Annihilated by the Lie-algebra element $K_2 + K_3$, the function $x_-$ is invariant under the action of the nilpotent Lie group $N \subset G$ generated by $\sigma_+$. Since $Y_{\rm pp}$ is $G$-invariant, it follows that the functions in (\ref{Ysc}) are $N$-invariant. The constant $\beta \in \mathbb{R}$ in (\ref{Ysc}) is arbitrary; it is the log-radial coordinate on the critical order-parameter orbit $G/N \cong \mathbb{R}^2 \setminus \{ 0 \}$. Sending the radius $\mathrm{e}^{\beta/2}$ to the zero boundary ($\beta \to -\infty$) of $G/N \cong \mathbb{R}^2 \setminus \{ 0 \}$ yields the $G$-symmetric solution $Y_{\rm pp}$.

Our construction of the solutions $Y_\mathrm{sc}$ was formulated for the restricted class of functions that extend to $\overline{C_+^4}$ as functions independent of $x_1$. This restriction singles out a boundary point $z = (x_1 + \mathrm{i}x_2) / x_3 = \mathrm{i}$ in the Poincar\'e disk model $|z| < 1$ of $\mathrm{H}^2$. The critical solutions $Y_{\rm sc}$ in the ray of solutions (\ref{Ysc}) parameterized by $\beta \in \mathbb{R}$ are constant on horocycles centered at that very boundary point. Of course, we could rotate the boundary point ($z \to \mathrm{e}^{\mathrm{i} \phi} z$) and obtain another ray of critical solutions which are invariant under a conjugate nilpotent subgroup. Varying the boundary point and the light-ray coordinate $\mathrm{e}^{\beta/2}$ generates the full critical order-parameter orbit $G/N \cong \mathbb{R}^2 \setminus \{ 0 \}$.

\begin{figure}[ht]
\centering
\begin{minipage}[c]{0.6\textwidth}
\includegraphics[width=0.9\textwidth]{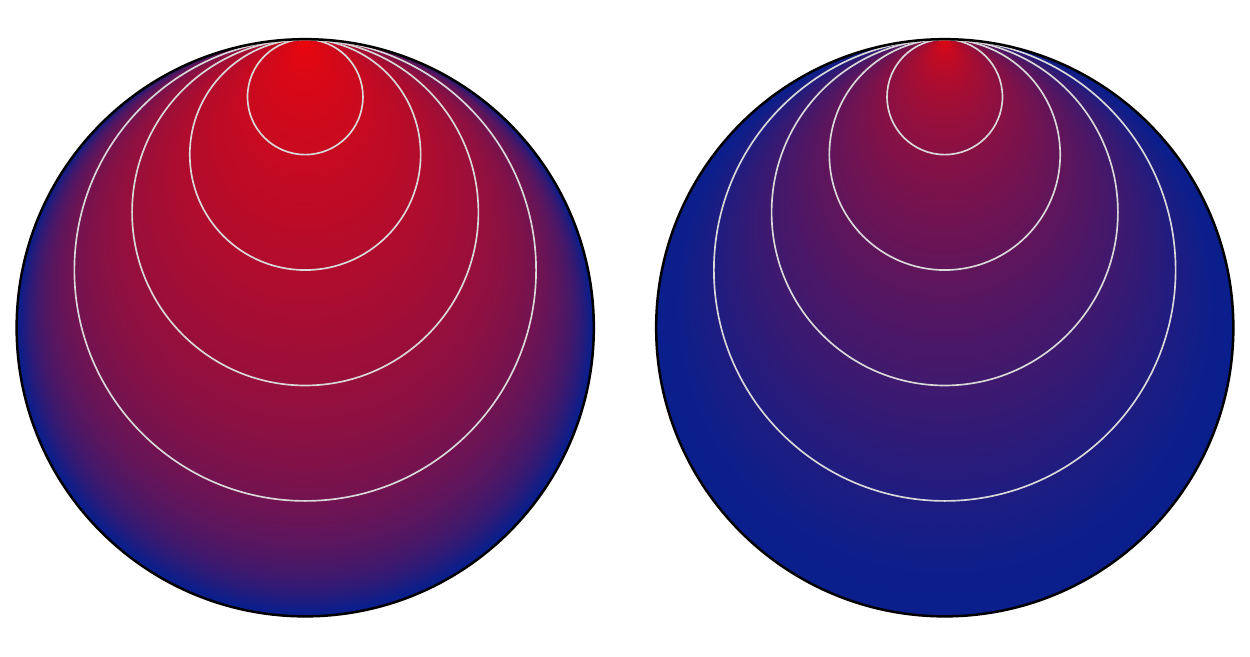}
\end{minipage}
\begin{minipage}{0.35\textwidth}
\caption{Sketch of two critical solutions $Y_{\rm sc}$ on $\mathrm{H}^2_{x_0}$ for fixed $x_0 = w_T^{-1} m > 0$ and different $\beta \in \mathbb{R}$. Their level sets are horocycles based at a common boundary point of $\mathrm{H}^2_{x_0}$ (here in the Poincar\'e disk model). Thus they are invariant under a nilpotent subgroup $N$ of $G = \mathrm{SU}(1,1)$. The decay is indicated again by a change of color from red to blue. The level sets of $Y_{\rm sc}$ are tangent to the boundary of the Poincar\'e disk; this differs from the metallic case, where they are contained entirely in the interior; cf.\ Fig.~\ref{Pic ac}.}
\end{minipage}
\label{Pic sc}
\end{figure}

In the following section, we show that the family of critical solutions given by (\ref{Ysc}) (and augmented by all the rotated rays of critical solutions) describes a singular continuous behavior of the observable $K_{n,\varepsilon}$.

\section{Mixture on $G/N$ and critical scaling of $K_{n,\varepsilon}$}\label{K sc}

We start from identity (\ref{spectral integral}):
\begin{equation}\label{spectral integral 3}
    K_{n,\varepsilon} = \int_X dx\, P(x)\, \langle Y_x,1\rangle_\varepsilon = \int_X dx\, P(x) \int d Q_n\, Y_x (Q_n),
\end{equation}
and first calculate the rightmost integral in the extremal Gibbs states $Y_x$ given by $Y_\mathrm{sc}$ for arbitrary $\beta \in \mathbb{R}$. As we shall see, this integral diverges for vanishing regulator $\varepsilon$ with a fractal exponent $\alpha = 1/2$ independent of $\beta$. While we have analyzed the AAT equation so far with the regulator $\varepsilon$ set to zero, the regulator must now be reintroduced in order to keep the integral finite. For fixed field coordinates (and $w = w_c$), the finite-$\varepsilon$ solutions approach the $\varepsilon = 0$ critical states pointwise as $\varepsilon\to 0$. This means that a small regulator $\varepsilon$ becomes relevant only in the large-field region. From Sect.~\ref{sect:spec} we know that $\varepsilon$ has the dynamical meaning of a death rate causing exponential decay of the return probability for times $t$ greater than $\varepsilon^{-1}$. Thus, the regulator acts effectively as a cutoff of the function $Y_\mathrm{sc}$ at the time scale $\mathrm{e}^\varphi/ w_T \sim \varepsilon^{-1}$. More accurately speaking, we expect a finite regulator $\varepsilon > 0$ to introduce a crossover function $f_{\mathrm{reg}} (\varepsilon\, \mathrm{e}^{\varphi} /w_T)$, with $f_{\mathrm{reg}}(0) = 1$ and rapid decay as its argument tends to $+\infty$.

So, we now look at
\begin{equation}\label{sc scaling integral}
    \langle Y_\mathrm{sc} , 1 \rangle_\varepsilon \approx \frac{1}{\pi w_T^2} \int_{\mathbb{R}} \!\! d\varphi \int_{\mathbb{R}} \!\! dm \int_{\mathbb{R}} \!\! db \; \mathrm{e}^{-2 w^2 m^2 + 2\mathrm{i} \frac{E}{w_T} m} \, F\left( - \sqrt{m^2 + b^2}\, \mathrm{e}^{-(\beta - \varphi)/2} \right)\, f_{\mathrm{reg}}\left(\frac{\varepsilon\, \mathrm{e}^\varphi}{w_T} \right).
\end{equation}
To begin the calculation, we note that the divergent behavior for $\varepsilon \to 0$ comes from the large-$\varphi$ region. There, the characteristic scale for the variable $m$ is given by $1/w$, which is small compared to the characteristic scale $\mathrm{e}^{(\varphi - \beta) / 2}$ for $b$. Consequently, the leading contributions to the integral come from the region $|b| \gg |m|$. Approximating $\sqrt{m^2 + b^2}$ by $|b|$, we obtain the integral
\begin{equation}
    \langle Y_\mathrm{sc} , 1 \rangle_\varepsilon \approx \frac{1}{\pi w_T^2} \int_{\mathbb{R}} \!\! d\varphi \int_\mathbb{R} \!\! db \; F\left(- |b|\, \mathrm{e}^{(\beta-\varphi)/2}\right)\, f_{\mathrm{reg}}\left(\frac{\varepsilon\, \mathrm{e}^\varphi}{w_T} \right) \int_\mathbb{R} \!\! dm \; \mathrm{e}^{-2 w^2 m^2 + 2\mathrm{i} \frac{E}{w_T} m} ,
\end{equation}
which is easily evaluated to give
\begin{equation}\label{K fractal scaling}
   \langle Y_\mathrm{sc} , 1 \rangle_\varepsilon \approx 2 \overline{\rho_n(E)} \left(w_T\, \mathrm{e}^\beta \varepsilon \right)^{-1/2}\; \int^0_{-\infty} \!\!\!\!\!\! du\, F(u) \; \int^\infty_{0}\!\! \frac{ds} {\sqrt{s}}\, f_{\mathrm{reg}}(s).
\end{equation}
The integral over $F$ is finite by the integrability of the ac-crossover profile, and the rightmost integral converges by the rapid decay of the regulator function $f_{\mathrm{reg}}$.

Having extracted the $\varepsilon$-scaling of the inner integral against the critical extremal Gibbs state $Y_{\rm sc}$, we turn to the average in (\ref{spectral integral 3}) over the order-parameter orbit $X$ in order to reconstruct the value in the physical state $Y_{\rm FL}$. For this purpose we have to specify the boundary weight function $P$ [subject to the requirement of invariance under $\mathrm{U}(1)_{\mathrm{i} \sigma_3}$]. To make the mixture over the non-compact space $X$ well-defined, $P$ must introduce a cutoff scale $\gamma = \gamma(w_c - w)$, which is to be removed (as $w \to w_c$) so as to reveal the true nature of the spectral measure at criticality:
\begin{equation}
    K_{n,\varepsilon} \equiv K^\mathrm{crit}_{n,\varepsilon} = \!\!\!\!\!\!\!\!
    \lim_{\substack{w \to w_c \\ \gamma(w_c-w)\to 0}} \!\!\!\!\!\!\!\!
    K_{n,\varepsilon}^{w,\gamma}, \qquad
    K_{n,\varepsilon}^{w,\gamma} = \int_X dx\, P_{\gamma}(x)\, \langle Y_x, 1 \rangle_\varepsilon .
\end{equation}
The specific form of the critical limit leading to $K^\mathrm{crit}_{n, \varepsilon}$ is determined below. 

We now specify the $G$-invariant measure $dx$ on the metallic order-parameter orbit $X = G/K$ in terms of polar coordinates given by the boost rapidity $\tau > 0$ and an angular variable $\vartheta$ for the boost direction:
\begin{equation}
    x(\tau,\vartheta) = k(\vartheta)\, g(\tau)\, K\in G/K, \quad dx = \frac{d\vartheta}{2\pi} \wedge \sinh\tau\, d\tau.
\end{equation}
[For our class of functions that extend to $x_1$-independent functions on $\overline{C_+^4}$, this takes the concrete form of $k(\vartheta) = \mathrm{e}^{\mathrm{i} \vartheta \sigma_3 /2}$ and $g(\tau) = \mathrm{e}^{\tau \sigma_1 /2}$.] Next, we make a convenient choice of $\mathrm{U}(1)_{\mathrm{i} \sigma_3}$-invariant boundary weight function,
\begin{equation}\label{choice Pgam}
    P_\gamma(x) = P_\gamma(\tau) = Z_\gamma^{-1} \mathrm{e}^{-2\gamma(\cosh\tau-1)} \quad (\gamma > 0),
\end{equation}
with normalization constant
\begin{equation}
    Z_\gamma = \int_0^\infty \!\!\!\! d\tau\, \sinh\tau\; \mathrm{e}^{-2\gamma(\cosh\tau-1)} = \frac{1}{2\gamma}.
\end{equation}
Let us recall here that our goal is to elucidate the critical scaling behavior of $K^\mathrm{crit}_{n, \varepsilon}$ and especially its scaling exponent. What matters for that is not the precise form of $P_\gamma$ but rather the introduction of a radial cutoff scale $\tau \sim -\ln\gamma$. Since $\gamma \ll 1$ (recall that $\gamma$ is eventually sent to zero), the boundary (or mixing) distribution simplifies at the asymptotic scale $\tau \gg 1$:
\begin{equation}\label{Pgam asymptotic}
   dx\, P_\gamma(x) \sim \frac{d\vartheta}{2\pi}\, \wedge \mathrm{d} \big(-\mathrm{e}^{-\gamma\, \mathrm{e}^{\tau}} \big).
\end{equation}

We now pause for a moment to connect this asymptotic form of $dx\, P_\gamma(x)$ with the development of Sect.~\ref{Grassmann in Yx}. There, by integrating out the Grassmann variables in $\mathcal{Q}_\infty$, we obtained a normalized distribution function $\widetilde{I}_\gamma (Q_\infty)$, Eq.~(\ref{Igam}),
with dependence on the microscopic boundary field $Q_\infty$. To arrive at $dx\, P_\gamma(x)$, the microscopic boundary data in $Q_\infty$ must be reduced to the thermodynamic labels ($x \in G/S$) of the extremal Gibbs states. Our explicit construction of symmetry-broken solutions was carried out for boosts $g = g(\tau)$ generated by the vector field $x_3\, \partial/\partial x_2 + x_2\, \partial/\partial x_3$, which act as translations of $\varphi$ in the coordinates (\ref{eq:horoscoord}). We may therefore construct the boundary weight along this family of symmetry-broken solutions by marginalizing out the massive variable $m_\infty$ and the horospherical coordinate $b_\infty$,
\begin{equation}
  \gamma^2 d\varphi_\infty \int_\mathbb{R} \frac{dm_\infty}{\sqrt{\pi}} \int_{\mathbb{R}} \frac{db_\infty}{\sqrt{\pi}} \; \mathrm{exp}\left(- \gamma\, (\mathrm{e}^{\varphi_\infty} + (m_\infty^2 + b_\infty^2)\, \mathrm{e}^{-\varphi_\infty}) \right) = d \big(-\mathrm{e}^{-\gamma\, \mathrm{e}^{\varphi_\infty}} \big),
\end{equation}
starting from the microscopic boundary weight function $\widetilde{I}_\gamma (Q_\infty)$. The boost directions due to $\mathrm{U}(1)_{\mathrm{i} \sigma_3}$-rotations ($z \to \mathrm{e}^{ \mathrm{i} \vartheta} z$; see the end of Sect.~\ref{sect:ac boost}) are incorporated by restoring the angular variable $\vartheta \equiv \vartheta_\infty$. While the boost parameter $\varphi_\infty$ is real-valued, the polar rapidity $\tau$ used to parameterize $G/K$ is restricted to positive values. This difference disappears in the limit $\gamma\to 0$ as the statistical weight gets concentrated at large positive $\varphi_\infty \sim - \ln \gamma$. Hence, after restoring the variable boost direction $\vartheta$, the induced boundary measure takes the asymptotic form (\ref{Pgam asymptotic}).

Now when the critical point is approached from below, what is the right $\gamma$-scaling in order for the $X$-average to select the critical extremal Gibbs states $Y_\mathrm{sc}$? The answer is quite natural: the mixing distribution should probe the critical field fluctuations, i.e., it should be controlled by the scale of the inverse correlation volume $C$. To be concrete, we recall from Sect.~\ref{sc crit}, Eq.~(\ref{Ysc}), that the critical states are obtained by resolving (or magnifying) the metallic order-parameter orbit around a diverging boost scale, which we now fix as  
\begin{equation}\label{eq:def-tauC}
    \tau_C = 2 \mathrm{W}\left( \frac{1}{2C} \right) 
\end{equation}
by inversion of the defining equation $C\,\tau_C\,\mathrm{e}^{\tau_C /2} = 1$. Given $\tau_C$, we introduce a critical scaling map $\Psi_C$ taking the metallic order-parameter orbit $G/K$ (in the approach to criticality) to the critical order-parameter orbit $G/N$ by
\begin{equation}
    \Psi_C(\tau,\vartheta) = \big( \beta = \tau - \tau_C,\, \vartheta \big) ,
\end{equation}
in the limit of $w \nearrow w_c$. By this map, the distribution $dx\, P_\gamma(x)$ is pushed forward from $G/K$ to $G/N$ as
\begin{equation}
    {\Psi_C}_\ast \big( dx\,P_\gamma(x)\big) \sim \frac{d\vartheta}{2\pi} \wedge \mathrm{d} \big(-\mathrm{e}^{-\gamma\,\mathrm{e}^{\beta + \tau_C}} \big).
\end{equation}
A non-degenerate critical limiting measure on $G/N$ is obtained only when 
\begin{equation}\label{eq:kappa}
   \gamma\,\mathrm{e}^{\tau_C} \to \kappa \in (0,\infty) .
\end{equation}
Adopting this scaling limit, we end up with
\begin{equation}\label{G/N measure}
  (\Psi_C)_\ast\big( dx\, P_\gamma(x)\big) \xrightarrow[\gamma\,\mathrm{e}^{\tau_C} \to \kappa] {}\frac{d\vartheta}{2\pi} \wedge \mathrm{d} \big(-\mathrm{e}^{-\kappa\, \mathrm{e}^\beta}\big) = \frac{d\vartheta}{2\pi} \wedge \kappa\, \mathrm{e}^\beta \mathrm{e}^{-\kappa\, \mathrm{e}^\beta} d\beta \equiv d\sigma_{\mathrm{sc}}(\beta, \vartheta).
\end{equation}

In summary, we have shown how the metallic mixing distribution on $G/K$ gets pushed forward to the critical order-parameter orbit $G/N$ by the critical limit of Sect.~\ref{sc crit}. For a consistency check, we note that the resulting distribution $d\sigma_{\rm sc}$ is normalized:
\begin{equation}
   \int_{G/N} \!\!\!\! d\sigma_\mathrm{sc}(\beta,\vartheta) = 1,
\end{equation}
as befits a mixing distribution. Note also that $d\sigma_{\rm sc}$ is invariant under $K = \mathrm{U}(1)_{\mathrm{i}\sigma_3}$.

Finally, we assemble the physical state $Y_{\rm FL}$ by putting together the mixture of the extremal states $Y_{\rm sc}$ over $G/N$. For that, we recall that the construction in Sect.~\ref{sc crit} was carried out for the special case of one fixed ray direction, say $\vartheta = 0$. The other critical rays are obtained by acting with $k(\vartheta)\in K$:
\begin{equation}
    Y_{\mathrm{sc},(\beta,\vartheta)} = {}^{k(\vartheta)} Y_{\mathrm{sc},(\beta,0)},
\end{equation}
where $Y_{\mathrm{sc},(\beta,0)}$ is a member of the ray of critical solutions from Sect.\ \ref{sc crit}. Now the integral in Eq.~(\ref{sc scaling integral}) does not depend on the choice of ray; i.e., although the integrand transforms non-trivially under $K$, the value of the integral is $K$-invariant. So, by the $K$-invariance of the distribution $d\sigma_{\rm sc}$ from Eq.~(\ref{G/N measure}), the $K$-average over all light-ray directions in Eq.~(\ref{spectral integral 3}) has a trivial effect and the integral over $\vartheta$ can be dropped. In conjunction with the double-scaling convergence of the metallic extremal states derived in Sect.~\ref{sc crit},
\begin{equation}
    {}^{k(\vartheta)}({}^{g(\tau_C + \beta)}Y_{\rm ac})\equiv Y_{\rm ac, (\tau_C + \beta,\vartheta)}
    \longrightarrow Y_{\mathrm{sc},(\beta,\vartheta)} \qquad (\mathrm{e}^{\beta/2} = C\, \tau\, \mathrm{e}^{\tau/2}, \;\; w \to w_c),
\end{equation}
the convergence of the mixing distributions in Eq.~(\ref{G/N measure}) now yields for 
\begin{equation}
    K^\mathrm{crit}_{n,\varepsilon} = \int_{G/N} \!\!\!\!\!\! d\sigma_\mathrm{sc} (\beta, \vartheta)\, \langle Y_{\mathrm{sc},(\beta, \vartheta)}, 1 \rangle_\varepsilon 
\end{equation}
the result
\begin{equation}\label{critical scaling K}
    K^\mathrm{crit}_{n,\varepsilon} \sim \frac{\overline{\rho_n(E)}} { \left( w_T\, \varepsilon \right)^{1/2}} \int_{-\infty}^\infty \!\!\!\! d\beta \; \kappa\, \mathrm{e}^{\beta/2}\, \mathrm{e}^{-\kappa\, \mathrm{e}^\beta} = \sqrt{\pi \kappa}\; 
    \frac{\overline{\rho_n(E)}} {\left(w_T\, \varepsilon \right)^{1/2}} ,
\end{equation}
where the scaling limit (\ref{eq:kappa}) was assumed. The values of $\kappa > 0$ and the constant factors omitted in (\ref{critical scaling K}) depend on the details of the chosen boundary weight function $P_\gamma$, whereas the exponent $1/2$ in $\varepsilon^{-1/2}$ does not.

We remark that the fixed-$\beta$ asymptotics of $\langle Y_{\mathrm{sc},(\beta, \vartheta)}, 1 \rangle_\varepsilon$ is non-uniform in $\beta$ for $\beta$ going to minus infinity. Matching $\overline{\rho_n(E)}\, (w_T\, \mathrm{e}^\beta \varepsilon)^{-1/2}$ to the pp-scaling $\overline{\rho_n(E)}\, \varepsilon^{-1}$ gives a crossover scale $\beta_c \sim - \ln(w_T/ \varepsilon)$ for the range $\beta \gg \beta_c$ of sc-scaling to be definitely valid. Since $d\sigma_{\rm sc} \sim \mathrm{e}^\beta\, d\beta$ for $\beta\to -\infty$, the total weight outside that range is $\mathcal{O}(\varepsilon/w_T)$. Even if one makes the conservative assumption of pp-type scaling $\mathcal{O} (\overline{\rho_n(E)}\, \varepsilon^{-1})$ throughout the entire interval $\beta\in (-\infty, \ln(\varepsilon/w_T))$, its overall contribution is only $\mathcal{O}(1)$ in $\varepsilon$, subleading to the $\mathcal{O}((\varepsilon/w_T)^{-1/2})$ contribution from the critical extremal Gibbs states with $\beta \gg \ln (\varepsilon/w_T)$.

The discussion above also leads to a direct interpretation of the log-radial coordinate $\beta$ for the critical order-parameter orbit $G/N$. Decomposing the $K_{n, \varepsilon}$-expression (\ref{K and D}), Sect.~\ref{sect:spec}, as an integral over $\beta$ and inserting for each fixed value of $\beta$ the contribution $\langle Y_\mathrm{sc} , 1 \rangle_\varepsilon \sim \left(w_T\, \mathrm{e}^\beta \varepsilon \right)^{-1/2}$ from the critical extremal state at $\beta$, we see that the corresponding $\beta$-component of the $D_n$-distribution has a characteristic width
\begin{equation}\label{d extremal}
    d_{\varepsilon,\beta} \sim (w_T\, \mathrm{e}^\beta \varepsilon)^{1/2} .
\end{equation}
Seen from this perspective, $\beta$ acquires the meaning of a log-scale coordinate for the $\beta$-resolved critical $D_n$-fluctuations: finite values of $\beta$ encode scale-related critical distributions, all with the same fractal exponent ($\alpha = 1/2$), while taking $\beta \to -\infty$ continuously lowers the width $d_{\varepsilon, \beta}$ down to zero to reach the pp-type $D_n$-distribution. As the crossover scale $\beta_c \sim \ln(\varepsilon/w_T)$ is approached from above, the characteristic width becomes $d_{\varepsilon, \beta} \sim \varepsilon$. Thus $\beta_c$ truly divides the $\beta$-axis into an sc-regime ($\beta \gg \beta_c)$ and a pp-regime ($\beta \ll \beta_c$), and the interpretation via the $\beta$-components of (the imaginary part of the self-energy) $D_n$ explains the appearance of the crossover scale $\beta_c$ from first principles. 
To complete the whole chain of argument, we recall from Sect.~\ref{sect:III-B} that the scaling behavior (\ref{critical scaling K}) in the mixed state [as well as the characteristic width (\ref{d extremal})] are indicative of a singular continuous spectrum. By the general analysis of Sect.~\ref{sect:III-C}, this implies that the radial marginal of the Fourier-Laplace transform has a symmetry-restored plateau for $x_3 \ll \sqrt{\kappa}\, (w_T\, \varepsilon)^{-1/2}$, crosses over to the fractal tail $\overline{\rho_n(E)} \sqrt{\kappa}\, (w_T\, x_3)^{-1/2}$ corresponding to the long-time behavior near $x_3 \sim \sqrt{\kappa}\, (w_T\, \varepsilon)^{-1/2}$ of the disorder-averaged return probability, and is finally cut off at $x_3 \sim \varepsilon^{-1}$.

As a consistency check we now verify that the critical extremal states $Y_{\rm sc}$ reproduce the tail $\overline{\rho_n(E)}\, (w_T\, x_3)^{-1/2}$. In the large-$\varphi$ regime relevant here, the first term in $x_3 = w_T^{-1} (\mathrm{e}^\varphi + (m^2+b^2)\, \mathrm{e}^{-\varphi})/2$ dominates, so that $x_3 \sim w_T^{-1} \mathrm{e}^\varphi$. Therefore, to rediscover the fractal tail of the disorder-averaged return probability in the extremal states $Y_{\rm sc}$, we should study their large $\varphi$ tail. As we have seen in the preceding part, integrating out $m$ and $b$ in $Y_\mathrm{sc}$ at large $\varphi$ [more precisely, at the typical scale $\varphi \sim - \ln (\varepsilon/w_T)$ set by the regulator $\varepsilon \to 0$] yields the $\varphi$-marginal
\begin{equation}
   M_\mathrm{sc,\beta}(\varphi)\, d\varphi \sim \frac{\overline{\rho_n(E)}} {w_T}\, \mathrm{e}^{(\varphi -\beta)/2}\, d\varphi .
\end{equation}
Converting to $x_3\sim (2w_T)^{-1} \mathrm{e}^\varphi$, we obtain
\begin{equation}\label{FL sc scaling}
    M_{\rm sc,\beta}(x_3)\,dx_3 \sim \overline{\rho_n(E)}\, (w_T\, \mathrm{e}^\beta x_3)^{-1/2}\, dx_3 \qquad (x_3\sim \varepsilon^{-1} \rightarrow \infty).
\end{equation}
Thus, the critical extremal states do reproduce the $x_3^{-1/2}$ asymptotic tail that we anticipated in Sect.~\ref{sect:III-C} and is seen for the physical state in the result (\ref{critical scaling K}). The same analysis gives the large-time behavior of the disorder-averaged return probability:
\begin{equation}\label{sc return prob}
    \frac{2\pi}{\Delta E} \mathbb{E} \big( \mathrm{Prob}_{n\rightarrow n}^{E,\Delta E}(t) \big) \sim \overline{\rho_n(E)}\, \sqrt{\kappa}\, (w_T\, t)^{-1/2} .
\end{equation}
(\ref{sc return prob}) is the final result of this work.

\section{Conclusion}\label{Conclusion}
\subsection{Discussion}

Initiated by Wegner \cite{Weg79a}, developed by Efetov \cite{Efe83}, and applied by many others, effective field theories of Anderson transitions stand apart (within the realm of condensed matter physics) in that their target spaces are Riemannian symmetric \emph{superspaces} \cite{Zi-JMP96}, typically over a non-compact base manifold. The present paper, well aware of the pros and cons of the Wegner-Efetov supersymmetry (SUSY) approach, makes an effort to avoid super-calculus as much as possible, at least for the hard core of our analysis. This avoidance comes with two price tags: (i) we had to consider a very special (though still informative) observable, the time-integrated return probability of Wegner's one-orbital model of type $A$, and (ii) we settled for the rough approximation scheme given by AAT self-consistency. Operating within these restrictions, we were able to eliminate all of the Grassmann variables of the SUSY formalism and ultimately keep to the familiar framework of ordinary analysis. It should be stressed that the latent presence of the Grassmann degrees of freedom still plays a decisive role via the differential operator (\ref{eq:def-Delta}) of Laplace-type. (Incidentally, the appearance of that Laplacian documents the true nature of Grassmann ``integration'' as \emph{differentiation}).

Building on our reduced framework of purely ``bosonic'' kind, we developed a scenario for Anderson transitions in high spatial dimensions. Our main prediction is a pattern of spontaneous symmetry breaking which differs markedly from what is known for conventional critical phenomena. That difference has its seed in the non-compactness of the field-theory symmetry group $G$, where $G = \mathrm{SU}(1,1)$ in our reduced case. For Anderson transitions at large, it is understood that $G$ spontaneously breaks down to a compact subgroup $K$ in the metallic phase [$K \cong \mathrm{U}(1)$ in our case] and is fully restored in the insulating phase. We predict that something novel happens at the phase boundary in between: the symmetry $G$ is spontaneously broken to a \emph{nilpotent} subgroup $N \subset G$. From the perspective of the Iwasawa decomposition $G = N A K$ of a non-compact semisimple Lie group $G$, this proposal is quite natural: the nilpotent group $N$ appears as the limit of pushing the compact group $K$ by elements of the ``radial'' Abelian group $A$. In summary, the proposed pattern of symmetry breaking (metallic to critical to insulating) is
\begin{equation}\label{SSB diagram}
    G/K \rightarrow G/N \rightarrow G/G.
\end{equation}
When the critical point is approached from the insulating side, this scenario implies that the $A$-degrees of freedom go critical while the $N$-degrees of freedom remain symmetry-unbroken.

We have argued that the $G/N$ symmetry-breaking scenario at criticality is indicative of a local spectral measure of singular continuous type. To detect the spectral type, we monitored the scaling of the integrated Fourier-Laplace transform $Y_{\rm FL}$ in the limit of a vanishing regulator $\varepsilon \to 0$. Key to our approach was the resolution of the physical state $Y_{\rm FL}$ as a mixture of extremal Gibbs states. The individual extremal states may break a $\mathrm{U}(1)_{\mathrm{i} \sigma_3}$-symmetry as highlighted in our title, even though their incoherent mixture $Y_{\rm FL}$ never does. Our scheme of calculation was to first compute the observable in the symmetry-broken, extremal states and afterwards assemble the mixture by zero-mode integration. The extremal states themselves were computed approximately by neglecting electron loops and appealing to the AAT self-consistency scheme. We emphasize that the critical extremal Gibbs states $Y_{\rm sc}$ are brought to the fore by taking the zero-mode mixture in a scaling limit that resolves the critical field fluctuations. While the field fluctuations are small deep in the metallic phase, they bulge farther and farther into the direction of target-space infinity (visualized as the boundary $|z| = 1$ of the Poincar\'e disk model $|z| < 1$ of $\mathrm{H}^2$) as the phase transition is approached, with their extent controlled by the inverse correlation-volume scale $C$.

In Sect.~\ref{K sc} we implemented the critical scaling limit by a suitable regularization of the mixing distribution on the metallic order-parameter orbit $G/K$. This was accomplished by a $\mathrm{U}(1)_{\mathrm{i} \sigma_3}$-invariant boundary weight function $P_\gamma$ (with a physical meaning as weakly absorbing boundary conditions of local absorption rate $w_T\, \gamma \to 0$) whose radial weight is concentrated at boost scales $\mathrm{e}^\tau \sim \gamma^{-1}$. In parallel, the critical extremal states arise by zooming into the metallic order-parameter orbit at the diverging boost scale $\mathrm{e}^{\tau_C}$ defined by Eq.~(\ref{eq:def-tauC}). By keeping the product $\gamma\, \mathrm{e}^{\tau_C}$ fixed in the approach to the critical point, the regularized boundary weight distribution on $G/K$ is pushed forward to a boundary-weight distribution $d\sigma_{\rm sc}$ on the critical orbit $G/N$ with zoom-in parameter $\beta$. An important feature of this distribution is its behavior near the adjoining tip $G/G$ (reached as $\beta \to -\infty$) representing the fully $G$-symmetric pp-type state. Its radial density vanishes there as $d\sigma_{\rm sc}\propto \mathrm{e}^\beta d\beta$, whereas the contribution of an individual extremal Gibbs state scales as $\mathrm{e}^{-\beta/2}$; see Eq.~(\ref{K fractal scaling}). Thus the enhancement due to the extremal states near the pp-tip is over-compensated by the vanishing of the boundary-weight distribution. It follows that the critical mixing distribution of extremal Gibbs states has no atom at the tip $\beta \to -\infty$ and no $\varepsilon^{-1}$ contribution survives in this mixture. Our observable of interest therefore exhibits the scaling behavior displayed in (\ref{critical scaling K}) and (\ref{sc return prob}), with the characteristic scaling exponent (in the variable $t \sim \varepsilon^{-1}$) being $\alpha = 1/2$. Intermediate between the exponents $\alpha = 0$ for absolutely continuous spectrum and $\alpha = 1$ for pure point spectrum, this value implies a spectrum of singular continuous type.

To finish this discussion, let us be clear about one subtle point: (i) The quantity we have computed is $K_{n, \varepsilon} (E)$ for $\varepsilon \to 0$. (ii) In order to relate this quantity to the return probability $\mathbb{E}(\mathrm{Prob}_{n \to n}^{E, \Delta E}(t))$ it is necessary to average $K_{n, \varepsilon}(E)$ over an energy interval $\Delta E$. (iii) As we have seen, a spectral measure of singular continuous type exists only at the critical point $w = w_c$ (or for an isolated energy $E = E_c$). (iv) To detect the presence of that singular continuous measure, one must choose the energy-averaging window $\Delta E$ to be of microscopic size, i.e.\ small enough to avoid contamination by the neighboring spectral type (pp). We have argued that the proper size to be used at the regularization scale $\varepsilon$ is $\Delta E \sim \varepsilon^{1/2}$. (Note that in numerical diagonalization studies one takes $\varepsilon \sim \mathrm{vol}^{-1}$.) (v) Adhering to this prescription, we find that $\mathbb{E}(\mathrm{Prob}_{n \to n}^{E, \Delta E}(t))$ has a plateau up to times $\sim \varepsilon^{-1/2}$, followed by a fractal tail $\sim t^{-1/2}$ up to the cutoff scale $\varepsilon^{-1}$. (vi) Our prediction of $\alpha = 1/2$ disagrees qualitatively with the critical-localization scenario found in Ref.~\cite{Tik19} by exact diagonalization of the Anderson model on random regular graphs. There, the disorder-averaged return probability does not develop an algebraic tail (as $w \to w_c$) but approaches a plateau with logarithmic speed. The discrepancy may be related to a different way of handling the energy projection and its critical scaling.

\subsection{Speculation}

If our ``field-theoretical'' AAT-analysis lives up to expectations, it ought to deliver the Lagrangian of a field integral that computes the extremal Gibbs states $Y_{\rm sc}$ at criticality. Such a Lagrangian was put forth in Ref.~\cite{Zi-23}. Armed with the additional insight from the present paper, we now comment on that proposal and put an eye-opening twist on it.

We start from our target-space cone $C^3$ parameterized by the horospherical coordinates of Eq.~(\ref{eq:horoscoord}). These come from a parabolic group $P_- = A N_-$ where $A = \exp(\mathbb{R}\, \sigma_1)$ and $N = \exp(\mathbb{R}\, \sigma_-)$ for $\sigma_- = (\sigma_2 - \mathrm{i} \sigma_3)/2$, acting by conjugation on
\begin{equation}
q_0(m) = w_T^{-1} \left( m \cdot \mathbf{1} + {\textstyle{\frac{1}{2}}} (1 + m^2)\, \sigma_3 - {\textstyle{\frac{1}{2}}} (1 - m^2)\, \mathrm{i} \sigma_2 \right) \in \sqrt{-1} \times \mathrm{Lie}\, \mathrm{U}(1,1) .
\end{equation}
Because $A$ normalizes $N_-$, one may reorganize the factors as $P_- = N_- A$. If the latter factorization is adopted with $n = \mathrm{e}^{c \sigma_-} \in N_-$ and $a = \mathrm{e}^{\varphi \sigma_1 / 2} \in A$, the continuum approximation for a simple cubic lattice $\mathbb{G}$ in $d$ dimensions yields
\begin{equation}
    {\textstyle{\frac{1}{2}}} w_V^2 \sum \mathrm{Tr}\, Q_j^2 + {\textstyle{\frac{1}{2}}} w_T^2 \sum A_{j j^\prime} \mathrm{Tr}\, Q_j Q_{j^\prime} \approx \int d^d x \left( 2 w^2 m^2 + (\nabla m - m \nabla \varphi)^2 + 2 \mathrm{e}^{2\varphi} (\nabla c)^2 \right) .
\end{equation}
The functional integration measure is formally given by $\prod_x d\varphi(x) \, \mathrm{e}^{\varphi(x)} dc(x)\, dm(x)$, and $x$ is a dimensionless position variable measured in units of the lattice constant, or UV-cutoff, $a_{\rm UV}$. Because $m_j$ occurs in $\sum A_{j j^\prime} \mathrm{Tr}\, Q_j Q_{j^\prime}$ for constant $\varphi_j$ as $\sum A_{j j^\prime} (m_j +m_{j^\prime})^2$, we staggered $m_j \to (-1)^{j_1 + \ldots + j_d}\, m_j$ before making the continuum approximation.
Now we must not forget the presence of the Grassmann variables in the supermatrix $\mathcal{Q}$. Their effect in the AAT scheme was succinctly captured by the presence of the Laplacian $\Delta$ acting on $Y$. In the field-theory setting sketched here, their main effect is to invert the volume density $\mathrm{e}^{\varphi(x)} \to \mathrm{e}^{-\varphi(x)}$, thereby effectively compactifying the target-space cone $C^3$ towards $+ \infty$ for $\varphi$. Given these basic facts, the reasoning of Ref.~\cite{Zi-23} went as follows.

Driven by the volume-compactification due to the latent Grassmann variables, the field $\varphi$ populates predominantly the positive axis extending towards $\varphi \to + \infty$. This has the effect of freezing out the $c$-field by making its stiffness $\sim \mathrm{e}^{2\varphi}$ very large. Now while that is true, it does not offer the optimal perspective, as it looks at the situation in suboptimal coordinates. In hindsight, the ``good'' coordinates are provided by the action on $q_0(m)$ of the \emph{opposite} parabolic group $P_+ = N_+ A$ with $N_+ = \exp(\mathbb{R}\, \sigma_+)$ for $\sigma_+ = (\sigma_2 + \mathrm{i}\sigma_3) / 2$. If we parameterize $N_+$ by $n = \mathrm{e}^{b \sigma_+}$ (\underline{not} the $b$ of our main text), the continuum Lagrangian above takes the equivalent form
\begin{equation}\label{eq:L-UV}
    \mathcal{L}_{\rm UV} = \mathcal{L}_0 + \mathcal{L}_2\,, \quad \mathcal{L}_0 = 2 w^2 m^2, \quad \mathcal{L}_2 = (\nabla m - m \nabla \varphi)^2 + 2 m^4 \mathrm{e}^{-2\varphi} (\nabla b)^2 .
\end{equation}
Note that the stiffness has changed from very large ($\sim \mathrm{e}^{2\varphi}$) for the $c$-field to very small ($\sim m^4 \mathrm{e}^{-2\varphi}$) for the $b$-field. Having done no more than a simple change of target-space coordinates, we do find that the situation is now easier to analyze. Indeed, in high space dimensions we expect the anomalous scaling dimensions of the fields to be small (if not zero), and we can get a first-order approximation to the renormalization-group flow by simple power counting:
\begin{equation}\label{eq:L-IR}
    \mathcal{L}_{\rm UV} \to \mathcal{L}_{\rm IR} = (a_{\rm IR} / a_{\rm UV})^d \mathcal{L}_0 + (a_{\rm IR} / a_{\rm UV})^{d-2} \mathcal{L}_2 .
\end{equation}
As we renormalize from the ultraviolet (UV) to the infrared (IR), the fluctuating stiffness $(a_{\rm IR} / a_{\rm UV})^{d-2}\, 2 m^4 \mathrm{e}^{- 2\varphi}$ of the $b$-field seems to grow as a positive power of the ratio of cutoff lengths. Yet, this growth is over-compensated by the shrinking amplitude of the field fluctations of the massive variable $m$. Approximating the factor $m^4$ by its average value, $m^4 \sim w^{-4} (a_{\rm IR} / a_{\rm UV})^{-2d}$, we see that the $b$-stiffness flows to zero rapidly as $(a_{\rm IR} / a_{\rm UV})^{-d-2}$. This means that the fluctuations of the $b$-field grow strong in the infrared, resulting in symmetry restoration for the nilpotent group $N_+$. Such a scenario is consensual for the pp-phase, where all symmetries (not just $N_+$) are restored. The novelty we propose is that, on approaching criticality from the pp-phase, $N_+$ remains unbroken all the way up to the phase transition, while the remaining symmetry degrees of freedom go critical.

What is the fate of the nonlinear vertex field $\mathrm{e}^\varphi$ at criticality? Since there exists competition between different effects [cf.\ Eqs.~(\ref{eq:L-UV}, \ref{eq:L-IR})], that question cannot be answered by simple power counting. Moreover, turning from the AAT scheme to the general field-theory context, our reduced ``bosonic'' framework with target-space cone $C^3$ becomes too narrow in several respects. For one, the Grassmann degrees of freedom, elegantly circumvented in the AAT scheme, take on a prominent role in the field theory. For another, our framework does not hold up under renormalization to the critical infrared regime, as the non-convex space $C^3$ is not stable under coarse graining.

In this plight, we may seek guidance from the Wegner model with $N \geq 2$ orbitals [cf.\ Eq.~(\ref{eq:N-geq-2})] and supermatrix field $\mathcal{Q} = N^{-1} \sum_{\alpha = 1}^N \Psi_\alpha \otimes \Psi_\alpha^\dagger \Sigma_3$ where $\Sigma_3 = \sigma_3 \otimes \mathbf{1}_{\rm SUSY}$. By the method of superbosonization \cite{LSZ-07}, the boson-boson block of $\mathcal{Q} \Sigma_3$ assumes values in the convex cone $C_+^4 \cong \mathrm{GL}(2,\mathbb{C}) / \mathrm{U}(2)$ of positive Hermitian $2 \times 2$ matrices, while the fermion-fermion block bosonizes to an element of the compact Lie group $\mathrm{U}(2)$ or some complex-analytic equivalent thereof. Prognosticating conservatively, further analysis of the strong-disorder regime will take us to something akin to the Wegner-Efetov nonlinear $\sigma$-model, albeit with fluctuating eigenvalues or diagonal matrix blocks (generalizing our massive variable $m$). What we can say for sure is that the nonlinear field $\mathrm{e}^\varphi$ must get promoted to a supermatrix $g$ in a Riemannian real form of the complex Lie supergroup $\mathrm{GL}(1|1)$. Assuming that the scenario uncovered here remains qualitatively correct, the physical state at criticality will be a super-mixture of extremal Gibbs states, each of which is obtained by \emph{gauging} the nonlinear SUSY model with respect to some nilpotent supergroup $\mathcal{N}_+$. [Indeed, restoration of the symmetry $\mathcal{N}_+$ means that its degrees of freedom organize as short-range correlated singlet states and thus become redundant, dropping out of the physical sector in the way of gauge degrees of freedom.] On symmetry grounds, the gauge-reduced effective theory will be a principal chiral model for the field $g$. Clearly, we are leaving many details open here. We hope to provide them in an update of Ref.~\cite{Zi-23} (unless AI beats us to the finish line.)

\bigskip\noindent{\bf Acknowledgment.} The research reported in this article was conceived, developed and finalized by two humans at a human pace (4 years total). The salary of the junior author came from the Deutsche Forschungsgemeinschaft, grant ZI 513/3-1 (01/24-10/26).

\bibliography{Wegner-AAT.bib}

\section{Appendices}

\subsection{Derivation of AAT integral equation}\label{deriv: AAT}

The derivation starts from the Fourier-Laplace transform $Y_{\rm FL}$ on a finite sublattice. To formulate the AAT approximation as a recursion, let $\mathcal{G}^{-1}_{k+1}$ be obtained from $K$ independent copies of $\mathcal{G}_k$ according to Eq.~(\ref{eq:AATprime}), $k$ denotes the recursion depth, and denote by $\mathcal{Y}_{\mathrm{FL}, k}$ the corresponding super-analytic extension of the Fourier-Laplace transform.
We begin by inserting the r.h.s. of the AAT equation for the matrix Green's function (\ref{eq:AATprime}) into the definition of $\mathcal{Y}_{\mathrm{FL},{k+1}}$ that follows immediately from (\ref{eq:def-Y}) :
\begin{equation}
\begin{gathered}
    \mathcal{Y}_{\mathrm{FL}, k+1} (\mathcal{Q}_n) := \mathbb{E}\left(\mathrm{e}^{\mathrm{i} \mathrm{Tr}(Q_{\mathrm{BB}_n}-Q_{\mathrm{FF}_n}) (\mathcal{G}_{k+1})^{-1}}\right) \\= \mathrm{e}^{\mathrm{i} \mathrm{Tr}\, (Q_{\text{BB}_n} - Q_{\text{FF}_n}) (E + \mathrm{i}\varepsilon \sigma_3)}\, \mathbb{E} \left( \mathrm{e}^{- \mathrm{i} \mathrm{Tr} \, (Q_{\text{BB}_n} - Q_{\text{FF}_n}) h_{n,n} - \mathrm{i} \sum \mathrm{Tr} \, (Q_{\text{BB}_n} - Q_{\text{FF}_n}) {\mathcal{G}_{n^\prime, k}} |h_{n,n^\prime}|^2 } \right) .
    \end{gathered}
\end{equation}
Here we have omitted the sum subscript $n^\prime (\neq n)$ over the $K$ adjacent sites to simplify the notation. Next, we recall the notion of supertrace in $\mathbb{Z}_2$-graded linear algebra:
\begin{equation}
	\STr Q = \Tr Q_\text{BB} - \Tr Q_\text{FF}
\end{equation}
and abbreviate:
\begin{equation}\label{AAT deriv:abrev Y with STR}
    \mathcal{Y}_{\mathrm{FL}, k+1}(\mathcal{Q}_n) = \mathrm{e}^{\mathrm{i} \STr\, \mathcal{Q}_n (E + \mathrm{i}\varepsilon \Sigma_3)}\, \mathbb{E} \left( \mathrm{e}^{- \mathrm{i} h_{n,n} \STr \, \mathcal{Q}_n - \mathrm{i} \sum |h_{n,n^\prime}|^2 \STr \, \mathcal{Q}_n ({\mathcal{G}_{n^\prime, k}})} \right) .
\end{equation}
In (\ref{AAT deriv:abrev Y with STR}) we have extended 
\begin{equation}
    \sigma_3\rightarrow \Sigma_3 = \sigma_3\otimes \mathbf{1} = \mathrm{diag}(1,-1,1,-1), \qquad \mathcal{G}_{n^\prime, k} \rightarrow {\mathcal{G}_{n^\prime, k}} \otimes \mathbf{1}\equiv {{\mathcal{G}_{n^\prime, k}}}
\end{equation}
from the space $\mathbb{C}^2$ to the $\mathbb{Z}_2$-graded space $\mathbb{C}^{2|2}$ by tensoring with the identity $\mathbf{1}$ on $\mathbb{C}^{1|1}$. Next we observe that the entries of the Hamiltonian matrix $h_{n,n}$ and $h_{n,n^\prime}$ in Eq.~(\ref{AAT deriv:abrev Y with STR}) occur in an exponential to at most quadratic order. Because they are Gaussian-distributed, their disorder average can be computed. We begin with the disorder average over the diagonal elements $h_{n,n}$. Furthermore, to prepare the second step of averaging over the off-diagonal elements $h_{n,n^\prime}$, we observe that the $K$ terms ($K$ is the branching number of the underlying lattice) in the sum on the r.h.s.\ over $n^\prime$ are all statistically independent and identical. We obtain:
\begin{equation}\label{eq:int-res}
    \mathcal{Y}_{\mathrm{FL}, k+1}(\mathcal{Q}_n) = \mathrm{e}^{\mathrm{i} \STr\, \mathcal{Q}_n (E\mathbf{1} + \mathrm{i}\varepsilon \Sigma_3) -\frac{1}{2} w_V^2 (\STr \, \mathcal{Q}_n)^2} \mathbb{E} \left( \mathrm{e}^{ - \mathrm{i} |h_{n,n^\prime}|^2 \STr \, \mathcal{Q}_n{\mathcal{G}_{n^\prime, k}}}  \right)^K ,
\end{equation}
where $n^\prime$ now labels one of the $K$ nearest neighbors of site $n$. Next we introduce an auxiliary supermatrix $\mathcal{Q}_{n^\prime}$ (the field values on the neighboring sites) by
\begin{equation}
    \mathcal{Q}_{n^\prime} \equiv \Phi_{n^\prime} \otimes \widetilde\Phi_{n^\prime} , \quad \Phi_{n^\prime} = \begin{pmatrix} u_{n^\prime} \cr v_{n^\prime} \cr \zeta_{n^\prime}\cr \omega_{n^\prime} \end{pmatrix} , \quad \widetilde{\Phi}_{n^\prime} = (\bar{u}_{n^\prime} \;\; - \bar{v}_{n^\prime} \;\; \bar{\zeta}_{n^\prime} \;\; \bar{\omega}_{n^\prime}) \quad (u_{n^\prime} , v_{n^\prime} \in \mathbb{C}).
\end{equation}
We then process the exponential under the expectation by a Gaussian Berezin integral:
\begin{equation}\label{eq: SUSY trick}
\begin{gathered}
    \mathrm{e}^{ -\mathrm{i} |h_{n,n^\prime}|^2 \STr \, \mathcal{Q}_n {\mathcal{G}_{n^\prime, k}} } = \mathrm{e}^{ -\mathrm{i} |h_{n,n^\prime}|^2 \widetilde{\Phi}_n {\mathcal{G}_{n^\prime, k}} \Phi_n } \\= - \pi^{-2} \int_{\mathbb{C}} d^2 u_{n^\prime} \int_{\mathbb{C}} d^2 v_{n^\prime} \frac{\partial^2}{\partial\zeta_{n^\prime} \partial\bar{\zeta}_{n^\prime}}\frac{\partial^2}{\partial\omega_{n^\prime} \partial\bar{\omega}_{n^\prime}}  \, \mathrm{e}^{\mathrm{i} \widetilde\Phi_{n^\prime} {\mathcal{G}_{n^\prime, k}^{-1}} \Phi_{n^\prime} + \mathrm{i} h_{n,n^\prime} \tilde{\Phi}_n \Phi_{n^\prime} + \mathrm{i} h_{n^\prime,n} \widetilde{\Phi}_{n^\prime} \Phi_n } .
    \end{gathered}
\end{equation}
The integral converges because the real part of the exponent is negative by our choice of $\Phi_{n^\prime}, \widetilde\Phi_{n^\prime}$. [Recall ${D_{n^\prime}} > 0$ from Eq.\ (\ref{eq:strucG}) by construction.] Now the reason for introducing Grassmann variables and continuing $Y_{\rm FL, k}$ to the superfunction $\mathcal{Y}_{\rm FL, k}$ (\ref{eq:SUSY-Y}) is transparent: The Gaussian integration over the complex variables $u_{n^\prime}, v_{n^\prime}$ produces a determinant $\Det({\mathcal{G}_{n^\prime, k}})$. To recover the l.h.s.\ of Eq.~(\ref{eq: SUSY trick}), this determinant has to cancel out. The introduction of Grassmann variables does this job: indeed, the Grassmann derivatives in (\ref{eq: SUSY trick}) yield the inverse determinant $\Det({\mathcal{G}_{n^\prime, k}})^{-1}$. Also, the integrand in Eq.~(\ref{eq: SUSY trick}) is Gaussian, and the disorder average over the Gaussian disorder can easily be performed. In summary, with the Grassmann variables, we recover the l.h.s.\ while retaining a simple form of the integrand.

We are now in a position to perform the average over the Gaussian-distributed off-diagonal elements $h_{n,n^\prime} = \overline{h_{n^\prime,n}}$:
\begin{equation}
    \mathbb{E} \left( \mathrm{e}^{\mathrm{i} h_{n,n^\prime} \widetilde{\Phi}_n \Phi_{n^\prime} + \mathrm{i} h_{n^\prime,n} \widetilde{\Phi}_{n^\prime} \Phi_n } \right)
    = \mathrm{e}^{ - w_T^2 \widetilde{\Phi}_n \Phi_{n^\prime} \widetilde{\Phi}_{n^\prime} \Phi_n} = \mathrm{e}^{ - w_T^2 \STr\, \mathcal{Q}_n \mathcal{Q}_{n^\prime}} .
\end{equation}
Assembling the various pieces yields
\begin{equation}
\begin{gathered}
    \mathbb{E} \left( \mathrm{e}^{ - \mathrm{i} |h_{n,n^\prime}|^2 \STr \, \mathcal{Q}_n {\mathcal{G}_{n^\prime, k}}}  \right)\\
    = - \pi^{-2} \int_{\mathbb{C}} d^2 u_{n^\prime} \int_{\mathbb{C}} d^2 v_{n^\prime} \frac{\partial^2}{\partial\zeta_{n^\prime} \partial\bar{\zeta}_{n^\prime}}\frac{\partial^2}{\partial\omega_{n^\prime} \partial\bar{\omega}_{n^\prime}}  \; \mathrm{e}^{ - w_T^2 \STr\, \mathcal{Q}_n \mathcal{Q}_{n^\prime}}
    \, \mathbb{E} \left(\mathrm{e}^{\mathrm{i} \widetilde\Phi_{n^\prime} ({\mathcal{G}_{n^\prime, k}})^{-1} \Phi_{n^\prime}} \right) .
    \end{gathered}
\end{equation}
Finally, we notice that the expectation value on the r.h.s.\ is the Fourier-Laplace transform at the preceding step of the recursion $\mathbb{E} \big( \mathrm{e}^{\mathrm{i} \widetilde\Phi_{n^\prime} ({\mathcal{G}_{n^\prime, k}})^{-1} \Phi_{n^\prime}} \big) = {\mathcal{Y}_{\mathrm{FL}, k}(\mathcal{Q}_{n^\prime})}$, and we recall the intermediate result (\ref{eq:int-res}). We then arrive at the AAT integral equation:
\begin{equation}
    \mathcal{Y}_{\mathrm{FL}, k+1}(\mathcal{Q}) = \mathrm{e}^{ \mathrm{i} \, \mathrm{STr}\, \mathcal{Q}(E + \mathrm{i} \varepsilon \Sigma_3 ) - \frac{1}{2} w_V^2 \mathrm{STr}\, \mathcal{Q}^2} \left( \int \mathcal{D Q}^\prime \, \mathrm{e}^{- w_T^2 \mathrm{STr}\, \mathcal{Q Q}^\prime}\ {\mathcal{Y}_{\mathrm{FL}, k}(\mathcal{Q}^\prime)} \right)^K ,
\end{equation}
where $\mathcal{Q}$ denotes the $4 \times 4$ supermatrix
\begin{equation}
    \mathcal{Q} := \rowfour{u}{v}{\zeta}{\omega} \otimes \columnfour{\bar{u}} {-\bar{v}} {\bar\zeta} {\bar\omega} \equiv \begin{pmatrix} Q_\mathrm{BB} &Q_\mathrm{BF} \cr Q_\mathrm{FB} &Q_\mathrm{FF} \end{pmatrix} ,
\end{equation}
which coincides with $Q_\mathrm{BB}$ in the Bose-Bose sector and $Q_\mathrm{FF}$ in the Fermi-Fermi sector, and we have used that $\mathcal{Q}$ has rank one: 
\begin{equation}
    (\mathrm{STr}\,\mathcal{Q})^2 = \mathrm{STr}\,\mathcal{Q}^2.
\end{equation}
Note that ${\mathcal{Y}_{\mathrm{FL}, k}} \equiv {\mathcal{Y}_{\mathrm{FL}, k}(\mathcal{Q})}$ remains a function of $Q_\mathrm{BB} - Q_\mathrm{FF}$, the difference of the diagonal blocks for all $k\in \mathbb{N}$. In the thermodynamic limit with a fixed boundary condition $x$, we look for a fixed point $\mathcal{Y}_x$ of this recursion. It provides the AAT approximation of the extremal Gibbs state on high dimensional lattices $\mathbb{G}$ centered around $x$ as explained in Sect.~\ref{sect:Gibbs}.

\subsection{Derivation of field theory on lattices $\mathbb{G}$}\label{App.Efetov SUSY}

With the observable 
\begin{equation}
    K_{n,\varepsilon} = \mathbb{E}(G_n(E + \mathrm{i}\varepsilon)\, G_n(E - \mathrm{i}\varepsilon)),
\end{equation}
the derivation of the field theory is very transparent. We begin by defining field-theoretical states that contain complex retarded ($+$) variables $u_j,\, j \in \mathbb{G}$ or complex advanced ($-$) variables $v_j,\, j \in \mathbb{G}$:
\begin{equation}\label{App.retarded advanced states}
\begin{gathered}
    \langle\bullet\rangle_{+} \equiv \mathrm{Det}(\varepsilon-\mathrm{i}(E-h)) \int \prod_{j\in\mathbb{G}} \frac{d^2 u_j}{\pi}\ \bullet\ \mathrm{e}^{\mathrm{i} \bar{u} \cdot (E + \mathrm{i}\varepsilon-h)u},\\
    \langle \bullet \rangle_{-} \equiv \Det(\varepsilon + \mathrm{i}(E-h)) \int \prod_{j\in\mathbb{G}} \frac{d^2 v_j}{\pi}\ \bullet\ \mathrm{e}^{-\mathrm{i} \bar{v} \cdot (E-\mathrm{i} \varepsilon - h)v},
\end{gathered}
\end{equation}
where 
\begin{equation}
    \bar{u}\cdot (E+\mathrm{i}\varepsilon-h)u \equiv \sum_{j,j^\prime} \bar{u}_j\, (E + \mathrm{i}\varepsilon - h)_{j,j^\prime}\, u_{j^\prime},
\end{equation}
and equivalently for $v$. The determinants in the expressions~(\ref{App.retarded advanced states}) ensure the normalization of the states. We note that, strictly speaking, the integrals are only well-defined on a finite sublattice of $\mathbb{G}$ (with certain boundary conditions) and the thermodynamic limit is taken after deriving the field theory representation. For notational simplicity we here omit this intermediate step. 

To perform the expectation value $\mathbb{E}$ with respect to the disorder, we rewrite them as Berezin integrals of Gaussian type. We use the identity:
\begin{equation}
    \Det(A) = \prod_{j}\frac{\partial^2}{\partial\zeta_j\partial\bar{\zeta}_j}\, \mathrm{e}^{\bar{\zeta} \cdot A\zeta}
\end{equation}
for suitable matrices $A$. We then arrive at
\begin{equation}
\begin{gathered}
    \langle\bullet\rangle_{+} \equiv \int \prod_{j} \frac{d^2 u_j}{\pi} \otimes \frac{\partial^2}{\partial\zeta_j\partial \bar{\zeta}_j} \bullet\ \mathrm{e}^{\mathrm{i} \bar{u} \cdot (E + \mathrm{i} \varepsilon - h)u - \mathrm{i} \bar{\zeta} \cdot (E + \mathrm{i} \varepsilon - h)\zeta},\\
    \langle \bullet \rangle_{-} \equiv \int \prod_{j} \frac{d^2 v_j}{\pi} \otimes \frac{\partial^2}{\partial\omega_j\partial \bar{\omega}_j} \bullet\ \mathrm{e}^{-\mathrm{i} \bar{v} \cdot (E - \mathrm{i} \varepsilon - h)v + \mathrm{i} \bar{\omega} \cdot (E - \mathrm{i} \varepsilon - h) \omega}.
\end{gathered}
\end{equation}
We observe that with the states $\langle\bullet\rangle_{\pm}$ one can express retarded and advanced Green's functions as a field-theoretical expectation value of local field operators:
\begin{equation}
\begin{gathered}
    \mathrm{i}G_n(E+\mathrm{i}\varepsilon)=\langle\bar{\zeta}_n\zeta_n\rangle_{+}\\
    -\mathrm{i}G_n(E-\mathrm{i}\varepsilon)=\langle\bar{\omega}_n\omega_n\rangle_{-}.
\end{gathered}
\end{equation}
Accordingly, we find
\begin{equation}
    K_{n,\varepsilon} = \mathbb{E}\left(\langle\bar{\zeta}_n\zeta_n\rangle_{+}\, \langle\bar{\omega}_n\omega_n\rangle_{-}\right).
\end{equation}
We now introduce the supervectors:
\begin{equation}
    \Phi_j=\begin{pmatrix}
    u_j \\ v_j \\ \zeta_j \\ \omega_j
    \end{pmatrix}, \quad
    \widetilde{\Phi}_j = \begin{pmatrix}
    \bar{u}_j & -\bar{v}_j & \bar{\zeta}_j & \bar{\omega}_j
    \end{pmatrix}.
\end{equation}
By transforming $\zeta\rightarrow -\zeta$, and abbreviating
\begin{equation}
    \int_{\Phi,\widetilde{\Phi}} \equiv \int \prod_{j} \left[ -\frac{d^2u_j}{\pi} \frac{d^2 v_j}{\pi} \otimes \frac{\partial^2}{\partial\zeta_j \partial \bar{\zeta}_j} \frac{\partial^2}{\partial\omega_j \partial \bar{\omega}_j} \right],
\end{equation}
we obtain for the product of the two states:
\begin{equation}
    \langle \bullet \rangle_{+}\, \langle \bullet \rangle_{-} = \int_{\Phi,\widetilde{\Phi}}\ \bullet\ \mathrm{e}^{\mathrm{i} \widetilde{\Phi} \cdot (E + \mathrm{i} \varepsilon \Sigma_3 - h) \Phi},
\end{equation}
where the product $\cdot$ now also contains the vector indices of $\Phi_j$, $\alpha,\beta$. Next we perform the disorder average $\mathbb{E}$ (we use the sum convention):
\begin{equation}
\begin{aligned}
    &\mathbb{E}\left(\mathrm{e}^{-\mathrm{i} \widetilde{\Phi}_{\alpha,j}h_{j,j^\prime} \Phi_{\alpha,j^\prime}}\right) = \mathrm{e}^{-\frac{1}{2} \widetilde{\Phi}_{\alpha,j} \Phi_{\alpha,j^\prime} \widetilde{\Phi}_{\beta,l} \Phi_{\beta,l^\prime} \mathbb{E}(h_{j,j^\prime} h_{l,l^\prime})}\\ &=\mathrm{e}^{-\frac{1}{2}w_V^2\,(-1)^{|\alpha|}\, (\Phi\otimes\widetilde{\Phi})_{\alpha\beta,j} (\Phi\otimes\widetilde{\Phi})_{\beta\alpha,j} - \frac{1}{2} w_T^2\, A_{j,j^\prime}\, (-1)^{|\alpha|}\, (\Phi\otimes\widetilde{\Phi})_{\alpha\beta,j} (\Phi\otimes\widetilde{\Phi})_{\beta\alpha,j^\prime}}\\ &=\mathrm{e}^{-\frac{1}{2}w_V^2\sum_{j}\, \mathrm{STr}\mathcal{Q}_j^2 - \frac{1}{2} w_T^2\, \sum_{jj^\prime} A_{jj^\prime} \mathrm{STr} \mathcal{Q}_j \mathcal{Q}_{j^\prime}}.
\end{aligned}
\end{equation}
Combining the above with the deterministic site-diagonal terms, we finally obtain:
\begin{equation}
    K_{n,\varepsilon} = \langle -\bar{\zeta}_n\zeta_n \bar{\omega}_n\omega_n\rangle,
\end{equation}
where the state reads:
\begin{equation}
    \langle\bullet_n\rangle = \int\prod_{j} \mathcal{D} \mathcal{Q}_j\ \bullet_n\ \mathrm{e}^{-S},
\end{equation}
and
\begin{align}
 \begin{split}
    &S = {S_{\mathrm reg}} + {S_V} + {S_T}, \quad
    {S_{\mathrm reg}} = \sum_{j} \mathrm{STr}\, (\varepsilon \Sigma_3) \mathcal{Q}_j \,, \cr &{S_V} = \sum_{j} \mathrm{STr} \left( - \mathrm{i} E \mathcal{Q}_j + {\textstyle{\frac{1}{2}}} w_V^2 \mathcal{Q}_j^2 \right) , \quad {S_T} = \frac{1}{2} w_T^2 \sum_{j j^\prime} A_{j j^\prime} \, \mathrm{STr}\, \mathcal{Q}_j \mathcal{Q}_{j^\prime}.
    \end{split}
\end{align}

\subsection{General theory of (extremal) Gibbs states}\label{Interlude}

For more details, we refer to a pedagogical presentation of the main physical ideas \cite{RG-PATH}, and three standard references  \cite{StatMech, RandFields, Conv}. In this section, we generalize the notation and look at a field theory with Hamiltonian function $H$ on a lattice $\mathbb{G}$. We assume that the theory undergoes spontaneous symmetry breaking in a certain parameter regime.
A field configuration is called $\omega$, and the single-site target space is called $\Omega$.  The field configuration lives in the product space: $\omega\in\Omega^\mathbb{G}$. For a finite sublattice $\Gamma\subset\mathbb{G}$, $H_\Gamma$ contains the interaction terms involving the field values in $\Gamma$, including their interactions with the fixed exterior configuration. The inverse tepmerature has been absorbed into $H_\Gamma$. We note that the Hamiltonian function $H_\Gamma$ evaluated on $\omega$, $H_\Gamma(\omega),$ depends on the restriction of $\omega$ to $\Gamma$ and the boundary values connected to $\Gamma$ via the specific interaction of the model (note that in general this could be the restriction to the whole complement $\omega_{\Gamma_c}$). The so-called finite-volume Gibbs state is obtained from the finite-volume partition function
\begin{equation}\label{Gibbs finite partition}
Z_\Gamma(\omega_{\Gamma^c})=\int\prod_{i\in\Gamma}d\mu(\omega_i)\exp\left(-H_\Gamma(\omega)\right),
\end{equation}
where $d\mu$ is some a priori measure on the single-site target space. In our supersymmetric field theory, the analogue of the a priori measure is the Berezin integration form (\ref{eq:Berezin-int}). [Strictly speaking, it is not a positive measure. Nevertheless, we expect an analogous structure of states and their decomposition also in the supersymmetric setting, especially after integrating out the Grassmann variables]. 

Note that after integrating over the field configurations $\omega_i,\ i\in\Gamma$, the expression (\ref{Gibbs finite partition}) only depends on the field on the complement $\omega_{\Gamma^c}$. Now from the partition function (\ref{Gibbs finite partition}) one constructs the finite volume Gibbs state $\pi_\Gamma(\omega,\bullet)$ on $\Gamma$ with boundary condition $\omega_{\Gamma^c}$. It is a functional on physical observables, i.e., measurable functions $\mathcal{O}:\Omega^\mathbb{G}\rightarrow \mathbb{R}$:
\begin{equation}\label{finite volume Gibbs measure}
\pi_\Gamma(\omega_{\Gamma^c},\mathcal{O})=\frac{1}{Z_\Gamma(\omega_{\Gamma^c})}\int\prod_{i\in\Gamma}d\mu(\omega_i)\,\mathcal{O}(\omega)\,\exp\left(-H_\Gamma(\omega)\right).
\end{equation}

We are now in a position to describe Gibbs states on the full configuration space $\Omega^\mathbb{G}$. They arise as a weak limit (in a suitable topology) $\Gamma\rightarrow\mathbb{G}$ of the finite volume states (\ref{finite volume Gibbs measure}), the \textit{thermodynamic limit}. 

In general, the infinite-volume Gibbs states are not unique. This is at the heart of the spontaneous symmetry-breaking mechanism.

Next, we describe the structure of the space $\mathcal{G}(\Pi)$ of infinite volume Gibbs states. $\Pi$ denotes the family of the finite volume states $\{\pi_\Gamma\}_{\Gamma\subset\mathbb{G}}$.
$\mathcal{G}(\Pi)$ is a convex space (every convex combination of two Gibbs states is again a Gibbs state). The states that cannot be written as a convex combination of other ones are called \textit{extremal Gibbs states}. Furthermore, any non-extremal Gibbs state can be written as a convex combination of extremal ones, and this combination is unique; thus $\mathcal{G}(\Pi)$ is a simplex.  

Now, in the phase where spontaneous symmetry breaking is absent, for instance, the insulating phase of the Wegner model, the space $\mathcal{G}$ only consists of fully symmetric elements. In the cases considered here, there is a unique such Gibbs state, which is therefore obviously extremal.

The situation changes when spontaneous symmetry breaking occurs. Now the extremal Gibbs states are non-unique and individually break the symmetry. Within a given symmetry orbit, the extremal Gibbs states are transformed into one another by the symmetry group. Non-extremal Gibbs states are convex combinations (mixtures with respect to a probability measure) of the extremal ones.  

To select an extremal Gibbs state in a field theory, one can proceed as follows. One imposes a collective boundary condition on the finite lattice $\Gamma$, which mimics the desired extremal Gibbs state (``alignment'' of all fields). When spontaneous symmetry breaking occurs, the extremal Gibbs state in the thermodynamic limit still retains information about the chosen boundary condition (the fields on the lattice $\mathbb{G}$ align collectively), whereas this information is lost in the thermodynamic limit when spontaneous symmetry breaking is absent. On the level of the AAT integral equation, Eq.~(\ref{eq:eqn4Y}) this means that, when spontaneous symmetry breaking occurs, the solutions $\mathcal{Y}_{x}(\mathcal{Q})$ depend on the boundary condition $x$, whereas they are independent of $x$ when spontaneous symmetry breaking is absent.

To finish this interlude on the theory of extremal Gibbs states, we give two short examples. The prototype example is the Ising model on $\mathbb{Z}^d,\ d\geq 2$. At low temperature, $\mathbb{Z}_2$ spontaneous symmetry breaking occurs, and there are two distinct extremal Gibbs states $\langle\bullet\rangle^+$ (spin up) and $\langle\bullet\rangle^-$ (spin down) \cite{Geo2011}. Hence the set of (translation-invariant) extremal Gibbs states can be identified with $\mathbb{Z}_2$. A general (translation-invariant) Gibbs state is the convex combination
\begin{equation}
\langle\bullet\rangle=t\langle\bullet\rangle^+ +(1-t)\langle\bullet\rangle^-,\ 0\leq t\leq 1.
\end{equation}
For the $\mathbb{Z}_2$ symmetric state, one chooses $t=1/2$. 
\begin{figure}[ht]
\centering
\begin{minipage}[c]{0.55\textwidth}
\includegraphics[width=0.4\textwidth]{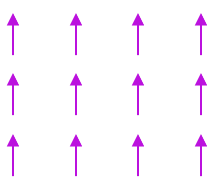}
\end{minipage}
\begin{minipage}{0.4\textwidth}
\caption{Sketch of (an extract of) a typical field configuration in the extremal Gibbs state $\langle\bullet\rangle^+$ for the Ising model.}
\end{minipage}
\label{Pic Ising}
\end{figure}

An example with a continuous but compact symmetry group is the $\mathrm{SO}(3)$ Heisenberg antiferromagnet on $\mathbb{Z}^d,\ d\geq 3$. At low temperature, the symmetry is spontaneously broken \cite{Dys78} and the Gibbs states are non-unique. The set of extremal Gibbs states can be identified with the order parameter space, the unit sphere $S^2$. Each extremal Gibbs state $\langle\bullet\rangle_\mathbf{n}$ exhibits long-range N{\'e}el order with staggered magnetization pointing along the unit vector $\mathbf{n}$. Any other Gibbs state is again a convex mixture of the extremal ones:
\begin{equation}
\langle\bullet\rangle_\mu=\int_{S^2}d\mu(\mathbf{n})\langle\bullet\rangle_\mathbf{n},
\end{equation}
where $\mu$ is an arbitrary probability measure on $S^2$. For the fully symmetric state, one chooses the rotation-invariant one on $S^2$.
\begin{figure}[ht]
\centering
\begin{minipage}[c]{0.55\textwidth}
\includegraphics[width=0.4\textwidth]{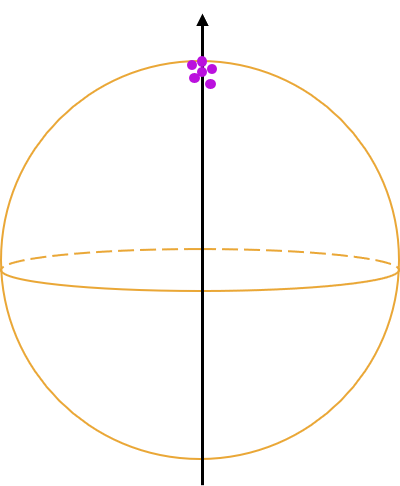}
\end{minipage}
\begin{minipage}{0.4\textwidth}
\caption{Sketch of a typical field configuration in the extremal Gibbs state $\langle\bullet\rangle_{\mathbf{n}=e_z}$ in the Heisenberg model. The field clusters around the north pole. When tuning the temperature, $T\nearrow T_c$, the clusters become larger until at $T=T_c$ they restore the symmetry.}
\end{minipage}
\label{Pic Heisenberg}
\end{figure}

One purpose of the present work is to generalize the compact case, such as the example above, to a situation where the symmetry group is non-compact.

\subsection{Expressing the Laplacian (\ref{eq:def-Delta}) in terms of $K_i, D$}\label{Laplace manipulation}

One proves the identity (\ref{eq:Delta-VF}) by a straightforward calculation. We start by expressing
\begin{equation}
 \Delta^\flat=K_3^2-K_2^2-K_1^2+D^2+D
\end{equation}
 in terms of the coordinate vector fields $\frac{\partial}{\partial x_i}\equiv\partial_{x_i}$.
\begin{equation}
\begin{aligned}
&K_3^2=(x_1\,\partial_{x_2}-x_2\,\partial_{x_1})\,(x_1\,\partial_{x_2}-x_2\,\partial_{x_1})=x_1^2\,\partial_{x_2}^2+x_2^2\,\partial_{x_1}^2-x_1\,\partial_{x_1}-x_2\,\partial_{x_2}-2x_1\,x_2\,\partial_{x_1}\,\partial_{x_2},\\&K_2^2=(x_3\,\partial_{x_1}+x_1\,\partial_{x_3})(x_3\,\partial_{x_1}+x_1\,\partial_{x_3})=x_3^2\,\partial_{x_1}^2+x_1^2\,\partial_{x_3}^2+x_3\,\partial_{x_3}+x_1\,\partial_{x_1}+2x_1\,x_3\,\partial_{x_1}\,\partial_{x_3},\\&
K_1^2=(x_2\,\partial_{x_3}+x_3\,\partial_{x_2})(x_2\,\partial_{x_3}+x_3\,\partial_{x_2})=x_2^2\,\partial_{x_3}^2+x_3^2\,\partial_{x_2}^2+x_2\,\partial_{x_2}+x_3\,\partial_{x_3}+2x_2\,x_3\,\partial_{x_2}\,\partial_{x_3},\\&
D^2+D=\left(\sum_{\mu=0}^3 x_\mu\,\partial_{x_\mu}\right)^2+\sum_{\mu=0}^3 x_\mu\,\partial_{x_\mu}=x_0^2\,\partial_{x_0}^2+x_1^2\,\partial_{x_1}^2+x_2^2\,\partial_{x_2}^2+x_3^2\,\partial_{x_3}^2+2x_0\,\partial_{x_0}+\\&+2x_1\,\partial_{x_1}+2x_2\,\partial_{x_2}+2x_3\,\partial_{x_3}
+2x_0\,x_1\,\partial_{x_0}\,\partial_{x_1}+2x_0\,x_2\,\partial_{x_0}\,\partial_{x_2}+2x_0\,x_3\,\partial_{x_0}\,\partial_{x_3}+\\&+2x_1\,x_2\,\partial_{x_1}\,\partial_{x_2}+2x_1\,x_3\,\partial_{x_1}\,\partial_{x_3}+2x_2\,x_3\,\partial_{x_2}\,\partial_{x_3},\\&\Delta^\flat=(x_3^2-x_2^2-x_1^2)\,(\partial_{x_3}^2-\partial_{x_2}^2-\partial_{x_1}^2)+x_0^2\,\partial_{x_0}^2+2x_0\,\partial_{x_0}+2x_0\,x_1\,\partial_{x_0}\,\partial_{x_1}+\\&+2x_0\,x_2\,\partial_{x_0}\,\partial_{x_2}+2x_0\,x_3\,\partial_{x_0}\,\partial_{x_3}.
\end{aligned}
\end{equation}
Next we rewrite the middle term of Eq.\ (\ref{eq:Delta-VF}):
\begin{equation}
\begin{aligned}
&2D\circ x_0\,\partial_{x_0}=2\left(\sum_{\mu=0}^3x_\mu\,\partial_{x_\mu}\right)\circ x_0\,\partial_{x_0}=2x_0^2\,\partial_{x_0}^2+2x_0\,\partial_{x_0}+2x_0\,x_1\,\partial_{x_0}\,\partial_{x_1}+\\&+2x_0\,x_2\,\partial_{x_0}\,\partial_{x_2}+2x_0\,x_3\,\partial_{x_0}\,\partial_{x_3}.
\end{aligned}
\end{equation}
Finally, we obtain:
\begin{equation}
\begin{aligned}
&\frac{1}{4}(x_3^2-x_2^2-x_1^2)^{-1}\left(\Delta^\flat-2D\circ x_0\,\partial_{x_0}+(x_0^2+x_1^2+x_2^2-x_3^2)\,\partial_{x_0}^2\right)\\&=\frac{1}{4}(x_3^2-x_2^2-x_1^2)^{-1}\,(x_3^2-x_2^2-x_1^2)\,(\partial_{x_3}^2-\partial_{x_2}^2-\partial_{x_1}^2-\partial_{x_0}^2)\\&=\frac{1}{4}\left(\partial_{x_3}^2-\partial_{x_0}^2-\partial_{x_1}^2-\partial_{x_2}^2\right)=\Delta.
\end{aligned}
\end{equation}

\subsection{Analysis of boundary terms from partial integration of $\Delta$}\label{App. boundary}
In the following, we anticipate the asymptotic form of the perturbations $\delta Y_\lambda$ derived in Sect.\ \ref{Harmonic analysis} and show explicitly that, for perturbations of this form, all boundary terms at infinity generated by the partial integration in Eq.\ (\ref{eq: linearized eq involving Delta_0 in MF coordinates}) vanish.
The asymptotic form of $\delta Y_\lambda$ for $p+q\to\infty$, $\lambda\ll 1$, while $p-q$ remains finite is (see Eq.~(\ref{delta Yasy})):
\begin{equation}
    \delta Y_\lambda(p,q)\sim\,-\frac{1}{\lambda}\sqrt{p+q}\,f_\lambda(p-q)\,\sin\big(\Phi_\lambda(p,q)\big),\qquad \Phi_\lambda(p,q)=\lambda\ln\left(\frac{2(p+q)}{|p-q|}\right)-a\lambda
\end{equation}
where $f_\lambda$ exhibits Gaussian decay when $|p^\prime-q^\prime|$ becomes large compared with its characteristic scale, and $a=2-2\ln(2)$. To investigate the boundary terms at infinity, we introduce a cutoff $M$ for the respective boundary term and subsequently send $M\to\infty$. By noting that we can restrict to $p,q>0$ ($p^\prime$, $q^\prime$ are the integration variables), and using the asymptotic expansion of modified Bessel functions:
\begin{equation}
    I_n(z)=\frac{\mathrm{e}^z}{\sqrt{2\pi z}}(1+\mathcal{O}(z^{-1})),\quad z\to\infty,
\end{equation}
these terms are
\begin{equation}\label{App. boundary terms}
  \frac{1}{\sqrt{4\pi w_T^2}}(pq)^{-\frac{1}{4}}\left( \int_0^\infty dq^\prime\,\frac{\mathrm{e}^{-w_T^2(\sqrt{pM}-\sqrt{qq^\prime})^2}}{(Mq^\prime)^\frac{1}{4}}\,\frac{\partial}{\partial q^\prime}\,\delta Y_\lambda(M,q^\prime)-\int_0^\infty dp^\prime\,\frac{\partial}{\partial p^\prime}\left\{\frac{\mathrm{e}^{-w_T^2(\sqrt{pp^\prime}-\sqrt{qM})^2}}{(p^\prime M)^\frac{1}{4}}\right\}\,\delta Y_\lambda(p^\prime, M)\right).
\end{equation}
We analyze the two summands in (\ref{App. boundary terms}) separately. In the first one, the relevant contribution to the integral comes from the region $q^\prime=\frac{p}{q}M+\mathcal{O}(M^\frac{1}{2}).$ Here the exponential kernel does not cause rapid decay. Since $\delta Y$ decays exponentially in $p^\prime-q^\prime$, it follows that $p=q\equiv x$, otherwise the boundary term vanishes. The same can be concluded from inspection of the second term in (\ref{App. boundary terms}). So, we can set $q^\prime=M-s$ in the first summand and $p^\prime=M+s$ in the second, where, actually, $s=\mathcal{O}(1)$ due to the exponential decay of $f$. Injecting these conclusions into (\ref{App. boundary terms}) and changing coordinates to $s$, we obtain
\begin{equation}\label{App. boundary terms 2}
    \frac{1}{\sqrt{4\pi w_T^2 x}}\left(-\int_{-\infty}^M ds \frac{\mathrm{e}^{-w_T^2 x(\sqrt{M}-\sqrt{M-s})^2}}{M^\frac{1}{4}(M-s)^\frac{1}{4}}\frac{\partial}{\partial s}\delta Y_\lambda(M,M-s)-\int^\infty_{-M} ds\frac{\partial}{\partial s}\left\{\frac{\mathrm{e}^{-w_T^2x(\sqrt{M}-\sqrt{M+s})^2}}{M^\frac{1}{4}(M+s)^\frac{1}{4}}\right\}\delta Y_\lambda(M+s,M)\right).
\end{equation}
We now expand the two summands in $M$. For the first, we note that
\begin{equation}
    \frac{\partial}{\partial s}\delta Y_\lambda(M,M-s)=-\frac{1}{\lambda}\,\sqrt{2M}\,\frac{\partial}{\partial s }\big\{f_\lambda(s)\,\sin\big(\Phi_\lambda(M,M-s)\big)\big\}+\mathcal{O}_s(M^{-\frac{1}{2}}),
\end{equation}
where the subscript $s$ indicates coefficients depending polynomially on the $\mathcal{O}(1)$ variable $s$,
and
\begin{equation}\label{App. kernel expansion}
    \frac{\mathrm{e}^{-w_T^2x(\sqrt{M}-\sqrt{M-s})^2}}{M^\frac{1}{4}(M-s)^\frac{1}{4}}=M^{-\frac{1}{2}}\left\{1+\mathcal{O}_s(M^{-1})\right\}.
\end{equation}
The first summand in (\ref{App. boundary terms 2}) then becomes:
\begin{equation}\label{int}
\begin{aligned}
    &\int_{-\infty}^M ds \frac{\mathrm{e}^{-w_T^2 x(\sqrt{M}-\sqrt{M-s})^2}}{M^\frac{1}{4}(M-s)^\frac{1}{4}}\frac{\partial}{\partial s}\delta Y_\lambda(M,M-s)= -\frac{\sqrt{2}}{\lambda}\int_{-\infty}^M ds\,\frac{\partial}{\partial s }\big\{f_\lambda(s)\,\sin\big(\Phi_\lambda(M,M-s)\big)\big\}+\mathcal{O}(M^{-1})\\&=-\frac{\sqrt{2}}{\lambda}\,f_\lambda(M)\,\sin\big(\Phi_\lambda(M,0)\big)+\mathcal{O}(M^{-1})\to 0.
    \end{aligned}
\end{equation}
Strictly speaking, the integrals in (\ref{int}) require a symmetric regularization of the logarithmic singularity at $s=0$. The singular contributions from $s>0$ and $s<0$ cancel. Note also that the Gaussian decay of $f_\lambda(s)$ restricts the relevant integration region to $s=\mathcal{O}(1)$, where the asymptotic expansion of $\delta Y_\lambda$ is valid. The contributions from the remaining integration domain, including the endpoint $s=M$, are exponentially suppressed and do not affect the result.
For the second summand in (\ref{App. boundary terms 2}) we note that
\begin{equation}
    \delta Y_\lambda(M+s,M)= -\frac{1}{\lambda}\sqrt{2M}\,f_\lambda(s)\,\sin\big(\Phi_\lambda(M+s,M)\big)+\mathcal{O}_s(M^{-\frac{1}{2}}),
\end{equation}
and
\begin{equation}
   \frac{\partial}{\partial s}\left\{\frac{\mathrm{e}^{-w_T^2x(\sqrt{M}-\sqrt{M+s})^2}}{M^\frac{1}{4}(M+s)^\frac{1}{4}}\right\}=\frac{\mathrm{e}^{-w_T^2x(\sqrt{M}-\sqrt{M+s})^2}}{M^\frac{1}{4}(M+s)^\frac{1}{4}}\left\{-\frac{1}{4(M+s)}-w_T^2x\left(1-\sqrt{\frac{M}{M+s}}\right)\right\}=\mathcal{O}_s(M^{-\frac{3}{2}}),
\end{equation}
so this summand vanishes as $M^{-1}$. For $\lambda\to 0$, $\lambda^{-1}\sin\big(\Phi_\lambda(p,q)\big)$ produces an additional logarithmic factor in $M$, so that the boundary terms still vanish.

\subsection{Calculation of $\mathcal{I}_\lambda$}\label{g integral in horospherical coordinates}
Here we calculate the remaining integral in Eq.~(\ref{Ilam}):
\begin{equation}\label{last g integral harmonic analysis}
\begin{aligned}
	 \mathcal{I}_\lambda(m,m^\prime) = 2 \int_0^\infty \!\!\!\! d\theta^\prime \sinh\theta^\prime \bigg(1 + w_T^2 \frac{\partial}{\partial w_T^2} \bigg) \, \mathrm{e}^{- w_T^2 \Tr q_0 Q^\prime} \, \varphi_\lambda (\theta^\prime) ,
	 \end{aligned}
\end{equation}
where 
\begin{equation}
    q_0 = \frac{1}{w_T} \big( m \cdot \mathbf{1} + |m|\, \sigma_3 \big). 
\end{equation}
An efficient way to calculate it is to change coordinates to $t:H^2\rightarrow\bbR$, and $\widetilde{b}:H^2\rightarrow\bbR$. These are connected to hyperbolic polar coordinates as follows:
\begin{equation}\label{App.coordinates}
\begin{gathered}
\cosh\theta+\sinh\theta \cos\phi=\mathrm{e}^t,\quad \cosh\theta=\cosh t+\frac{1}{2}\mathrm{e}^t \widetilde{b}^2,\\ \frac{1}{\pi}\sinh\theta\,d\theta\,d\phi=\frac{1}{\pi}\mathrm{e}^t\,dt\,d\widetilde{b}.
\end{gathered}
\end{equation}
We recall the integral expression for the zonal spherical functions:
\begin{equation}\label{zonal spherical function polar}
	 \varphi_{\lambda}(\theta)=\int_0^{2\pi}\frac{\mathrm{d}\phi}{2\pi}(\cosh\theta+\sinh\theta\,\cos\phi)^{-\frac{1}{2}-\mathrm{i}\lambda}.
\end{equation}
By inserting the integral formula (\ref{zonal spherical function polar}) into Eq.~(\ref{last g integral harmonic analysis}), and using that the integral kernel is independent of $\phi$, we can exchange $\phi$ integration with the multiplication by the integral kernel. It follows that we can express the whole integral in the coordinates (\ref{App.coordinates}). We arrive at:
\begin{equation}
\begin{aligned}
	& \mathcal{I}_\lambda(m,m^\prime)=\frac{1}{\pi}\int_\bbR\int_\bbR \mathrm{e}^{t^\prime}dt^\prime d\widetilde{b}^\prime\big(1+w_T^2\frac{\partial}{\partial w_T^2}\big) \mathrm{e}^{- w_T^2 \Tr q_0 Q^\prime}\times\\& \times\mathrm{e}^{-\frac{t^\prime}{2}-\mathrm{i}\lambda t^\prime}.
	 \end{aligned}
\end{equation}
We now perform the integrals:
\begin{equation}
\begin{aligned}
\mathcal{I}_\lambda(m,m^\prime)&= \frac{1}{\sqrt{\pi}}\left(1+\frac{\partial}{\partial s}\bigg|_{s=1}\right)\int_\mathbb{R}dt^\prime \frac{1}{\sqrt{s |mm^\prime|}}\mathrm{e}^{-2 s mm^\prime-2 s |mm^\prime|\cosh t^\prime}\cos(\lambda t^\prime)\\&=\frac{2}{\sqrt{\pi}}\left(1+\frac{\partial}{\partial s}\bigg|_{s=1}\right)\frac{1}{\sqrt{s |mm^\prime|}}\mathrm{e}^{-2 s mm^\prime}K_{\mathrm{i}\lambda}(2 s |mm^\prime|),
\end{aligned}
\end{equation}
where $K$ denotes the modified Bessel function of the second kind, and the parametric derivative with respect to $s$ is performed for fixed $m, m^\prime, t^\prime$ . Finally, performing the parameter derivative yields the desired result~(\ref{eq: final reduced m equation 1}):
\begin{equation}
\begin{aligned}
\mathcal{I}_\lambda(m,m^\prime) = \frac{\mathrm{e}^{-2 m m^\prime}}{\sqrt{\pi | m m^\prime |}}
    \bigg( (1-4 mm^\prime) K_{i\lambda}(2 |mm^\prime|) - 2 |mm^\prime| \big( K_{1+i\lambda} + K_{1-i\lambda} \big) (2 |mm^\prime|) \bigg)
\end{aligned}
\end{equation}

\subsection{Scaling of $\lambda^\ast$}\label{app:Scaling gamma ast}

We here argue how $\lambda^\ast$ scales with the disorder $w$. For $\lambda\in\mathbb{R}$ the eigenvalues $\nu(\lambda)$ of the linearized operator (\ref{eq:eqn4deltaYBB}) at disorder $w$, as a function of the harmonic parameter $\lambda$, have a maximum at $\lambda=0$. Consequently, we have for small $\lambda$:
\begin{equation}
\nu(\lambda)=\nu(0)-\frac{1}{2}\,|\nu''(0)|\,\lambda^2+...
\end{equation}
Now $\lambda^\ast$ fulfills:
\begin{equation}\label{gamma ast}
\lambda^\ast \approx \sqrt{2}\sqrt{\frac{\nu(0)-1}{|\nu''(0)|}}.
\end{equation}
In principle, we could calculate $\lambda^\ast$ exactly from the result for the eigenvalues $\nu(\lambda)$ from the harmonic analysis. However, here we are interested only in the scaling, which can be anticipated with a faster argument. For $w\nearrow w_c$, the denominator in Eq.\ (\ref{gamma ast}) is nonzero (by assuming a non-degenerate maximum), whereas the numerator goes to zero.  From the harmonic analysis, we know that (up to a logarithmic correction)
\begin{equation}
\nu(0)\sim \frac{K}{w}.
\end{equation}
For $w_c-w\ll 1$ we expand $\nu(0)$ around $w_c$ and obtain:
\begin{equation}
\nu(0)\approx 1+\frac{w_c-w}{w_c}.
\end{equation}
For $\lambda^\ast$, we find the scaling:
\begin{equation}
\lambda^\ast\sim\sqrt{w_c-w},\quad w_c-w\ll 1.
\end{equation}

\end{document}